\documentclass[
10pt,
showpacs,preprintnumbers,nofootinbib,
 amsmath,amssymb,
 aps,
twocolumn,groupedaddress,superscriptaddress,
floatfix,
showkeys
]{revtex4-1}

\usepackage[labelfont={small},subrefformat=parens,caption=false]{subfig}
\usepackage[dvipsnames]{xcolor}
\usepackage{amsfonts,amsmath,amssymb,bm}
\usepackage{graphicx}
\usepackage[utf8]{inputenc}
\usepackage{xspace}
\usepackage{isotope}
\usepackage{xparse}
\usepackage{aas_macros}

\usepackage[pdfpagelabels, pdfencoding=auto, psdextra]{hyperref}
\hypersetup{%
 pdfsubject=Paper,
 pdfkeywords={nuclear physics} {Bayesian} {chiral EFT} {nuclear matter},
 unicode = true,
 breaklinks = true,
 colorlinks = true,
 linkcolor = blue,
 citecolor = blue,
 menucolor = blue,
 citecolor = blue,
 urlcolor = blue
}

\graphicspath{{./figures/}}

\usepackage{silence}
\makeatletter
\newcommand\newsubcommand[3]{\newcommand#1{#2\sc@sub{#3}}}
\def\sc@sub#1{\def\sc@thesub{#1}\@ifnextchar_{\sc@mergesubs}{_{\sc@thesub}}}
\def\sc@mergesubs_#1{_{\sc@thesub#1}}

\newcommand\newsupcommand[3]{\newcommand#1{#2\sc@sup{#3}}}
\def\sc@sup#1{\def\sc@thesup{#1}\@ifnextchar^{\sc@mergesups}{^{\sc@thesup}}}
\def\sc@mergesups^#1{^{\sc@thesup#1}}
\makeatother

\DeclareMathAlphabet{\mathbcal}{OMS}{cmsy}{b}{n}

\newcommand{\ordervec}{\vec}

\newcommand{\inputvec}{\mathbf}

\newsubcommand{\ckvec}{\ordervec{c}}{k}

\newsubcommand{\bkvec}{\ordervec{b}}{k}
\newsubcommand{\ckvecset}{\ordervec{\inputvec{c}}}{k}

\newsubcommand{\ckvecapprox}{\mathbf{c}'}{k}
\newsubcommand{\ckvecapproxset}{\mathbf{C}'}{k}

\newsubcommand{\bkvecapprox}{\mathbf{b}'}{k}
\newsubcommand{\bkvecset}{\mathbf{B}}{k}
\newsubcommand{\bkvecapproxset}{\mathbf{B}'}{k}

\newcommand{\genobs}{y}

\newsubcommand{\genobsvec}{\ordervec{\genobs}}{k}
\newsubcommand{\genobsvecset}{\ordervec{\inputvec{\genobs}}}{k}

\newsubcommand{\akvec}{\mathbf{a}}{k}

\newsubcommand{\akvecapprox}{\mathbf{a}'}{k}
\newsubcommand{\akvecset}{\mathbf{A}}{k}
\newsubcommand{\akvecapproxset}{\mathbf{A}'}{k}

{}  % Remove the definition from the Physics package

\newcommand{\thetavec}{\bm{\theta}}

\def\diffd{\mathrm{d}}  % Upright differentials
\DeclareDocumentCommand\differential{ o g d() }{ % Differential 'd'
    \IfNoValueTF{#2}{
        \IfNoValueTF{#3}
            {\diffd\IfNoValueTF{#1}{}{^{#1}}}
            {\mathinner{\diffd\IfNoValueTF{#1}{}{^{#1}}\argopen(#3\argclose)}}
        }
        {\mathinner{\diffd\IfNoValueTF{#1}{}{^{#1}}#2} \IfNoValueTF{#3}{}{(#3)}}
    }
\newcommand{\pathd}{\mathcal{D}}  % differential symbol for path integrals

\DeclareDocumentCommand\pathdifferential{ o g d() }{ % Path 'D'
    \IfNoValueTF{#2}{
        \IfNoValueTF{#3}
            {\pathd\IfNoValueTF{#1}{}{^{#1}}}
            {\mathinner{\pathd\IfNoValueTF{#1}{}{^{#1}}\argopen(#3\argclose)}}
        }
        {\mathinner{\pathd\IfNoValueTF{#1}{}{^{#1}}#2} \IfNoValueTF{#3}{}{(#3)}}
    }

\newcommand{\Hhat}{H}
\newcommand{\Ohat}{O}

\newcommand{\Vhat}{V}

\newcommand{\myvec}{\boldsymbol}

\newcommand{\psifull}{\Psi}
\newcommand{\psifullsub}{\Psi^\mathrm{sub}}
\newcommand{\psifullasym}{\Psi^\mathrm{asym}}
\newcommand{\tildepsifullasym}{\tilde{\Psi}^\mathrm{asym}}
\newcommand{\psifd}{\psi} 
\newcommand{\psifdsub}{\psi^\mathrm{sub}} 
\newcommand{\psifdasym}{\psi^\mathrm{asym}}  
\newcommand{\tildepsifdasym}{\tilde{\psi}^\mathrm{asym}}
\newcommand{\psifree}{\phi}
\newcommand{\psidimer}{\varphi}
\newcommand{\deltabar}{\overline{\delta}}

\newcommand{\Pini}{P_\mathrm{in}}
\newcommand{\vecPf}{\myvec{P}_\mathrm{out}}
\newcommand{\vecPini}{\myvec{P}_\mathrm{in}}
\newcommand{\bsren}{\xi}
\newcommand{\LAMtwo}{\Lambda_2}
\newcommand{\perm}{\hat{\mathcal{P}}}
\newcommand{\varfun}{\mathcal{F}}
\newcommand{\varconsfun}{\mathcal{G}}
\DeclareMathOperator\erf{Erf}
\DeclareMathOperator\erfi{Erfi}
\DeclareMathOperator\erfc{Erfc}

\newcommand{\ex}{\text{ex}}
\newcommand{\lambdaLM}{\lambda_{\text{L}}}
\newcommand{\Nproj}{N_{\text{proj}}}
\newcommand{\lambdacut}{\Lambda_{\text{cut}}}

\usepackage{amsmath,environ}
\NewEnviron{EqS}{%
\begin{equation}\begin{split}
  \BODY
\end{split}\end{equation}
}
\NewEnviron{EqSn}{%
\begin{equation*}\begin{split}
  \BODY
\end{split}\end{equation*}
}
\begin{document}

\title{Fast emulation of quantum three-body scattering}

\author{Xilin Zhang}
\email{zhangx@frib.msu.edu}
\affiliation{Facility for Rare Isotope Beams, Michigan State University, MI 48824, USA}
\affiliation{Department of Physics, The Ohio State University, Columbus, OH 43210, USA}

\author{R.~J. Furnstahl}
\email{furnstahl.1@osu.edu}
\affiliation{Department of Physics, The Ohio State University, Columbus, OH
43210, USA}

\date{October 8, 2021}

\begin{abstract}
We develop a class of emulators for solving quantum three-body scattering problems. 
They are based on  combining the variational method for scattering observables and the recently proposed eigenvector continuation concept. 
The emulators are first trained by the exact scattering solutions of the governing Hamiltonian at a small number of points in its parameter space, and then employed to make interpolations and extrapolations in that space. Through a schematic nuclear-physics model with finite-range two and three-body interactions, we demonstrate the emulators to be extremely accurate and efficient. 
The computing time for emulation is on the scale of milliseconds (on a laptop), with relative errors ranging from $10^{-13}$ to $10^{-4}$ depending on the case. The emulators also require little memory. 
We argue that these emulators can be generalized to even more challenging scattering problems. Furthermore, this general strategy may be applicable for building the same type of emulators in other fields, wherever variational methods can be developed for evaluating physical models.
\end{abstract}

\maketitle

\section{Introduction}

Methods to solve quantum three-body scattering problems, i.e., the Faddeev approach~\cite{Glockle:1983,Gloeckle:1995jg,Nielsen:2001zz,Deltuva:2012kt}, are well developed but are complex and computationally expensive in terms of both time  and memory costs.
This is a severe barrier to applications that require many evaluations in the parameter space of  the governing Hamiltonian $H$, such as Bayesian uncertainty quantification.
An alternative to direct solutions of the three-body equations for each parameter set is to use \emph{emulators}.
These are models trained on the full solutions for a small number of parameter sets that provide predictions elsewhere in the parameter space for a small fraction of the computational time and memory requirements.
In this work, we demonstrate effective emulation of three-body scattering by generalizing the eigenvector continuation (EC) method originally proposed for bound-state calculations~\cite{Frame:2019jsw}.

Emulators based on Gaussian processes (GP)~\cite{Mackay:1998introduction, rasmussen2006gaussian} and on neural networks (e.g.,~\cite{Bogojeski2020,2020MNRAS.494.2465B}) have been widely used in physics and other fields.
A different type of emulator is based on EC applied to variational calculations.
In short, the eigenvectors from the training solutions can be used as an extremely effective basis for variational estimates.
In nuclear physics, fast and accurate EC emulators have been developed recently for few- and many-body ground states~\cite{Konig:2019adq,Ekstrom:2019lss,Wesolowski:2021cni}, excited states~\cite{Franzke:2021ofs}, and transitions~\cite{Wesolowski:2021cni}; for shell-model calculations~\cite{Yoshida:2021jbl}; and, of particular relevance here, for two-body scattering~\cite{Furnstahl:2020abp,Melendez:2021lyq,Drischler:2021qoy}.
They are particularly efficacious for Bayesian statistical analyses, where extensive sampling in the parameter space is typically needed for parameter estimation, propagation of errors to observables, and experimental design.

An extension of EC emulators to three-body scattering would have broad impact.
Potential applications in nuclear physics include parameter estimation for chiral effective field theory~\cite{Epelbaum:2008ga,Hebeler:2020ocj}---which is the foundation for modern {\it ab initio} calculations---and for phenomenological reaction models needed in analyzing data from rare isotope facilities~\cite{King:2019sax}; analyzing Lattice QCD simulations of three-nucleon or hadron systems~\cite{Eliyahu:2019nkz}; and facilitating the generalization of a new approach for {\it ab initio} scattering calculations~\cite{Zhang:2020rhz} to three-cluster systems. 

Considered in a even broader context, three-body scattering, and particle scattering from a two-particle bound system in particular, represents a process of one particle probing a compound target. 
The EC emulators with proper generalization for the compound systems could  expedite scattering data analyses and the extraction of the structure information for a wide range of targets.
This study is the first step towards constructing EC emulators for general reactions including inelastic and breakup processes. More generally, as discussed in the end of this paper, a complete set of EC emulators would enable a new type of workflow
in nuclear physics and beyond. 

The variational method for quantum ground states is well-known from introductory courses in quantum mechanics.
Less familiar are the variational approaches for scattering observables due to Kohn, Schwinger, and others.
The Kohn variational principle (KVP) is based on a stationary functional at fixed energy.
It does not provide bounds to S-matrix elements but nevertheless has been effectively applied, particular in atomic physics~\cite{nesbet1980variational}.
As with any variational approach, the key to success is choosing a good ansatz or basis for the trial function.
The secret of EC is that the subspace covered by the wave functions of interest when sweeping through the parameters is a much smaller space than the full Hilbert space of wave functions. 
This subspace is well-spanned by a small subset of these wave functions (the training points), which comprise the EC trial function.
The effectiveness of EC emulation applying the KVP has been demonstrated for two-body scattering observables~\cite{Furnstahl:2020abp}, and for $R$~matrix theory calculations of fusion observables~\cite{Bai:2021xok}.

Here we extend the EC with KVP to three-body scattering and give a proof-of-principle demonstration for spin-zero bosons with separable potentials. 
The relative errors for emulating $S$ matrix elements and the computing time and memory costs for each emulation are 
summarized in Table~\ref{tab:results_summary} for three different cases. The time cost is generally milliseconds on a laptop, and the memory costs are tiny. 
When the Hamiltonian's parametric dependence can be factorized from individual potential operators (the ``linear" case in the table), e.g., with linear coupling strengths for the potential, the EC emulator works at its full advantage with an extremely small error. 
The accuracy deteriorates  when varying the interaction operators (``nonlinear-1"), e.g., the ranges of interactions, and varying the structure of the subclusters (``nonlinear-2"), but they are still far better than needed in most nuclear physics applications. 
For emulating more complicated \emph{realistic} calculations, the costs and accuracy would be similar. This is elaborated further in Sec.~\ref{sec:summary}.

In comparison, solving three-body scattering Faddeev equations is computationally expensive. For example, nucleon-deuteron ($N$-$d$) scattering is generally computed using supercomputers~\cite{Gloeckle:1995jg,Witala:2012te,2016CoPhC.204..121P}.%
\footnote{In Ref.~\cite{2016CoPhC.204..121P}, the run time was of order $10^3$ seconds for a single calculation on a personal computer. To achieve more complete calculation by including three-nucleon interactions, it would require tens of GB up to a TB of memory to simply store the extra interaction  ~\cite{Hebeler:2020ocj}. Note that trained emulators do not need to store these physical details.}.
Performing such calculations  many thousands or even millions of times, as required in (Bayesian) data analysis and experimental design, is so expensive that no such calculation has appeared to date. The emulators developed  here can solve this bottleneck issue.

\begin{table}[!h]
    \centering
    \begin{tabular}{ | c | c|  c | c|  } 
     \hline   
  EC emulators    &  $S$ Relative error &  Time & Memory \\ \hline 
  linear\footnote{Note that the dependence of the scattering observable on these parameters can be highly nonlinear even in the ``linear" case.}  & $10^{-14}\sim 10^{-13}$ & ms & $<$ MB \\ \hline   
  nonlinear-1   &
   $10^{-6}\sim 10^{-5}$   & ms & MB \\ \hline
  nonlinear-2   
       & $10^{-4}$ & ms & 10s MB \\ \hline   
    \end{tabular}
    \caption{A summary of the accuracy, and  time and memory costs of the emulators that we have developed for varying the interaction parameters. The values listed here are representative and reflect their means for test points in the parameter space and for the typical numbers of trainings we employed in this work. Details can be found in Sec.~\ref{sec:3body_emulators}.}
    \label{tab:results_summary}
\end{table}

In Sec.~\ref{sec:EC_emulators}, we elaborate on the formulation of emulators combining EC and KVP. 
The necessary elements of three-body scattering and the KVP are reviewed in Sec.~\ref{sec:three-body_scattering} and exemplary results for the EC emulators are shown in Sec.~\ref{sec:3body_emulators}.
Our summary and outlook is in Sec.~\ref{sec:summary}.
Additional details on conventions, three-body scattering wave functions, and EC emulators are given in the appendices and Supplementary Material (SM).
The self-contained set of codes that can be used to generate our results will be made public~\cite{BUQEYEgithub}.

\section{Variational-plus-EC emulation}\label{sec:EC_emulators}

The general emulator strategy we use starts with a functional $\varfun_{\thetavec}[\psifull]$ characterizing the physical system of interest.
Here $\psifull$ specifies a state of the system and $\thetavec$ is a vector of parameters to be varied for emulation. 
For example, $\thetavec$ could determine the Hamiltonian and $\psifull$ could be the ground-state eigenvector or a scattering state solution at a specified energy.%
\footnote{It should be emphasized that the $\psifull$ are not limited to wave functions, but can be general objects that are variates for the functionals. In Ref.~\cite{Melendez:2021lyq}, the object is the scattering $K$-matrix.} 
The functional is stationary at the exact $\psifull$, but not necessarily an extremum.
The insight from EC, or the reduced basis method~\cite{Rheinboldt1993849,ChenRBA2017} more generally, is in constructing a trial basis for $\psifull$ that enables the relevant physical observables to be calculated more efficiently for many instances of $\thetavec$ than possible by direct solution.

That is, EC suggests that a very effective way  of choosing a variational trial function $\psifull_t$ for solving the problem at $\thetavec$ is to use  
\begin{align}
    \psifull_t = \sum_{i=1}^{N_b} c_i \,\psifull^{(i)} , \label{eq:construct_trial_psi}
\end{align}
where the $\psifull^{(i)}$ are exact solutions determined at $N_b$ parameter vectors
$\thetavec_i$ for $i = 1, 2,...,N_b$. 
If there are constraints on the trial state in the form  $\varconsfun[\psifull_t] = 0 $, they can be enforced with the Lagrange multiplier method; 
the final functional we need to work with is $\varfun_{\thetavec}[\psifull_t] + \lambdaLM \varconsfun[\psifull_t]$. 
To find the stationary point with $\psifull_t$, we need to find $c_i $ and $\lambdaLM$ that make the first derivatives zero, i.e., 
\begin{align}
    \frac{\partial \bigl(\varfun_{\thetavec}[\psifull_t] + \lambdaLM \varconsfun[\psifull_t] \bigr) }{ \partial c_i } & = 0  , \\ 
    \varconsfun[\psifull_t] & = 0 .
\end{align}
Knowing the $c_i$, we can compute  $\psifull_t$ and evaluate the functional to get estimates of  quantities of interest, such as the energy.
The efficiency of EC originates from $N_b$ being a small number in practice to get high accuracy, much smaller than needed for a typical basis spanning the full Hilbert space.

\subsection{Many-body bound state emulators}

In  few- and many-body bound-state calculations, the functional for estimating the ground-state energy is 
\begin{align}
    \varfun_{\thetavec}[\psifull_t] & = \langle \psifull_t | \Hhat(\thetavec) |\psifull_t \rangle  ,  \\
     \varconsfun[\psifull_t]  &= \langle \psifull_t| \psifull_t\rangle - 1 ,
\end{align}  
The resulting equation for evaluating the EC emulator is a $N_b$-dimensional generalized eigenvalue problem, which can be solved extremely fast and with little memory for small $N_b$. 
In the previous bound-state emulator studies~\cite{Konig:2019adq, Ekstrom:2019lss} treating chiral potentials and their to-be-fitted low-energy constants (LECs),  the $\thetavec$ dependence of the potential term in $\Hhat$ is linear, i.e., $\Vhat(\thetavec) = \sum_{n=1}^{N} \theta_n\,\Ohat_{n} $. 
The coefficients of the resulted linear equations depend on $\langle \psifull^{(i)} |\Ohat_n | \psifull^{(j)}  \rangle$  and $\langle \psifull^{(i)} | \psifull^{(j)}  \rangle$. These matrix elements, even though possibly costly to evaluate, can be computed and saved at the emulator-training stage and tabulated for use in the emulating stage. 
The resulting emulators are extremely fast and accurate~\cite{Frame:2017fah, Konig:2019adq,Ekstrom:2019lss}.  

\subsection{Two-body scattering emulator} \label{subsec:2body_scattering_emulator}

For the two-body scattering calculation, the KVP functional for estimating the $S$ matrix takes the form (and similarly for the $T$ or $K$ matrices), 
\begin{align}
    \varfun_{\thetavec}[ \psifull_{t}] & =  S_t - \frac{i}{\mathcal{C}} \langle \psifull_t| E- H(\thetavec)| \psifull_t\rangle  .  \label{eq:varFunc_2bd}
\end{align}
$S_t$ is the $S$-matrix of the scattering wave function $\psifull_t$, which can be inferred from the asymptotic behavior of $\psifull_t$  at large separation; i.e., in a particular partial wave $\ell$ channel,
\begin{align} \label{eq:asymptotic_S_ell}
    \psifull_t(r) \underset{r\rightarrow\infty}{\longrightarrow}
     \frac{1}{\sqrt{v}} 
     \bigl(-\hat h_\ell^{(-)}(pr) + S_t\, \hat h_\ell^{(+)}(pr) \bigr)
      \;.
\end{align}
$\hat h_\ell^{(\pm)}$ are the outgoing and incoming spherical waves~\cite{Descouvemont:2015xoa}, $v$ is the relative velocity, and the coefficient $\mathcal{C} = 1$ for this particular normalization of the scattering wave function. 
In this case 
\begin{align}
    \varconsfun[\psifull_t] = \sum_{i=1}^{N_b} c_i  -1 .
\end{align}

The functional with the EC trial wave function becomes 
\begin{align}
    \varfun_{\thetavec} & = \sum_{i=1}^{N_b} c_i S^{(i)} - \sum_{i,j=1}^{N_b} c_i c_j \delta\tilde{U}_{i,j} , \\
    \delta{U}_{i,j} & \equiv \frac{1}{2i\mathcal{C}} \langle \psifull^{(i)} | H - E | \psifull^{(j)} \rangle . \\ 
    \delta\tilde{U}_{i,j} & = \delta{U}_{i,j} + \delta{U}_{j,i} .
\end{align}
$S^{(i)}$ is the $S$ matrix at the $i$th training point. We then get a set of linear equations 
for finding the stationary point of the functional:
\begin{align}
    2 \sum_{j=1}^{N_b} \delta\tilde{U}_{i,j} c_j + \lambdaLM & = S^{(i)} \ \text{with}\  i = 1,,, N_b \label{eq:scatt_emulator_eqs1} \\      
    \sum_{i=1}^{N_b} c_i  & = 1 , \label{eq:scatt_emulator_eqs2}
\end{align}
which can be directly solved using a linear equation solver. 
The equations can become ill-conditioned as $N_b$ increases; 
here we precondition them using nugget regularization~\cite{Furnstahl:2020abp}. 

As with the bound-state emulators, if the $\thetavec$ dependence in $H(\thetavec)$ is  factorized from individual potential operators, i.e., $\Vhat(\thetavec) = \sum_{n=1}^{N} \theta_n \Ohat_{n} $, the $\langle \psifull^{(i)} | \Ohat_n  | \psifull^{(j)} \rangle $ pieces in $ \delta{U}_{i,j}$ can be computed and stored at the training stage and directly reused when emulating. These emulators can be extremely fast and needs little memory, because the \emph{only} cost is solving the low-dimension linear equations. 
In Sec.~\ref{sec:3body_emulators} we explore ways to construct efficient EC emulators when the parameter factorization condition is not met.

\section{Three-body scattering} \label{sec:three-body_scattering}

\subsection{Exact solution of Faddeev scattering equations }
\label{subsec:FaddeevFormalism}
\begin{figure}
    \centering
    \includegraphics[width=0.3\textwidth]{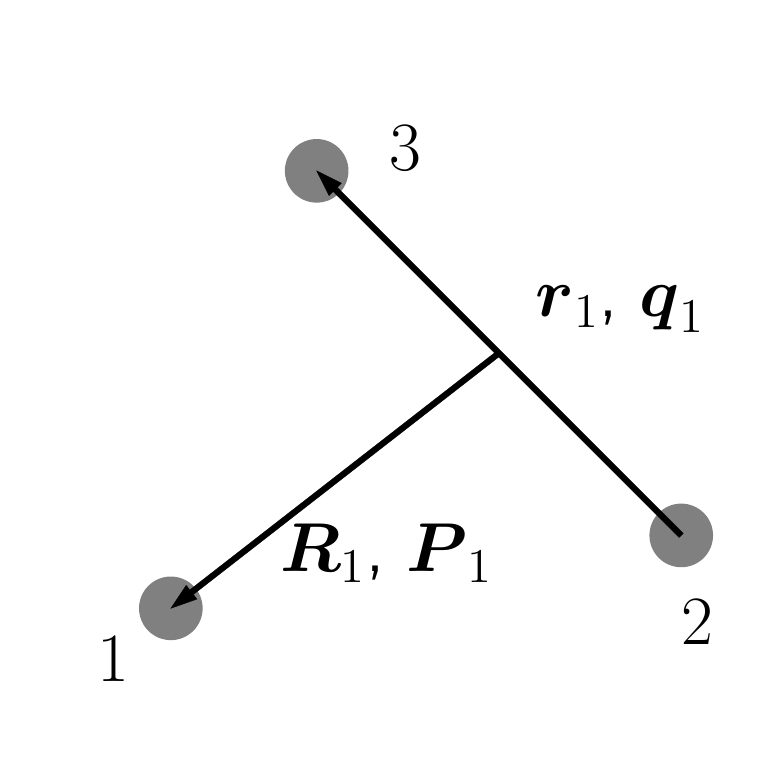}
    \caption{A schematic layout of a 3-body system with a particular choice of coordinate and momentum variables for describing the internal dynamics.}
    \label{fig:3-body_layout}
\end{figure}

There are three \emph{equivalent} sets of  kinetic variables describing a three-body system. In Fig.~\ref{fig:3-body_layout}, we consider particles `2' and `3' a dimer system and particle `1' a spectator. The variables $\myvec{r}_1$ and $\myvec{q}_1$ are the relative coordinate and momentum vector between `2' and `3', while $\myvec{R}_1$ and $\myvec{P}_1$ are those between the spectator and the center-of-mass (CM) of the dimer. The other two sets of coordinates arise from permutations of the particle indices. See Appendix~\ref{app:conventions} for further details.  The notation conventions in Ref.~\cite{Glockle:1983} will be followed here generally, except that the assignments of momentum variables are different: 
$P \leftrightarrow q_{\text{Ref.\,\cite{Glockle:1983}}}$, $q \leftrightarrow p_{\text{Ref.\,\cite{Glockle:1983}}}$. 

In this work, we focus on the case where there is a bound state in each of the three dimer systems, and study the scattering between the spectator and the dimers below the dimer-breakup thresholds. We further assume for now that the particles are distinguishable spinless bosons and then consider the case of identical bosons in the next subsection. 

The full Hamiltonian for the three-body system in its CM frame is 
\begin{align}
    H  = H_0  + \sum_{\alpha=1}^3  V_\alpha + V_4 .  
\end{align}
$H_0$ is the full kinetic energy operator. In the momentum-space representation, $H_0 = \frac{P_\beta^2}{2\mu^\beta} + \frac{q_\beta^2}{2\mu_\beta}$, with $\beta = 1$, $2$, \emph{or} $3$, and with $\mu_\beta $ and $\mu^\beta $ the reduced masses for the relative motions described by $\myvec{q}_\beta$ and $\myvec{P}_\beta$, respectively. $V_\alpha$ is the two-body interaction (e.g., $V_1$ is for particle 2 and 3), and $V_4$ the three-body interaction. The sum of all these potentials is $V$.   

To calculate the transition amplitudes and wave functions---both are required for training the EC emulators---one can directly solve the three-body Faddeev equations in  coordinate space at a given scattering energy~\cite{Faddeev:1960su} by treating them as coupled partial differential equations. The scattering phase shift~\cite{PhysRevLett.33.1350,Lazauskas:2019rfb} is then extracted from the scattering wave function solution.  

Here, we take a different route by starting with the transition amplitudes and working in momentum space. 
The  transition operators describing particle-dimer scatterings are usually defined through~\cite{Glockle:1983} 
\begin{align}
    U_{\beta\alpha} |\psifree_\alpha \rangle & = (V - V_\beta) |\psifull_\alpha^{(+)} \rangle  . \label{eq:U_def}
\end{align}
 The indices $\alpha, \beta = 1, 2, 3$ label the reaction channels in which spectators $\alpha$ and $\beta$ scatter from the associated dimer. 
 $| \psifree_\alpha \rangle$ and $|\psifull_\alpha^{(+)} \rangle$ are the asymptotic and full scattering states in the $\alpha$-channel, respectively.  That is, they are the eigenstates of $H_0 + V_\alpha$ and $H_0 + V$. 
 For example, $ |\psifree_1\rangle = |\boldsymbol{P}_1 \rangle |\psidimer_B \rangle $ with $|\boldsymbol{P}_1 \rangle$ the scattering plane wave between 1 and the 23 dimer, and $|\psidimer_B \rangle$ the dimer bound state.  
 The on-shell transition amplitudes between  the $\alpha$ and $\beta$ channels then take the form $\langle \psifree_\beta | U_{\beta\alpha}|\psifree_\alpha\rangle$.  For later use, a $U_{4\alpha}$ operator is defined in the same way, i.e., with $\beta$ set to 4 in Eq.~\eqref{eq:U_def}.

These operators satisfy the so-called  AGS~\cite{Alt:1967fx,Glockle:1983} coupled integral equations (equivalent to the Faddeev equations~\cite{Faddeev:1960su,Deltuva:2012kt}): 
\begin{align}
    U_{\beta\alpha} 
    & = \deltabar_{\beta\alpha} G_0^{-1} + t_4 + \sum_{\gamma=1}^3  \left( \deltabar_{\beta\gamma} + t_4 G_0\right) t_\gamma G_0 U_{\gamma \alpha} . \label{eq:UcoupledW3body3} 
\end{align}
 Here $G_0 \equiv 1/(E+ i0^+ - H_0)$ is the free Green's function,  $t_\gamma$ and $t_4$  the full $T$-matrix associated with $V_\gamma$ and $V_4$ interactions, and $\deltabar_{\beta\gamma} \equiv 1 - \delta_{\beta\gamma}$ (i.e., only nonzero when $\beta \neq \gamma$).

A separable  two-body potential is employed in this work to simplify the full calculation:  
\begin{align}
    V_\gamma & =  \lambda_\gamma  \int d \myvec{P}_\gamma |\myvec{P}_\gamma,g_\gamma  \rangle \langle \myvec{P}_\gamma,g_\gamma   |   , \label{eq:2-body_pot_def}
\end{align}
where $  \langle \myvec{q}_\gamma | g_\gamma \rangle  = g_\gamma(\myvec{q}_\gamma) $, usually called form factors, depend only on the inter-particle motions within the corresponding dimers. The state normalization conventions are detailed in Appendix~\ref{app:conventions}. 
The asymptotic state in channel $\gamma$ then has an analytical form~\cite{Watsonbook1967,Glockle:1983}: %
\begin{align}
    | \psifree_\gamma \rangle = \bsren_\gamma G_0 (E) | \myvec{P}_\gamma, g_\gamma \rangle , \label{eq:asympdimer1}
\end{align}
with $E= \frac{P_\gamma^2}{ 2\mu^\gamma } - B_\gamma$ the total on-shell energy and $B_\gamma$ the dimer binding energy. 
The parameter $\bsren_\gamma$ guarantees the proper normalization of the dimer bound state. 
The $T$-matrix associated with $V_\gamma$, which showed up in Eq.~\eqref{eq:UcoupledW3body3}, becomes~\cite{Watsonbook1967, Glockle:1983}
\begin{align}
    t_\gamma(E) & =    \int d \myvec{P}_\gamma\, \tau_\gamma\bigl(E-\frac{P_\gamma^2}{2\mu^\gamma}\bigr) |\myvec{P}_\gamma,g_\gamma  \rangle \langle \myvec{P}_\gamma,g_\gamma |  ,  \notag \\ 
    \frac{1}{\tau_\gamma(e_2)} & = \frac{1}{\lambda_\gamma} - \int d \myvec{q}_\gamma \frac{|g_\gamma(\myvec{q}_\gamma) |^2}{e_2-\frac{q_\gamma^2}{2 \mu_\gamma}}  . 
\end{align}
The variable $e_2$ is the energy of the two scattering particles; in the three-body setting it turns into $E-P_\gamma^2/(2\mu^\gamma)$.  (Note that $\{\lambda_\gamma, \tau_\gamma, g_\gamma, \bsren_\gamma\} \rightarrow \{\lambda, \tau, g, \bsren\}$  when discussing the identical bosons later.)
We also take a separable form for the three-body interaction~\cite{Phillips:1966zza}: $ V_4  = \lambda_4 | g_4 \rangle \langle g_4 |$. 
$t_4$ is also separable: $ t_4  = \tau_4 | g_4 \rangle \langle g_4 |$ with
\begin{align}
    \tau_4^{-1} & = \left[\lambda_4^{-1}- \langle g_4|G_0(E)| g_4 \rangle \right]  .  \label{eq:tau4}
\end{align}

Based on the separable potentials and  Eq.~\eqref{eq:asympdimer1},  Eq.~\eqref{eq:UcoupledW3body3} is then transformed to the so-called Lovelace equations~\cite{Lovelace:1964mq,Glockle:1983} ($X_{\beta\alpha}(\myvec{P}_\beta, \myvec{P}_\alpha)  \equiv \langle \psifree_\beta | U_{\beta\alpha} | \psifree_\alpha \rangle  $),
\begin{widetext}
\begin{align}
 X_{\beta\alpha}(\myvec{P}_\beta, \myvec{P}_\alpha)  &  =   Z_{\beta\alpha}(\myvec{P}_\beta, \myvec{P}_\alpha) + Z_{\beta 4}(\myvec{P}_\beta)\, \tau_4(E)\, Z_{4 \alpha}(\myvec{P}_\alpha) \notag \\ 
    & +  \sum_{\gamma} \int d\myvec{P}_\gamma  \ 
   \left[ Z_{\beta\gamma}(\myvec{P}_\beta,\myvec{P}_\gamma) + Z_{\beta 4}(\myvec{P}_\beta)\, \tau_4(E)\, Z_{4 \gamma}(\myvec{P}_\gamma) \right]\, \hat{\tau}_\gamma(E-\frac{P_\gamma^2}{2\mu^\gamma})\,
    X_{\gamma \alpha}(\myvec{P}_\gamma, \myvec{P}_\alpha) , \label{eq:UcoupledSepW3body}  
\end{align}    
with
\begin{align}
   Z_{\beta\alpha}(\myvec{P}_\beta, \myvec{P}_\alpha) & \equiv  \deltabar_{\beta\alpha}  \langle \psifree_\beta |G_0^{-1} | \psifree_\alpha \rangle   ,  \
 Z_{\beta 4}(\myvec{P}_\beta)   \equiv \langle \hat{g}_\beta, \myvec{P}_\beta | G_0(E) | g_4 \rangle   , \
    Z_{4 \alpha}(\myvec{P}_\alpha)   \equiv \langle g_4 | G_0(E)| \hat{g}_\alpha, \myvec{P}_\alpha \rangle   .  
\end{align}
\end{widetext}
To keep the notation succinct, the following redefinitions have been made:
\begin{align}
    | \hat{g}_\alpha \rangle & \equiv \bsren_\alpha | g_\alpha \rangle  \  \text{and} \  
    \hat{\tau}_\alpha(e_2)   \equiv \bsren^{-2}_\alpha \, \tau_\alpha(e_2) .
\end{align}

Given the $X$s as the solution of Eq.~\eqref{eq:UcoupledSepW3body}, we can infer the full scattering  wave function $|\psifull_\alpha^{(+)} \rangle$ for an arbitrary incoming channel $\alpha$. First, $|\psifull_\alpha^{(+)} \rangle$ is decomposed into three different Faddeev components (FCs)~\cite{Glockle:1983}: 
\begin{align}
   |\psifull_\alpha^{(+)} \rangle  = \,&  G_0 \left(V_4 +  \sum_{\mu=1}^3  V_\mu \right) | \psifull_\alpha^{(+)} \rangle \equiv  \sum_\mu  | \psifd_{\alpha, \mu } \rangle  . \label{eq:FCdef1}
\end{align}
For separable potentials, the $\mu^{\text{th}}$ FC  is then  
\begin{align}
|\psifd_{\alpha,\mu}\rangle 
     = \, &  \delta_{\alpha\mu} (1+G_0 t_4)\, G_0 |\myvec{P}_\alpha, \hat{g}_\alpha \rangle \notag \\ 
    & + (1+G_0 t_4)\, G_0 t_\mu G_0 U_{\mu\alpha} G_0 |\myvec{P}_\alpha, \hat{g}_\alpha \rangle .
    \label{eq:FC_distinguishable}
\end{align}
Derivations can be found in Appendix~\ref{app:Faddeev}. Also see Ref.~\cite{Watsonbook1967} for the case without a three-body interaction.

\subsection{Identical bosons and s-wave scattering}\label{subsec:id_bosons}

When the particles are distinguishable, for a fixed initial channel, say 1, there exist three different scattering processes: one elastic, $1\to 1$, and two inelastic (known as rearrangement), $1\to 2$ and $1\to 3$. 
For identical particles, however, only elastic scattering exists below the break-up threshold. 
To compute its amplitude, all the transitions in the distinguishable case need to be summed coherently (see Sec.~(1.7) in Ref.~\cite{Thomas:1977ph}):
\begin{align}
X(\vecPf, \vecPini)  \equiv   X_\mathrm{d}(\vecPf, \vecPini) + 2 X_{\ex} (\vecPf,\vecPini)  , 
\end{align}
with $X_\mathrm{d} \equiv X_{11}$ labeling all the elastic amplitudes and $X_{\ex} \equiv X_{12}$ labeling all the  inelastic ones. 
Similarly, all the $Z_{\alpha\beta}$ with $\alpha \neq \beta$ are defined as $Z_{\ex}$. Other   quantities, such as the two-body interaction strengths $\lambda$ and the associated form factors $g$, do not have channel dependence either and will not carry channel indices anymore.

Based on  Eq.~\eqref{eq:UcoupledSepW3body}, we can find the equation for $X$: 
\begin{align}
   X(\vecPf, \vecPini) 
    = \, &  \overline{Z}(\vecPf, \vecPini)
    \notag \\
   +  &  \int d\myvec{P} \ \overline{Z}(\vecPf,\myvec{P}) \hat{\tau}(E-\frac{P^2}{2\mu^1}) X(\myvec{P}, \vecPini)   ,  \label{eq:UcoupledSepW3bodyID}
\end{align}
where 
\begin{align}
 \overline{Z}(\myvec{P}', \myvec{P} ) \equiv 2  Z_{\ex}(\myvec{P}', \myvec{P}) + 3  Z_{14}(\myvec{P}')\tau_4(E) Z_{41}(\myvec{P}) .
\end{align}
The symmetrized FCs are now:
\begin{align}
    |\psifd_\mu \rangle = \sum_{\alpha} |\psifd_{\alpha,\mu} \rangle . 
\end{align}
Since the particles are identical, it is evident~\cite{Glockle:1983} (or c.f.~Eq.~\eqref{eq:FC_distinguishable}) that the $_\mu\langle \myvec{P},\myvec{q} | \psifd_\mu \rangle $ have the same analytical forms in terms of the variable $\myvec{P}$ and $\myvec{q}$ for $\mu = 1, 2, 3$ and so do the $_\mu\langle \myvec{R},\myvec{r} | \psifd_\mu \rangle$ in terms of $\myvec{R}$ and $\myvec{r}$. (The basis states $_\mu\langle \myvec{P},\myvec{q} |$ and $_\mu\langle \myvec{R},\myvec{r} |$ are discussed in Appendix~\ref{app:conventions}.) The full scattering wave function, in the representation of a given coordinate systems (e.g., with particle 1 as the spectator), can be constructed from one FC via  
\begin{align}
 \langle \myvec{R}_1, \myvec{r}_1 | \psifull^{(+)}_{\vecPini} \rangle 
   & = \langle \myvec{R}_1,\myvec{r}_1|\psifd_{\vecPini,1}\rangle 
   + \null _2\langle \myvec{R}_2,\myvec{r}_2|\psifd_{\vecPini,2}\rangle \notag \\ 
   & \quad \null  +  \null _{3}\langle \myvec{R}_3,\myvec{r}_3|\psifd_{\vecPini,3}\rangle   \notag \\ 
    & = 
   \langle \myvec{R}_1,\myvec{r}_1|\psifd_{\vecPini,1}\rangle 
   + \null   _{1}\langle \myvec{R}_2,\myvec{r}_2|\psifd_{\vecPini,1}\rangle \notag \\ 
   & \quad \null  +  \null _{1}\langle \myvec{R}_3,\myvec{r}_3|\psifd_{\vecPini,1}\rangle  , 
\end{align}
with $\myvec{R}_{2,3}$ and $ \myvec{r}_{2,3}$ determined by  $\myvec{R}_1$ and $\myvec{r}_1$  through Eq.~\eqref{eq:kinRrrels1}. The first step uses the fact that $|\myvec{R}_\mu, \myvec{r}_\mu \rangle_\mu$ with $\mu = 1, 2, 3$ are the same states.   The  $\vecPini$ subscripts in the wave function and FCs are kept to identify the asymptotic momentum. They should not be confused with the channel index, e.g., $\alpha$ in Eq.~\eqref{eq:FCdef1}. 

In particular, we focus on the s-wave particle-dimer scattering. 
The dimer bound state is spinless as well. 
This three-body configuration has zero total angular momentum and is fully decoupled from any other states. 
$X$s, $Z$s, and Eq.~\eqref{eq:UcoupledSepW3bodyID} can be projected into this subspace by properly averaging over the angular dependence. 
We use a simplified notation such that whenever their momentum or coordinate variables are reduced to the corresponding magnitudes, they have implicitly been projected to this subspace, e.g., $Z_{\ex} (P', P) \equiv 1/2 \int d(\cos\theta)\, Z_{\ex} (\myvec{P}', \myvec{P}) $ with $\theta$ the relative angle between the two momenta. 

The full wave function and the FCs in both momentum and coordinate  space are discussed  in Appendix~\ref{app:Faddeev}.  
Here we just note as an example that when $R_1$  becomes much larger than  the interaction range, the first FC in the partial wave basis has the asymptotic form 
\begin{align}
\langle {R}_1, {r}_1|\psifd_{\vecPini,1}\rangle 
      \overset{R_1\to \infty}{\longrightarrow} \frac{\mathcal{N}}{\sqrt{v} } \frac{u_B({r}_1) }{ r_1 R_1} 
     \bigl(-e^{-i\Pini R_1} + S\, e^{i\Pini R_1} \bigr) , \label{eq:WFbelowbreakupAsymWo3BV}
\end{align}
where $v\equiv \Pini/\mu^1$ is the relative velocity and
\begin{align}
  \mathcal{N}\equiv \frac{\sqrt{v}}{i 2\sqrt{2} \pi \Pini} ,  \  \psidimer_B({r}_1) =  \frac{u_B(r_1)}{r_1} \frac{1}{\sqrt{4\pi}} ,  \label{eq:WFbelowbreakupAsymWo3BV_2}   
\end{align}
with
$ \psidimer_B$ the dimer bound-state wave function. 
In momentum space, $\psidimer_{B}(q_1) =  \hat{g}({q_1})/\left(-B -\frac{q_1^2}{2\mu_1} \right)$, but in coordinate space $\psidimer_B(r_1)$ needs to be computed numerically.
The $S$-matrix $S = e^{2i\delta_0} $ is related to $X$ via $X(\Pini,\Pini) = \frac{(-)}{4\pi^2\mu^1 } \frac{S-1}{2i \Pini}$. 

Since  $_\mu\langle \myvec{R},\myvec{r} | \psifd_{\vecPini,\mu} \rangle $ has the same analytical forms for all the $\mu$'s,  $|\psifd_{\vecPini,2} \rangle$ and $|\psifd_{\vecPini,3}\rangle$  in  coordinate space have the same oscillating behavior as the first FC's shown in Eq.~\eqref{eq:WFbelowbreakupAsymWo3BV} but with $R_{2,3} \to +\infty$ and  $r_{2,3}$ being fixed respectively.  According to Eq.~\eqref{eq:kinRrrels1},  with $R_1 \to \infty$, $R_2$, $r_2$, $R_3$ and $ r_3 \to \infty$  as well. 
Thus, the other two FCs, $ \langle {R}_1, {r}_1|\psifd_{\vecPini,2} \rangle$ and $ \langle {R}_1, {r}_1|\psifd_{\vecPini,3}\rangle$ approach zero, since the $\psidimer_B({r}_{2})$ and $\psidimer_B({r}_{3})$ terms in the two FC asymptotic forms behave as $\exp(-r_2 \sqrt{2 \mu_1 B} )$ and $\exp(-r_3 \sqrt{2 \mu_1 B} )$ respectively. Therefore, the full scattering wave function has three different asymptotic regions,  determined by $R_\mu \to \infty $ and $r_\mu$ fixed for $\mu = 1, 2, 3$. In each of these regions, Eq.~\eqref{eq:WFbelowbreakupAsymWo3BV} with $\{R_1, r_1\} \to \{R_\mu, r_\mu\}$ is the asymptotic form.

The oscillating behavior of each FC gives rise to the singularities in its momentum space representation, $_\mu\langle \myvec{P}_\mu,\myvec{q}_\mu|\psifd_{\vecPini,\mu}\rangle $. All the three sets of singularities are present in the full wave function in its momentum space representation.  To facilitate later discussion, we isolate these singularities by subtracting the singular pieces from the FCs, i.e.,  $ |\psifdsub_{\vecPini,1} \rangle \equiv | \psifd_{\vecPini,1}\rangle  - | \psifdasym_{\vecPini, 1}   \rangle $, with 

\begin{widetext}
\begin{align}
   \langle {P}_1, {q}_1|\psifdasym_{\vecPini, 1}\rangle 
   & \equiv \frac{4\pi \hat{g}({q}_1)}{E+i\epsilon-\frac{\Pini^2}{2\mu^1}-\frac{q_1^2}{2\mu_1} }   \left[ \frac{\delta( {P}_1 - \Pini)}{4\pi \Pini^2} +  \frac{2\mu^1}{\Pini^2- P_1^{2} + i 0^+ } \frac{\Pini^2 + \tilde{\Lambda}^2}{P_1^2 + \tilde{\Lambda}^2} X(\Pini, \Pini) \right] ,  \label{eq:FD_asymp_reg_pspace} 
\end{align}
\end{widetext}
which has singularities at $P_1 = \Pini$.
The subtracted FC, $|\psifdsub_{\vecPini,1} \rangle $, is smooth along the real  axis in the $P_1$ complex plane. The behavior of $|\psifdasym_{\vecPini, 1}\rangle $ at large $R_1$ (see Eq.~\eqref{eq:FD_asymp_reg_rspace}) is the same as Eq.~\eqref{eq:WFbelowbreakupAsymWo3BV}. However, because  the $\frac{\Pini^2 + \tilde{\Lambda}^2}{P_1^2 + \tilde{\Lambda}^2}$ factor regularizes the large $P_1$ behavior in Eq.~\eqref{eq:FD_asymp_reg_pspace},  $|\psifdasym_{\vecPini, 1}\rangle $ is finite at $R_1 = 0$, satisfying the boundary condition for a physical full wave function. In contrast, Eq.~\eqref{eq:WFbelowbreakupAsymWo3BV} diverges at $R_1 = 0$.

\subsection{Kohn variational approach}\label{subsec:KohnFormalism}
In this section, we show that the functional
\begin{align}
    \varfun[\psifull_{t} ] \equiv S_t - \frac{i}{3\mathcal{N}^2}    \langle \psifull_t| E- H| \psifull_t\rangle   \label{eq:varFunc} 
\end{align}
 can be used to estimate the s-wave scattering $S$-matrix at the functional's stationary point with a second-order error in terms of the wave function error. It takes the same form as Eq.~\eqref{eq:varFunc_2bd} with $\mathcal{C} = 3 \mathcal{N}^2$, but the overlap integrals in computing $\langle \psifull_t| E- H| \psifull_t\rangle $ are much more involved.  Similar functionals for the three-body system can also be found in previous studies, such as \cite{Kohn:1948col, Kievsky:1997zf}. 

It is clear that when $\psifull_t $ is the exact solution $\psifull$, the functional returns the exact $S$-matrix because the second term vanishes. 
The nontrivial step is to demonstrate that when $\psifull_t = \psifull$,  
\begin{align}
 \frac{\delta \varfun[\psifull_{t} ]}{\delta \psifull_t} = 0 , \  \text{so that} \      \delta \varfun[\psifull_{t} ] = O\left[ (\delta\psifull_t)^2 \right] .   
\end{align}
Let us first redefine 
$  u(R_1, r_1)  \equiv     R_1 r_1 \langle R_1, r_1|\psifd_{\vecPini,1}\rangle $. Its behavior with  $R_1 \to +\infty$ can be inferred from Eq.~\eqref{eq:WFbelowbreakupAsymWo3BV} directly, while  $u_B(r_1 \to \infty) =  0 $.   At small particle separations~\cite{PhysRevLett.33.1350}, 
\begin{align}
    u(R_1 = 0, r_1) = & \, u(R_1, r_1 = 0) = 0  . \label{eq:WF_at_small_R_r_1}
\end{align}
Now we consider the variation of the FC. At large $R_1$, 
\begin{align}
    \delta u \overset{R_1\to \infty}{=} & \mathcal{N}  u_B(r_1) \frac{1}{\sqrt{v}}  \delta S\, e^{i\Pini R_1}  .  \label{eq:var_on_scatt_WF}
\end{align} 
Note that the dimer bound-state wave function, $u_B$, is not varied  (in   Eq.~(5.9) of Ref.~\cite{Kohn:1948col}, the deuteron wave function is not varied).

The trick of integrating by parts is repeatedly used to compute 
$\delta \left[ \langle \psifull_t| E- H| \psifull_t\rangle \right]$. For an individual FC, the following equation holds: 
\begin{align}
  &  \langle \psifd_{\vecPini,1} | E- H| \delta  \psifd_{\vecPini,1}\rangle - \langle \delta \psifd_{\vecPini,1} | E- H|   \psifd_{\vecPini,1}\rangle \notag \\ 
    = &  \int d r_1  \frac{1}{2\mu^1} \left.\left( u\frac{\partial \delta u}{\partial R_1} - \delta u\frac{\partial u}{\partial R_1} \right)\right\vert^{R_1=\infty}= - i \delta S \mathcal{N}^2 . \label{eq:var_der_1}
\end{align}
The contribution of the $ \frac{1}{2\mu_1}\frac{\partial^2} {\partial  r_1^2}$ piece in the derivation, in particular the surface term, is zero because $u$ and $\delta u$ are zero at $r_1 = 0 $ and $\infty$. 
Furthermore, we also have 
\begin{align}
    & \langle \psifd_{\vecPini,2} | E- H| \delta  \psifd_{\vecPini,1}\rangle - \langle \delta \psifd_{\vecPini,2} | E- H|   \psifd_{\vecPini,1}\rangle  \notag \\ 
   = &  \int d r_1  \frac{1}{2\mu^1} \left.\left( \tilde{u}\frac{\partial \delta u}{\partial R_1} - \delta \tilde{u}\frac{\partial u}{\partial R_1} \right)\right\vert^{R_1=\infty}  = 0 . \label{eq:FDinterference2}
\end{align}
Here, $ \tilde{u}(r_1, R_1)\equiv \frac{r_1 R_1}{r_2 R_2} u(r_2, R_2)$.  Similar to the derivation of Eq.~\eqref{eq:var_der_1}, the  $ \frac{1}{2\mu_1}\frac{\partial^2} {\partial  r_1^2}$ piece does not contribute since $\tilde{u}, \delta \tilde{u} = 0 $ at $r_1 = 0 $ and $\infty$ as well. The important difference between the two derivations is that when $R_1 \to \infty$ with $r_1$ fixed, $\frac{r_1 R_1}{r_2 R_2} \propto \frac{r_1}{R_1}$ and thus the corresponding surface term becomes zero here.

By using  $(E-H) |\psifull_{\vecPini}^{(+)} \rangle = 0 $ and combining the above two relationships, which also holds for other FCs, we get 
\begin{align}
    & \sum_{\mu,\nu=1}^3 \left[\langle \psifd_{\vecPini,\mu} | E- H| \delta  \psifd_{\vecPini,\nu}\rangle - \langle \delta \psifd_{\vecPini,\mu} | E- H|   \psifd_{\vecPini,\nu}\rangle \right] \notag \\ 
    &= \langle \psifull_{\vecPini}^{(+)} | E- H| \delta  \psifull_{\vecPini}^{(+)}\rangle   = -3 i \mathcal{N}^2 \delta S  \label{eq:var_der_2}
\end{align}
This, together with  $\langle \delta \psifull_{\vecPini}^{(+)} | E- H| \psifull_{\vecPini}^{(+)}\rangle = 0 $, gives $ \delta \varfun[\psifull_{t} ]/\delta \psifull_t = 0 $ when $\psifull_t$ is the exact solution $\psifull_{\vecPini}^{(+)}$.

In Eqs.~\eqref{eq:var_der_1} and~\eqref{eq:FDinterference2}, the Coulomb interactions and the centrifugal terms, if present in $H_0$, would be cancelled out on their left sides. Therefore, the functional in Eq.~\eqref{eq:varFunc} applies to scattering in general partial waves and  with the Coulomb interaction between the projectile and target, just as Eq.~\eqref{eq:varFunc_2bd} can be used to estimate the scattering $S$ matrix in all these cases. 

The functional in  Eq.~\eqref{eq:varFunc} also works for higher-body elastic scattering, e.g., a particle scattering off a three-particle bound state (trimer)---with the proper definition of $\mathcal{N}$ depending on the normalization of the wave function. In general, the asymptotic wave function at large target-projectile separation will have the same factorized form as Eq.~\eqref{eq:WFbelowbreakupAsymWo3BV} but with $\psidimer_B$ changed to the product of the projectile and target bound-state wave functions (e.g., three-particle bound state in particle-trimer scattering). The derivation follows Eq.~\eqref{eq:var_der_1} and Eq.~\eqref{eq:var_der_2} and  the desired functional takes the form of Eq.~\eqref{eq:varFunc}. A general proof can be found in Sec.~V of Ref.~\cite{Gerjuoy:1983un}. 

The construction of emulators for particle-dimer scattering directly follows the discussion in Sec.~\ref{subsec:2body_scattering_emulator}. 

The next section discusses the testing of the EC emulators. The focus is on s-wave scattering, but we expect similar results for the other general scattering cases.

\section{Three-body scattering emulators} \label{sec:3body_emulators}
Gaussian forms  are employed for the two- and three-body potential form factors in this work.  A series of analytical formulas related to these form factors can be found in the SM.
 For the two-body interaction, 
\begin{align}
    g(\myvec{q})  = \frac{C}{\Lambda_2} \exp\left[-\frac{q^2}{2 \Lambda_2^2}\right] \ \text{with} \ C^2 \equiv \frac{\LAMtwo}{4 \pi^\frac{3}{2}\mu_1  }  . \label{eq:gauFF2def}
\end{align}  
Here  $C$  is chosen such that at low energy, the scattering length and effective range terms are related to $\LAMtwo$ and the interaction strength parameter $\lambda$ (see Eq.~\eqref{eq:2-body_pot_def}) via Eqs.~\eqref{eq:a0_for_Gau} and~\eqref{eq:r0_for_Gau} in the SM. Again, $\lambda$ and $g$ have no channel indices for a system of identical particles. 

The form factor of the three-body interaction is 
\begin{align}
    \langle \myvec{P}_1, \myvec{q}_1 | g_4 \rangle & = 
    \frac{1}{\sqrt{M \Lambda_4^4}} \exp\left[-\frac{M E_4}{2 \Lambda_4^2}\right]  ,  
\end{align}
with $E_4  \equiv \frac{P_1^2}{2\mu^1} + \frac{q_1^2}{2\mu_1}$. Evidently,  
\begin{align}
_2\langle \myvec{P}_2, \myvec{q}_2 | g_4 \rangle = \null _3\langle \myvec{P}_3, \myvec{q}_3 | g_4 \rangle = \null _1\langle \myvec{P}_1, \myvec{q}_1 | g_4 \rangle  ,   
\end{align}
i.e., $V_4$ takes the same  form in all the coordinate systems.  

We now turn to  the EC emulators that we have developed for three different cases: (1) varying the strength of only the three-body potential, $\lambda_4$; (2) varying both $\Lambda_4$ and $\lambda_4$ but not $\LAMtwo$; and (3) varying, $\lambda_4$, $\Lambda_4$ and $\LAMtwo$. 
The two-body binding energy is fixed throughout, considering that in most data analysis cases it should be known or measured separately or determined by examining the threshold location in the particle-dimer scattering cross sections.
The three cases will be discussed separately owing to their different characteristics. Our focus will be on three main aspects: the emulation accuracy and time and memory costs. 
 A brief discussion on the computing costs at the \emph{training} stage is provided at the end. 

Two distinct parameter sets are explored. The first one, called the \emph{nucleon} case, has two-body binding energy $B=2$ MeV and  the range of interactions of order 1\,fm such that $\LAMtwo$ and $\Lambda_4$ are in the range of several 100\,MeV. 
This set mimics $N$-$d$ scattering. 
The second one, called the \emph{nuclear} case, has $B=10$ MeV and multiple-fermi interaction ranges with $\LAMtwo$ and $\Lambda_4$ less than 100\,MeV. In all the cases, the particle mass is fixed to be the nucleon mass, 940\,MeV. The  nucleon case is elaborated below, while  the nuclear case, as presented in the SM,  has similar results in general. 

\subsection{Vary \texorpdfstring{$\lambda_4$}{lambda4}: the ``linear'' case} 
\label{subsec:emulator_1_dim}

\begin{figure*}[t]
    \centering
    \includegraphics[width=0.9\textwidth]{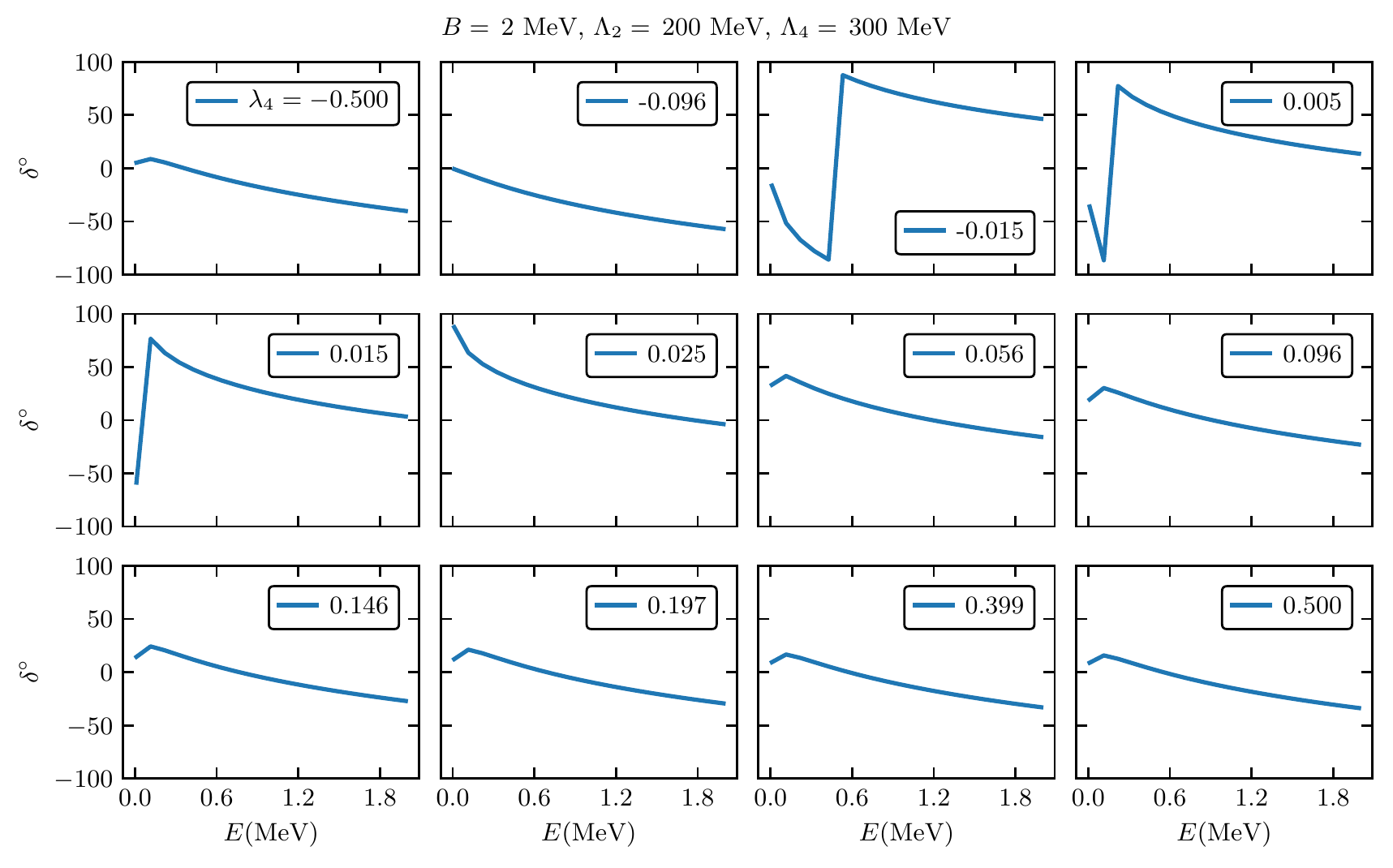}
%    \vspace*{-10pt}
    \caption{Particle-dimer s-wave scattering phase shifts vs.\ scattering energy $E$ for different $\lambda_4$ values (three-body interaction strength). The two-body interaction has fixed $\Lambda_2 = 200\,$MeV, and $\lambda$ is determined by the dimer binding energy $B=2\,$MeV.
    $\Lambda_4$ is fixed at 300\,MeV.}    \label{fig:3body_B_2_LAM2_200_LAM4_300_gau}
%\end{figure*}
%
\vspace*{15pt}
%\begin{figure*}
%    \centering
    \includegraphics[width=0.9\textwidth]{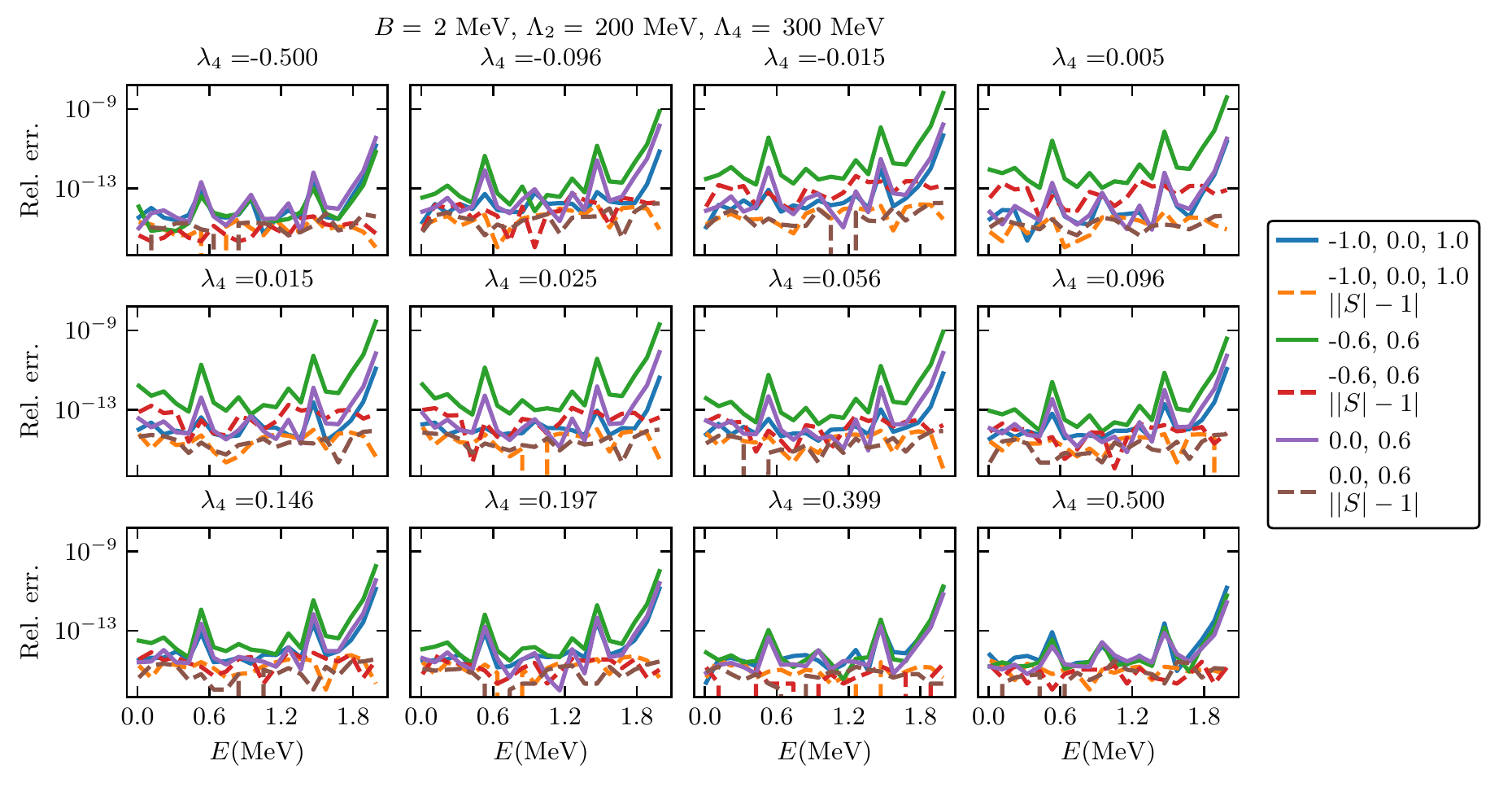}
%    \vspace*{-15pt}
    \caption{The EC emulator errors for various $\lambda_4$ values (shown at the top of each panel), for which the exact phase shifts have been plotted in Fig.~\ref{fig:3body_B_2_LAM2_200_LAM4_300_gau}. Three different training sets have been used for constructing the EC emulators. Their $\lambda_4$ values are shown in the legends. The solid curves are the relative errors of the $S$ matrix, while the dashed curves plot $||S|-1|$, i.e., the unitarity violation.}
    \label{fig:EmuErr_3body_B_2_LAM2_200_LAM4_300_gau}
\end{figure*}

Figure~\ref{fig:3body_B_2_LAM2_200_LAM4_300_gau} shows the particle-dimer scattering phase shifts for the nucleon case, for 12 representative $\lambda_4$ values with $\LAMtwo = 200$ and $\Lambda_4 = 300$ MeV. The phase shifts are sensitive to $\lambda_4$ when $|\lambda_4| 	\lessapprox 0.1$, but beyond that the dependence becomes much weaker. This can be understood from Eq.~\eqref{eq:tau4}; we can make the estimate, 
\begin{align}
    \langle g_4|G_0(E)| g_4 \rangle & = -\frac{4\pi^3}{3\sqrt{3}} \left(\frac{M}{\Lambda_4^2} \right)^2 \int_0^\infty d E_4 \frac{E_4^2\, e^{-\frac{M E_4}{\Lambda_4^2}}}{E_4 - E^+} \notag \\ 
     & \sim - \frac{4\pi^3}{3\sqrt{3}} \approx  - 20  ,   
\end{align}
which is independent of $M$ and $\Lambda_4$.  
Thus when $|\lambda_4| \gg \frac{1}{20}$, $\tau_4$ becomes independent of $\lambda_4$.  Therefore, we always choose $|\lambda_4| \leq 0.5$ when defining our parameter space.  

Three  sets of training points have been tried: (1) $\lambda_4 = \{-1, 0, 1 \}$, (2) $\lambda_4 = \{-0.6, 0.6 \}$, and (3) $\lambda_4 = \{0, 0.6 \}$.  The errors of thus trained emulators at the 12 points from Fig.~\ref{fig:3body_B_2_LAM2_200_LAM4_300_gau} are shown in Fig.~\ref{fig:EmuErr_3body_B_2_LAM2_200_LAM4_300_gau}, with solid and dashed lines for the relative errors of $S$ and the unitarity-violation error ($||S| - 1 |$), respectively. The errors are tiny:  the $S$ errors are mostly on the order of $10^{-13}$ and increase to $10^{-10}$ near the breaking-up threshold. For the third training, the test points with $\lambda_4 < 0$ should be considered as extrapolations, while the other tests points as interpolations. The errors are similar for both interpolations and extrapolations. The unitarity-violation errors are generally smaller than the $S$ relative errors, which is also true  in all the tests discussed in this work.

Besides the great accuracy, the other benefit of the emulator is its extreme efficiency. As mentioned in Sec.~\ref{sec:EC_emulators}, since $\lambda_4$ is a linear parameter in $V_4$, the $\delta U$ matrix can be computed once for a particular $\lambda_4$ value and then stored in the training period. At the emulating stage, the stored (\emph{low}-dimensional) $\delta U$ can be rescaled quickly to get the correct $\delta U $ for the emulation point. The computational cost of the emulator is $10^{-3}$ seconds, which is for solving the $N_b$-dimensional linear equations in Eqs.~\eqref{eq:scatt_emulator_eqs1} and \eqref{eq:scatt_emulator_eqs2}. The memory costs during emulation are tiny. For example, the memory for  storing $\delta U$ with $N_b = 2 $ and $3$ is negligible (cf.\ Table~\ref{tab:results_summary}).  

Such a scenario applies directly  to the fitting of the low-energy constants in chiral three-nucleon interactions to measurements of  $N$-$d$ scattering. 
Thus the EC emulators could play a key role in a full Bayesian treatment of these interactions.  
The generalization of the EC emulator to the energy region above the breakup-threshold,  including for both elastic scattering and break-up reactions, will be desirable as well for this purpose and perhaps will help to resolve the long-standing $A_y$ puzzle~\cite{Hebeler:2020ocj}. 

\subsection{Vary \texorpdfstring{$\lambda_4$ and $\Lambda_4$}{lambda4 and Lambda4}} \label{subsec:emulator_2_dim}

\begin{figure*}[t]
    \centering
    \includegraphics[width=0.9\textwidth]{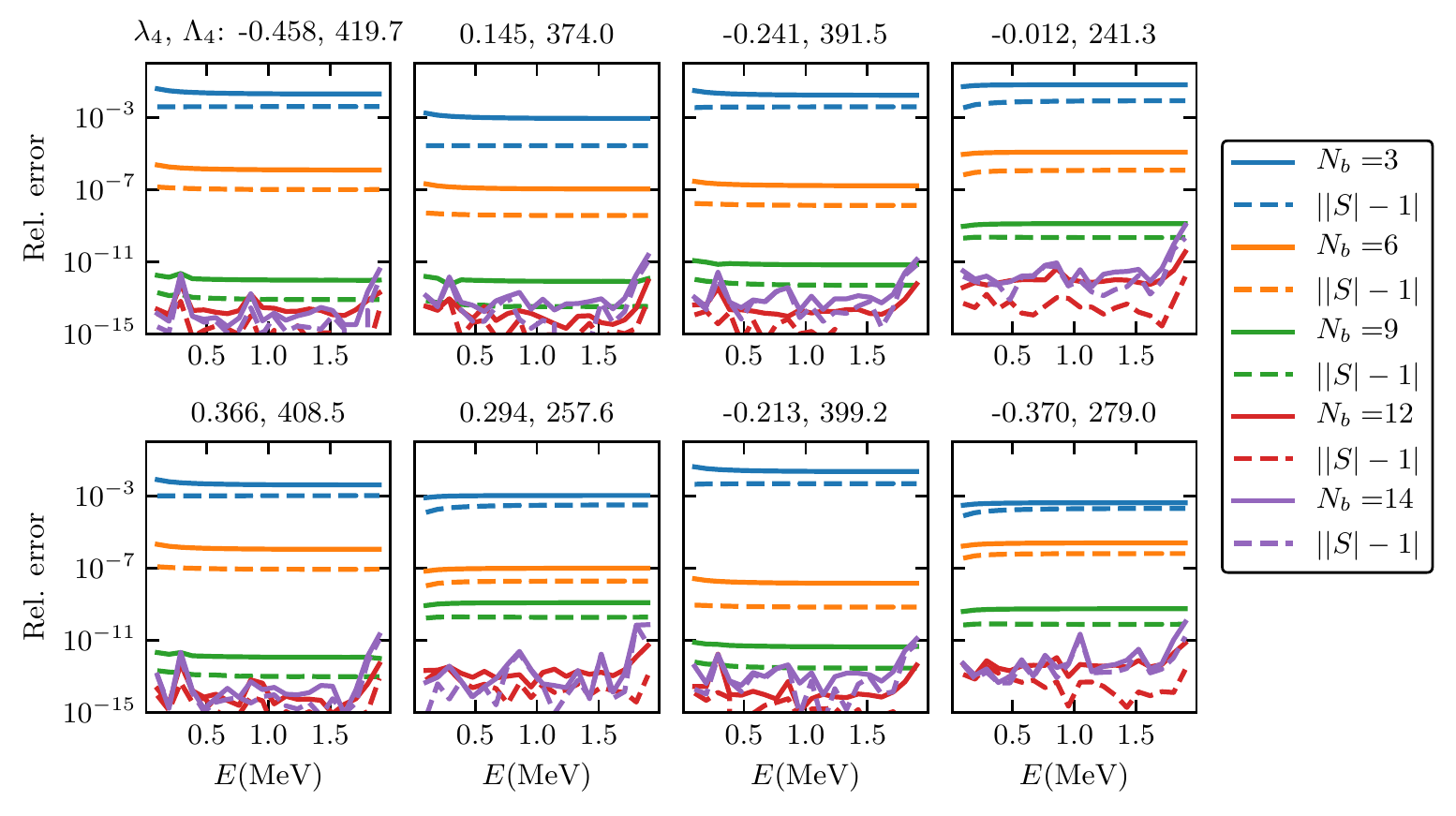}
    \caption{The EC emulator errors for a few test points in the two-dimensional parameter space for the nucleon case.  The solid and dashed curves in the same color are the $S$ matrix relative errors and the unitarity violation errors respectively with a fixed $N_b$ (see the legend). Here $B=2$ MeV and $\LAMtwo=200$ MeV. $\Lambda_4$ varies between 200 and 500 MeV, with $|\lambda_4| \leq 0.5$. }
    \label{fig:Emulator_RelError_Expamples_B_2_LAM2_200_Gau}
\end{figure*}

To further challenge the emulators, we now test them in a two-dimensional parameter space.  
The (nucleon case) parameter space is chosen with $ |\lambda_4| \leq 0.5$ and $ 200 \leq \Lambda_4 \leq 500$ MeV  and with  $\LAMtwo = 200$ MeV  fixed.

\begin{figure*}
    \centering
    \includegraphics[width=0.9\textwidth]{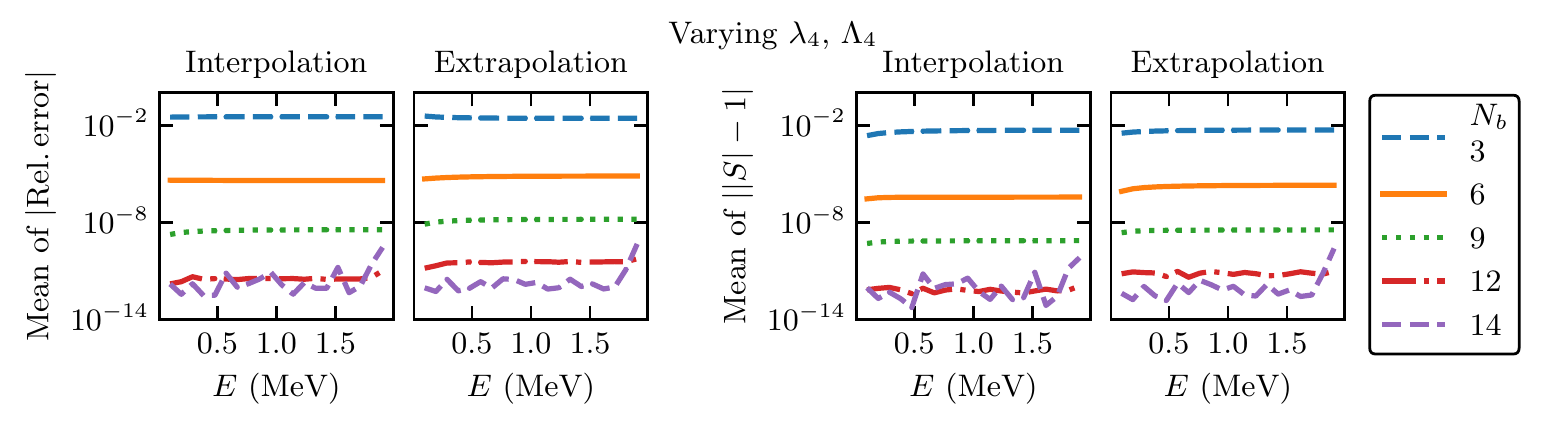}
    \caption{The mean of the EC emulator errors (the left two for $S$ and the right two for the unitarity violation) in the two-dimensional parameter space for the same five training sets as in Fig.~\ref{fig:Emulator_RelError_Expamples_B_2_LAM2_200_Gau} (see legend). In total, 200 points have been sampled and tested. $\LAMtwo$ is fixed at 200 MeV. The number of interpolation points grows with $N_b$. For $N_b = 14$, the interpolation and extrapolations have a similar number of points. }
    \label{fig:Emulator_RelError_B_2_LAM2_200_Gau}
\end{figure*}

In Fig.~\ref{fig:Emulator_RelError_Expamples_B_2_LAM2_200_Gau}, the emulator errors at eight representative points in the parameter space are shown. On the top of each panel are the $\lambda_4$ and $\Lambda_4$ values of the test point.   The training points are chosen using Latin hypercube sampling~\cite{doi:10.1080/01621459.1993.10476423}. 
The solid and dashed lines are the $S$ errors and the unitarity violation errors, respectively. 
In general, the latter are always smaller than the former. 
With six training points, the errors are of the order of $10^{-6}$.  (Note that in solving the linear equations in Eqs.~\eqref{eq:scatt_emulator_eqs1} and \eqref{eq:scatt_emulator_eqs2}, a simple nugget $\sim 10^{-20}$ is added to the coefficient matrix to mitigate the ill-conditioning problem~\cite{Furnstahl:2020abp}.) 

To verify that these errors represent the global emulation accuracy,
200 test points are uniformly sampled in this two-dimensional parameter space. The mean values of the $S$ errors and their unitarity violations of the test samples are plotted in 
Fig.~\ref{fig:Emulator_RelError_B_2_LAM2_200_Gau}. 
The resulting mean values at different $N_b$ are consistent with what is seen in Fig.~\ref{fig:Emulator_RelError_Expamples_B_2_LAM2_200_Gau}.

\begin{figure*}
    \centering
    \includegraphics[width=0.9\textwidth]{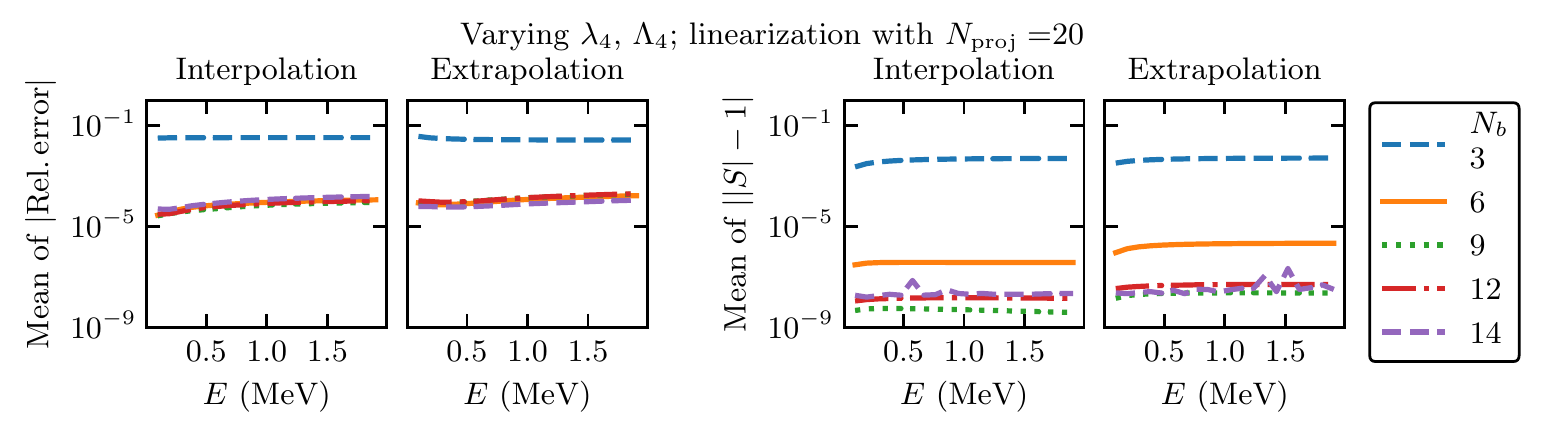}
    \caption{The mean of the errors of the EC emulators in the two-dimensional parameter space, which uses the Lagrange functions for the Legendre polynomials at $\Nproj$ order to project the three-body interaction. $\Nproj$ is fixed at 20.  The training sets and the test point sample are the same as those in Fig.~\ref{fig:Emulator_RelError_B_2_LAM2_200_Gau}. }
    \label{fig:EmulatorUsingLagProj_RelError_B_2_LAM2_200_Nproj_20_Gau}
\end{figure*}

The $\Lambda_4$ dependence cannot be factorized from the three-body potential operator, so the EC emulators require re-computing the $\delta U$ matrix at all emulation points. 
The cost is minimized in the present work because of the use of separable potentials. 
In realistic calculations, the cost could be substantial and slows down the EC emulators significantly.  To reduce this cost, our first approach  is to decompose the varying potentials into a linear combination of basis potentials. 
Here, we work directly with the interaction form factor,  $g_4(P, q)$, as a function of two variables in momentum space. (For non-separable potentials, the potentials need to be decomposed instead.) 
The form factor is  decomposed into the products of the Lagrange functions of  Legendre polynomials at order $\Nproj$~\cite{BAYE20151}:  
\begin{align}
    g_4(P, q) & = \sum_{m,n=0}^{\Nproj-1} f_{nm} L_n(x_p) L_m(x_q) , \label{eq:g4decomp_Lag} \\ 
    f_{nm} & = g_4(P_n, q_m)  . 
\end{align}
In the numerical calculations, $P$ and $q$ are restricted to be between 0 and a large momentum cut-off $\lambdacut$. The $x_p$ and $x_q$ variables used in the Lagrange functions are simply $x_p \equiv 2 P/\lambdacut-1$ and $x_q \equiv 2 q/\lambdacut-1$. The coefficients $f_{m,n}$ are the values of $g(P,q)$ at the mesh points---defined  as the zeros of the Legendre polynomial that generates the Lagrange functions~\cite{BAYE20151}. For the potential basis, $L_n(x)$ can be other functions, such as orthogonal polynomials, but inferring $f_{nm}$  could be more involved and thus  increase the emulator time cost. 

Figure~\ref{fig:EmulatorUsingLagProj_RelError_B_2_LAM2_200_Nproj_20_Gau}  shows the mean of the errors of the EC emulator with the $g_4$ linearization for the same test-point sample as used in Fig.~\ref{fig:Emulator_RelError_B_2_LAM2_200_Gau}.
The relative errors for $S$ have increased to $10^{-4}$---still useful for most data analysis---while the unitarity violation has smaller errors in general.  The emulator cost is about $10^{-3}$ to $10^{-2}$ seconds per evaluation on a laptop computer.

Unsurprisingly, by increasing $\Nproj$ large enough (up to $50$ here), the decomposition error reduces below the error of variational calculation; the new emulator errors become similar to the original EC errors in Figs.~\ref{fig:Emulator_RelError_B_2_LAM2_200_Gau}.  
However, the computational costs of training emulators grows with $\Nproj^2$ (although such training calculations can be easily parallelized). The right balance between the training cost and the emulation accuracy has to be determined for
each application of this emulator.

\begin{figure*}
    \centering
    \includegraphics[width=0.9\textwidth]{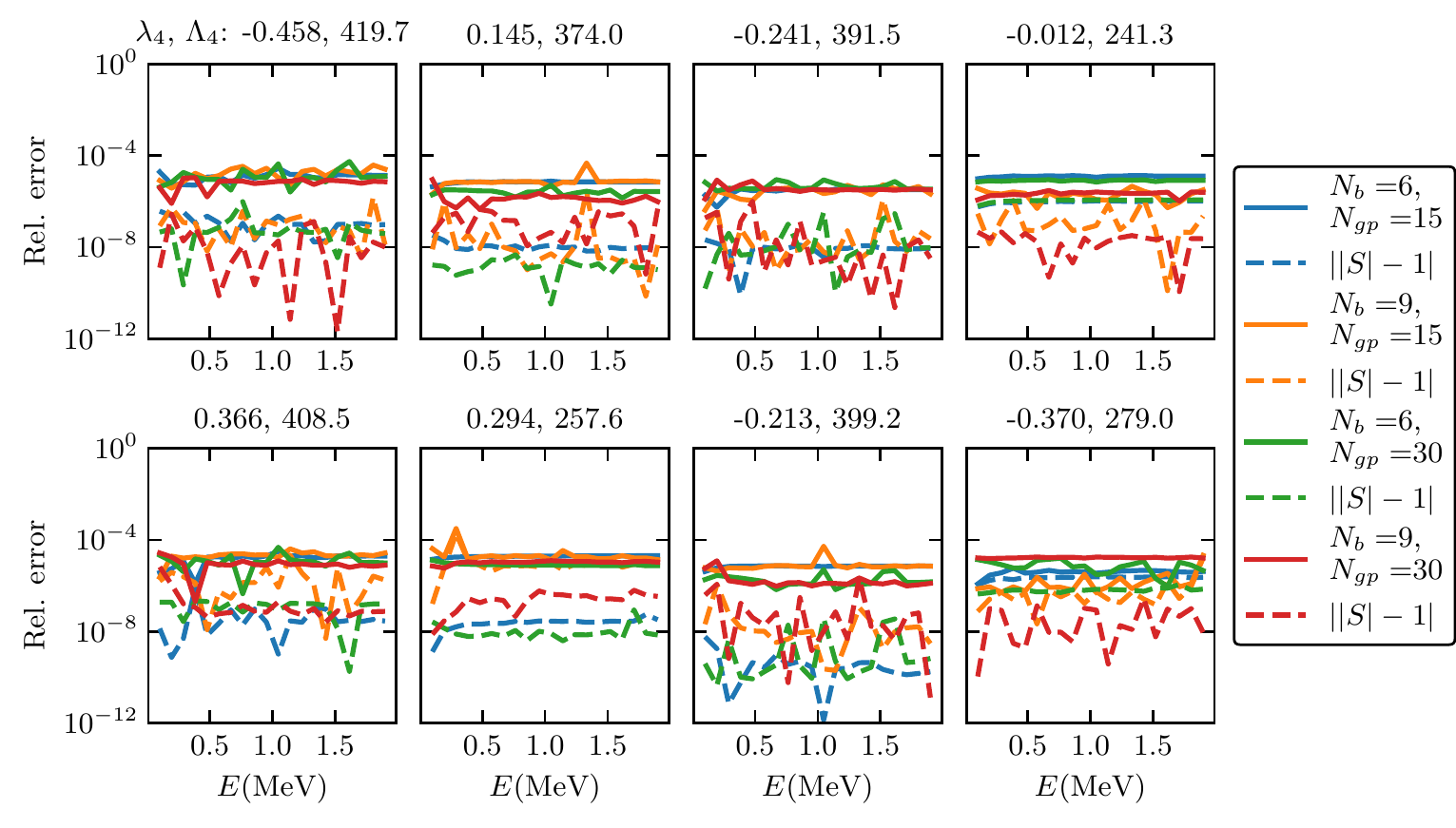}
    \caption{The GP-EC emulator errors for a few test points (the same as those in Fig.~\ref{fig:Emulator_RelError_Expamples_B_2_LAM2_200_Gau}) in the two-dimensional parameter space. The emulators implement GP interpolation within the EC emulators.  The solid and dashed curves in the same color are the $S$  errors and the unitarity violation errors, respectively, with a pair of fixed $N_b$ and $N_{gp}$ (see the legend).}
    \label{fig:EmulatorInEmulators_RelError_Expamples_B_2_LAM2_200_Gau}
\end{figure*}

\begin{figure*}
    \centering
    \includegraphics[width=0.9\textwidth]{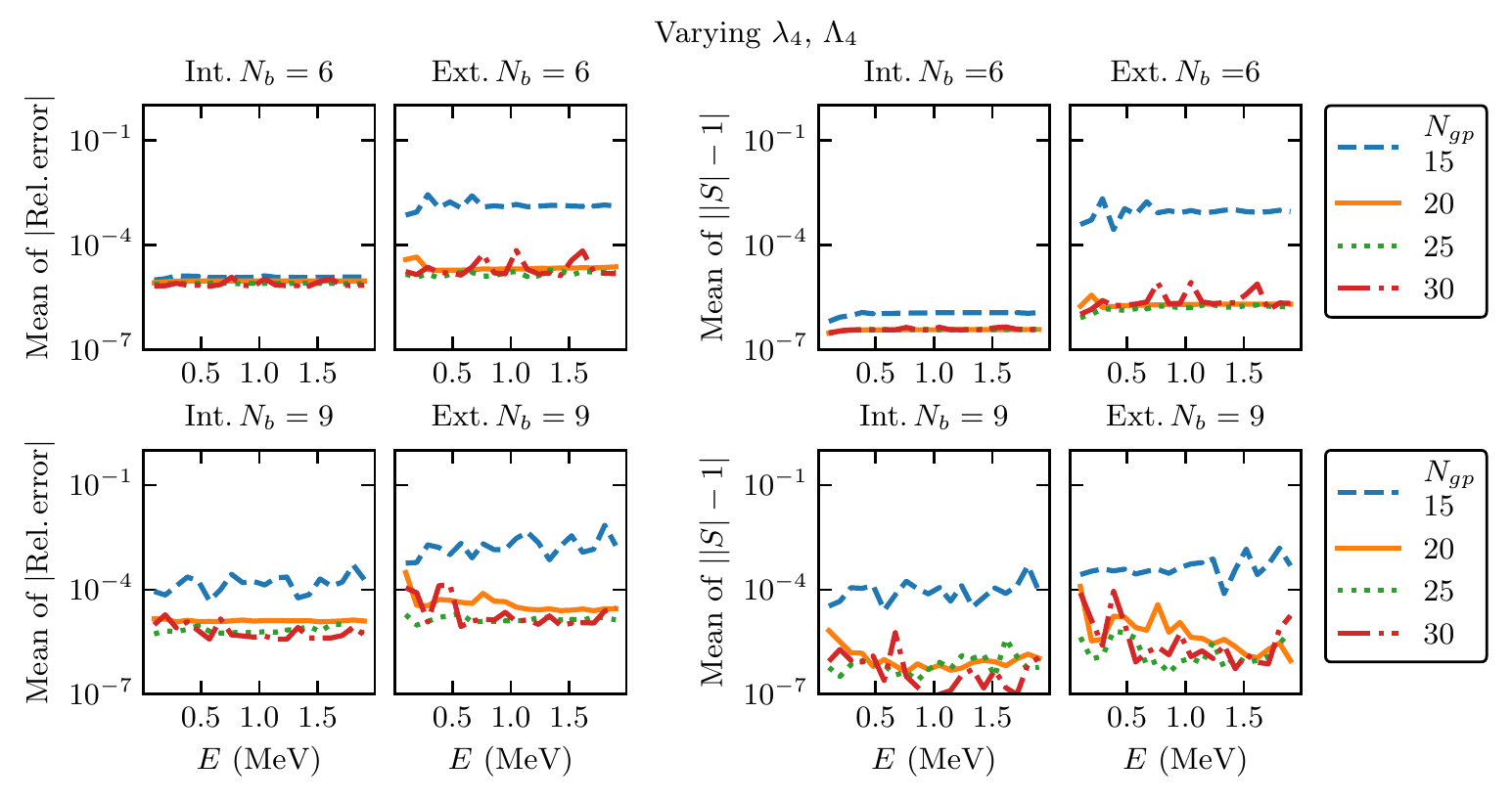}
    \caption{The mean  of the GP-EC emulator errors. The emulators implement GP interpolation within the EC emulators. The top (bottom) panel has $N_b = 6$ ($9$). Four different sizes of GP training sets from 15 to 30 are used in all the panels.}
    \label{fig:EmulatorInEmulator_RelError_B_2_LAM2_200_Nb_6_9_scalarGP_Gau}
\end{figure*}

A second approach to deal with the non-factorizable parametric dependence is to use  Gaussian processes (GPs)~\cite{Mackay:1998introduction} to interpolate $\delta U$ matrix elements in the parameter space. 
We call this a GP-EC emulator (it is also mentioned as the ``nonlinear-1'' case in  Table~\ref{tab:results_summary}).
Even though the parameter space is two-dimensional, the GP only needs to be trained in the $\Lambda_4$ dimension, while the $\lambda_4$ dependence in $\delta U$ is linear and thus can be interpolated simply by  rescaling. 
Note the $\lambda_4$ dependence of the $S$ matrix is still nonlinear. In short, when interpolating $\delta U$ instead of the final observables using the GP, the dimension of the interpolating space can be reduced by eliminating the  dimension with linear dependence. 

At the training stage, the GP is trained for each individual matrix element separately in the current work.  The GPy package~\cite{gpy2014} is utilized; the default model optimization is applied to get the ``best'' GP interpolants. It is worth noting that principle component analysis can be used in the GP training~\cite{10.2307/27640080} when the number of $\delta U$ matrix elements becomes large. 

In Fig.~\ref{fig:EmulatorInEmulators_RelError_Expamples_B_2_LAM2_200_Gau},  different combinations of the number of EC training points $N_b$ and the number of the GP training points $N_{gp}$ are explored; the emulator errors for the 8 different test points---the same as in Fig.~\ref{fig:Emulator_RelError_Expamples_B_2_LAM2_200_Gau}--- are shown there. 
With $N_b = 9 $ and $N_{gp} =30$, the errors are on the order of $10^{-5}$ (and the unitarity violation errors are even smaller). 
 Fig.~\ref{fig:EmulatorInEmulator_RelError_B_2_LAM2_200_Nb_6_9_scalarGP_Gau}  shows the mean values of these errors with increasing $N_{gp}$,  for the same test point sample as used  in Fig.~\ref{fig:Emulator_RelError_B_2_LAM2_200_Gau}. The left two and right two panels show the $S$ errors and the unitarity violation errors, respectively, while the top and bottom panels have $N_b  = 6$ and $9$ separately. 
The interpolation and extrapolation errors are similar and below $10^{-5}$ when $N_{gp} \sim 20$; increasing $N_{gp}$ further does not further reduce the  errors. This indicates that the GP-EC emulator errors are dominated by the GP interpolation errors. New ways to reduce GP errors  will be explored in the future. 

Meanwhile, the computational costs at the emulating stage include those for the trained GP to make predictions for $\delta U$ matrix elements and for solving the EC emulator linear equations. The total time cost is still only milliseconds. The memory costs for storing the trained GP emulators are a few MBs (c.f.~Table~\ref{tab:results_summary}), while those for solving Eqs.~\eqref{eq:scatt_emulator_eqs1} and~\eqref{eq:scatt_emulator_eqs2} are much smaller. The costs for training the emulators are detailed in Sec.~\ref{subsec:trainingcost}.

The emulators  studied in this section are suitable for  fitting three-body interactions in  three-cluster models that describe nuclear elastic scattering and reactions.  
The potential-linearization trick is not restricted to the Lagrange function basis. If the variation of the interaction range in the parameter space is small, other forms of basis may be available for reducing $\Nproj$. 
The GP-EC emulator method can be applied as well. It will be interesting to compare them in realistic data fittings.  

\subsection{Vary both \texorpdfstring{$V_4$ and $V_{1,2,3}$}{V4 and V123}: the ``nonlinear-2'' case} \label{subsec:emulator_3_dim}

There exist situations where constraining two-body interactions between reactions subsystems requires input from three-body systems.  For example, neutron-nucleus interactions are typically inferred from $d$-nucleus scatterings, because the direct experimental measurements of neutron-nucleus interactions are difficult. 
With this possibility in mind, we construct emulators for varying  $\lambda_4$, $\Lambda_4$, and $\LAMtwo$ at the same time. 
The two-body binding energy $B$ is always fixed ($\lambda$ depends on $\LAMtwo$ as a result), as explained earlier. 

Simply generalizing the EC emulators, however, faces a difficulty: the scattering wave functions of the training points and  emulation points generally have   different dimer bound-state wave functions. 
On the other hand, the functional in Eq.~\eqref{eq:varFunc} requires  the dimer bound-state wave function to be fixed while varying the trial wave function (see Eq.~\eqref{eq:var_on_scatt_WF}). 

To solve this problem, the dimer bound-state wave function within the training wave function in the  asymptotic region is substituted by the dimer bound state of the emulation point.
Meanwhile, in the interaction region the training wave functions are kept intact. 
The trial wave function built out of the linear combination of the new training wave functions (now labeled as $\tilde{\psifull}^{(i)}$)  can be directly plugged into the functional in Eq.~\eqref{eq:varFunc}. Each \emph{modified} training wave function is now
\begin{align}
    \tilde{\psifull}^{(i)} & \equiv  \sum_\mu \left(\psifd_{(i),\mu} - \psifdasym_{(i),\mu} +  \tildepsifdasym_{(i),\mu}\right) , 
   \label{eq:modified_basis_def}
\end{align}
with the asymptotic FC $\psifdasym_{(i),\mu} $ defined in Eq.~\eqref{eq:FD_asymp_reg_pspace} (in momentum space) and Eq.~\eqref{eq:FD_asymp_reg_rspace} (in coordinate space) and  the modified asymptotic FC $\tildepsifdasym_{(i),\mu}$ defined in the same way. 
The difference between $\psifdasym_{(i),\mu} $ and $\tildepsifdasym_{(i),\mu}$ is in their dimer bound-state wave functions: $\psidimer_{(i),B}$ determined by the two-body potential of the training point and  $\psidimer_B$ determined by the emulation point.

Meanwhile, at small $R_\mu$ the full wave functions $\psifull_{(i)}$ and the two asymptotic wave functions $\sum_i \psifdasym_{(i),\mu}$ and $\sum_i \tildepsifdasym_{(i),\mu}$ have the correct behavior, as shown in Eq.~\eqref{eq:WF_at_small_R_r_1}. Thus, the modified training wave functions $\tilde{\psifull}^{(i)}$ have the correct behavior as well. 

In short, the $\tilde{\psifull}^{(i)}$ now satisfy the requirements for the trial wave functions in the variational calculations (c.f.\  Sec.~\ref{subsec:KohnFormalism}), even though they are not  eigenstates of the training Hamiltonian anymore. Based on these satisfied boundary conditions, we get an interesting and useful corollary, by repeating the derivations leading to Eq.~\eqref{eq:var_der_2}:  
\begin{align}
   &  \langle \tilde{\psifull}_{\vecPini}^{(i)} | E - H |  \tilde{\psifull}_{\vecPini}^{(j)}\rangle - (i \leftrightarrow j)
   =  3 i \mathcal{N}^2 \left(S^{(i)} -S^{(j)} \right)   .  \label{eq:formula_check_deltaU} 
\end{align}
This property guarantees that the EC emulator errors are exactly zero for the emulation point that is the same as one of the training points. Note that the corollary is also valid  in the previous two emulator cases.

Equation~\eqref{eq:deltaU_varying_V2body} provides the working formula for computing $\delta U$. 
When the two-body interactions are not varied, the matrix element in Eq.~\eqref{eq:deltaU_varying_V2body} becomes a linear combination of the matrix elements of $V_4^{(j)}$ and $V_4$ at the emulation point. When varying $V_{1,2,3}$, a significant number of extra steps need to be taken (see the SM for more details). The numerical violation of Eq.~\eqref{eq:formula_check_deltaU} is used here to check the accuracy of the numerical overlap integrals in computing $\delta U$. 
Interestingly, all the overlap integrals  are only nonzero at finite range.  The range is determined by the ranges of the interactions as well as the size of the dimer bound state, which is consistent with the physical intuition for a particle scattering off a compound system.

\begin{figure*}
    \centering
    \includegraphics[width=\textwidth]{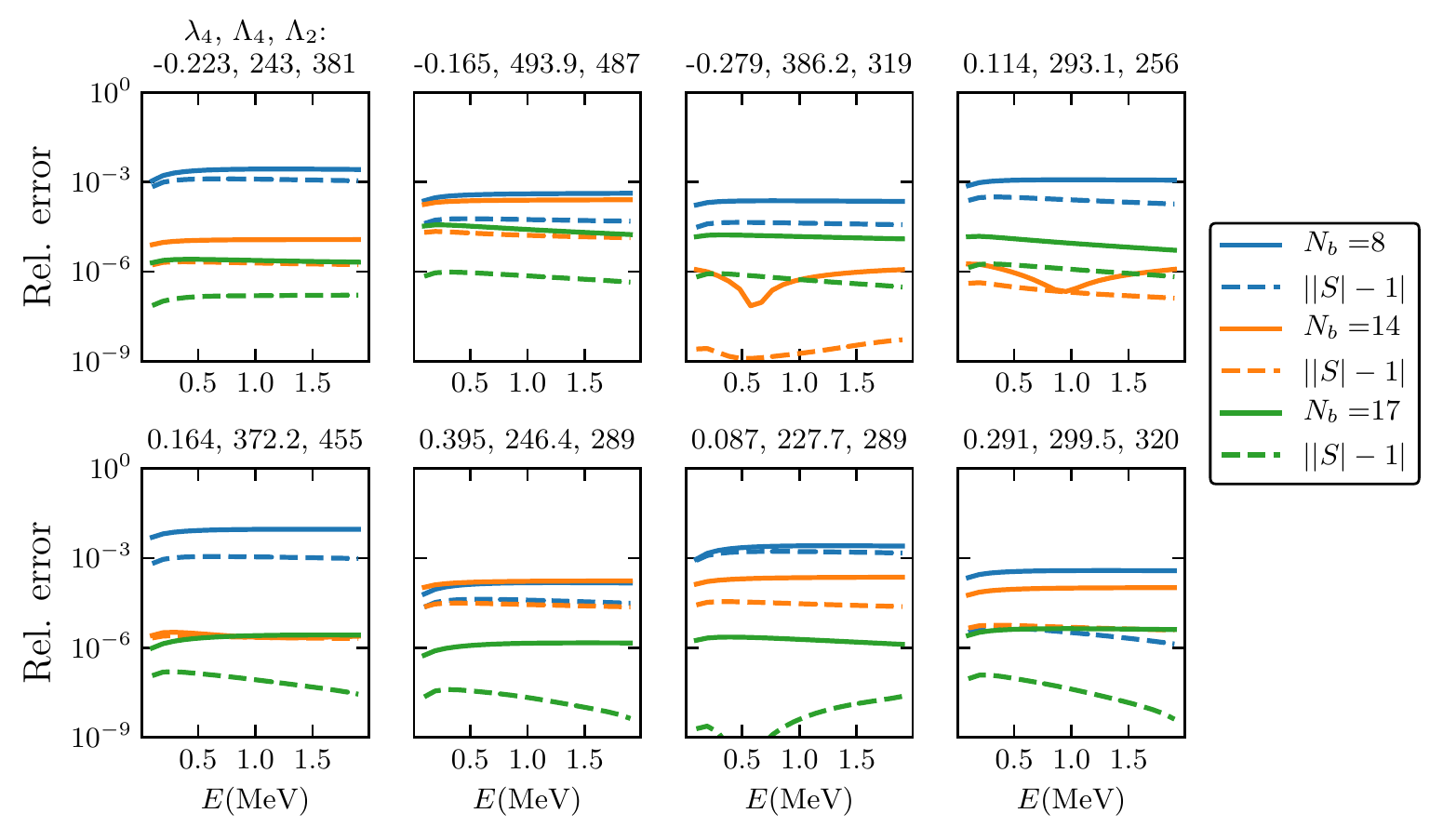}
    \caption{The EC emulator errors for a few test points in the three-dimensional parameter space with increasing $N_b$.   The solid and dashed curves in the same color are the $S$ matrix's relative errors and the unitarity violation errors respectively with a fixed $N_b$ (see legend).  }
    \label{fig:Emulator_RelError_Expamples_B_2_Gau}
\end{figure*}

\begin{figure*}
    \centering
    \includegraphics[width=\textwidth]{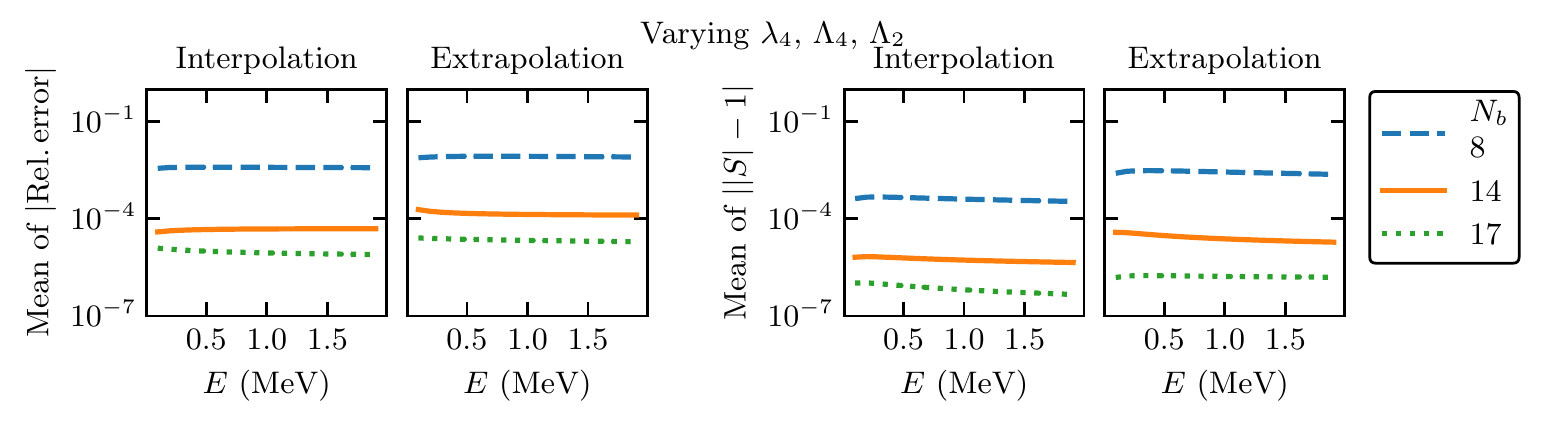}
    \caption{The mean  of the EC emulator errors in the three-dimensional parameter space with the same training sets as in Fig.~\ref{fig:Emulator_RelError_Expamples_B_2_Gau} (see legend). }
    \label{fig:Emulator_RelError_B_2_Gau}
\end{figure*}

Similar to Fig.~\ref{fig:Emulator_RelError_Expamples_B_2_LAM2_200_Gau}, Fig.~\ref{fig:Emulator_RelError_Expamples_B_2_Gau} shows the errors of the EC emulator for 8 different test points but now in the three-dimensional parameter space. 
The  space is defined by $ 200 \leq \{\LAMtwo,  \Lambda_4\} \leq 500$ MeV and $ |\lambda_4| \leq 0.5 $. With only $N_b = 8$, the $S$ errors are no worse than one percent; the unitarity violation errors are even smaller. When $N_b = 17$, the $S$ errors are reduced to at most $10^{-5}$. 

In Fig.~\ref{fig:Emulator_RelError_B_2_Gau}, 1000 test points in the three-dimensional parameter space are sampled uniformly. The test points enclosed by the training points are again considered as interpolations, while the others are extrapolations. 
The plot shows the mean of the errors for both samples, which are consistent with those seen for the 8 test points in Fig.~\ref{fig:Emulator_RelError_Expamples_B_2_Gau}. The interpolation and extrapolation errors are similar when $N_b$ becomes large enough. 

It turns out that the emulator errors are dictated by our current numerical error in computing the overlap integrals in Eq.~\eqref{eq:deltaU_varying_V2body}, which are more difficult than the calculations in the previous sections. By using Eq.~\eqref{eq:formula_check_deltaU}, we found the integral errors and the emulator errors  have similar magnitudes. 

\begin{figure*}[tbh]
    \centering
    \includegraphics[width=0.9\textwidth]{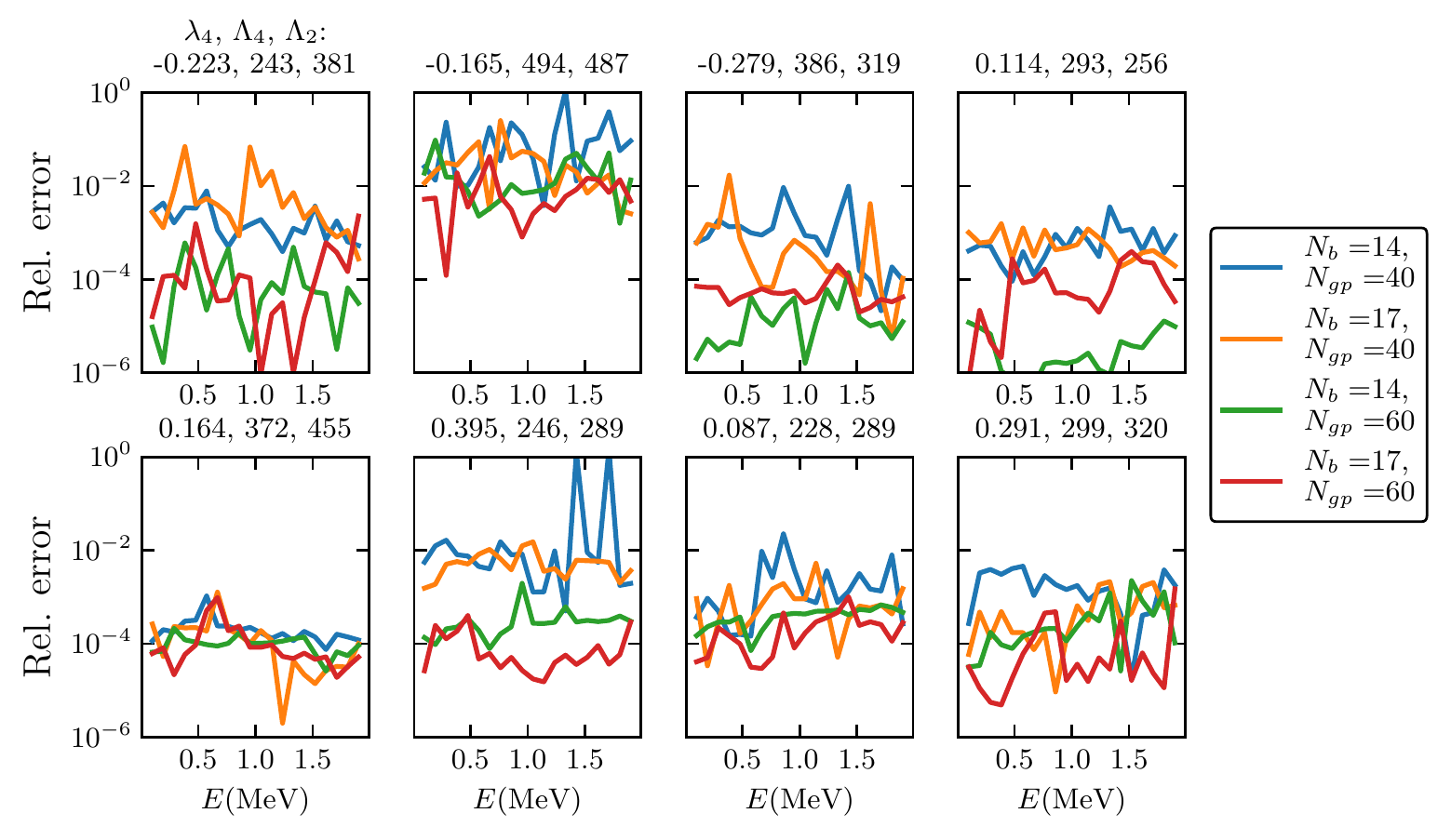}
    \caption{The GP-EC emulator relative errors for the same test points as in Fig.~\ref{fig:Emulator_RelError_Expamples_B_2_Gau}, in the three-dimensional parameter space. Four different combinations of $N_b$ and $N_{gp}$ are explored (see the legend). }
    \label{fig:EmulatorInEmulator_RelError_Expamples_B_2_Gau}
% \end{figure*}
\vspace*{10pt}
% \begin{figure*}
%     \centering
    \includegraphics[width=0.9\textwidth]{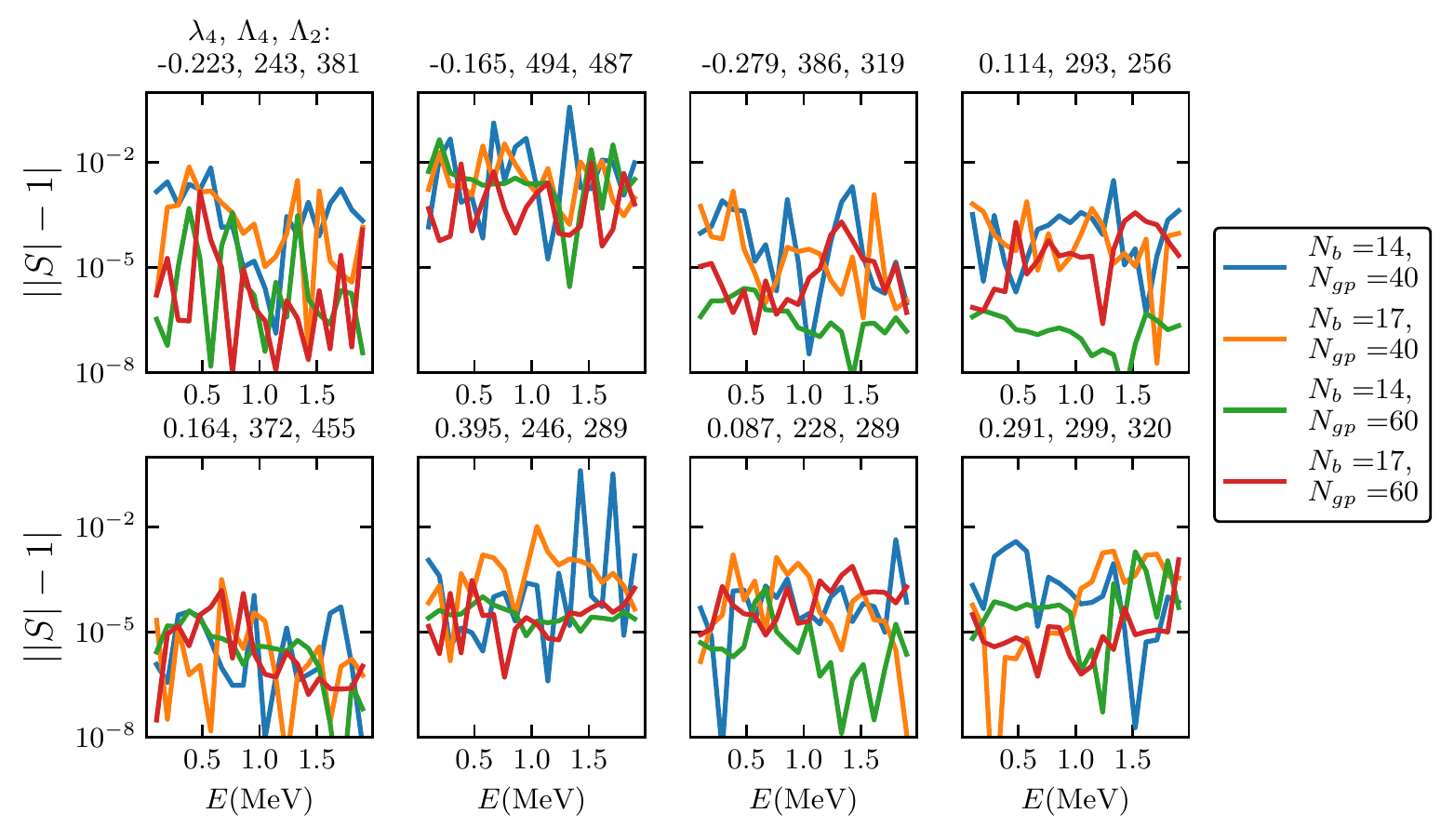}
    \caption{The same as  Fig.~\ref{fig:EmulatorInEmulator_RelError_Expamples_B_2_Gau}, but for the unitarity violation errors of the GP-EC emulators. } 
    \label{fig:EmulatorInEmulator_UnitarityError_Expamples_B_2_Gau}
\end{figure*}

Even though these EC emulators have excellent accuracy, the calculation of $\delta U$  needs to be performed at each emulation point because of the extra modification  applied to the training wave functions. This slows down the emulators dramatically. Therefore, the GP-EC emulator approach is employed here.  For the GP-based emulation---again by using the GPy package~\cite{gpy2014}---of the $\delta U$, the effective dimension of the parameter space is reduced to either 1 or 2, by taking advantage of the  presence of the linear parameter dependencies in $\delta U$.

\begin{figure*}[tbh]
    \centering
    \includegraphics[width=0.9\textwidth]{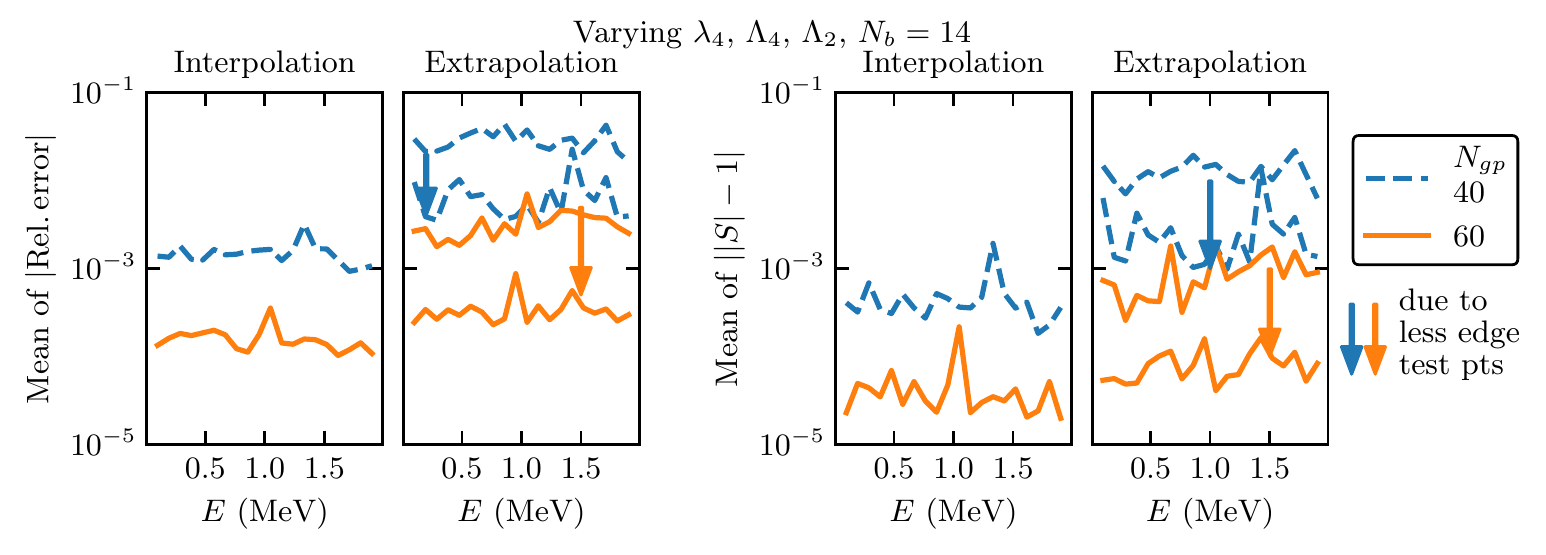}
    \caption{
    The mean of the GP-EC emulator errors vs $E$  in the three-dimensional parameter space, with $N_b = 14$ and different $N_{gp}$.  Two different calculations are performed: the first based on the same 1000 test point sample as used in Fig.~\ref{fig:Emulator_RelError_B_2_Gau} with $\LAMtwo$ and $\Lambda_4$ in $[200, 500]$ MeV (318 interpolation and  682 extrapolation points) and the second with $\LAMtwo$ and $\Lambda_4$ in a smaller $[230, 470]$ MeV range (312 interpolation and 325 extrapolation points). As can be seen, reducing the number of the edge test points reduces the errors in the extrapolation sample to a similar level as those in the interpolation sample. 
    }
    \label{fig:EmulatorInEmulator_RelError_B_2_Nb_14_scalarGP_Gau}
\vspace*{10pt}
% 
% 
% \begin{figure*}
%    \centering
    \includegraphics[width=0.9\textwidth]{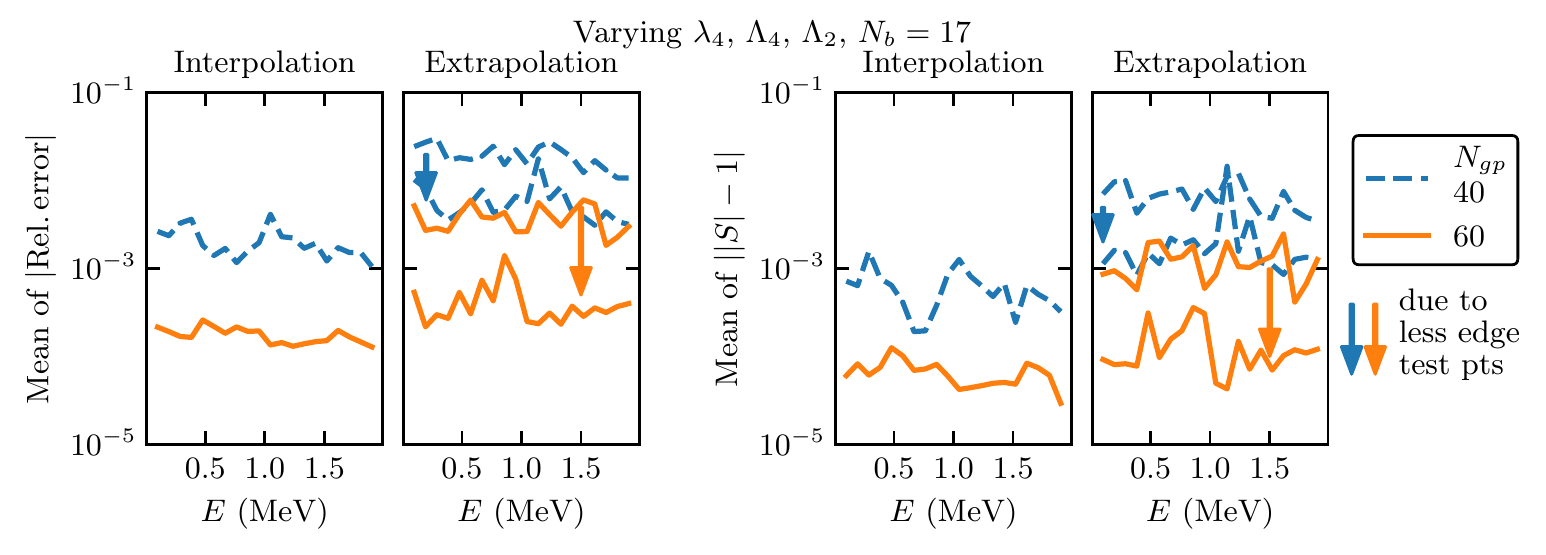}
    \caption{The same as  Fig.~\ref{fig:EmulatorInEmulator_RelError_B_2_Nb_14_scalarGP_Gau}, but with $N_b = 17$. For the full (smaller) sample, there are 322 (302) interpolation and 678 (335) extrapolation points. Similar to Fig.~\ref{fig:EmulatorInEmulator_RelError_B_2_Nb_14_scalarGP_Gau}, reducing the number of the edge test points reduces the errors in the extrapolation sample to the same level as those in the interpolation sample.}
    \label{fig:EmulatorInEmulator_RelError_B_2_Nb_17_scalarGP_Gau}
\end{figure*}

Similar to Fig.~\ref{fig:Emulator_RelError_Expamples_B_2_LAM2_200_Gau}, Figs.~\ref{fig:EmulatorInEmulator_RelError_Expamples_B_2_Gau} and~\ref{fig:EmulatorInEmulator_UnitarityError_Expamples_B_2_Gau}  show the errors of the GP-EC emulator with four different combinations of $N_b$ and $N_{gp}$ for the same 8 test points as shown in Fig~\ref{fig:Emulator_RelError_Expamples_B_2_Gau}. By comparing Figs.~\ref{fig:EmulatorInEmulator_RelError_Expamples_B_2_Gau} and~\ref{fig:Emulator_RelError_Expamples_B_2_Gau}, we can see that the  GP-EC emulator errors are dominated by those of the GP emulations. Therefore, the most significant accuracy improvement comes from increasing $N_{gp}$, not from  increasing $N_b$ with fixed $N_{gp}$. 
Another important observation can be made: the errors for the test points near the edge of the parameter space (called ``edge test points''), such as the one for panel (2), are significantly larger than those for the interior points. This is due to the large extrapolation errors of the GP emulation.

Figures~\ref{fig:EmulatorInEmulator_RelError_B_2_Nb_14_scalarGP_Gau} and~\ref{fig:EmulatorInEmulator_RelError_B_2_Nb_17_scalarGP_Gau} show the mean of the errors for $N_b = 14$ and $N_b = 17$, respectively. 
The interpolation and extrapolation points are defined in the same way as in Sec.~\ref{subsec:emulator_2_dim}. 
Increasing $N_{gp}$ from 40 to 60 reduces the errors from $10^{-3}$ ($10^{-2}$) to $10^{-4}$ ($10^{-3}$) for the interpolation (extrapolation) points. 
We further exclude the edge test points  by restricting $\LAMtwo$ and $\Lambda_4$ to a narrower range, $[230, 470]$ MeV. 
The mean of the errors for the interpolations are not changed, but for the extrapolations the mean gets reduced to a similar level as that of the interpolation points, which
is consistent with the observation we made for  Fig.~\ref{fig:EmulatorInEmulator_RelError_Expamples_B_2_Gau}. 
It also suggests that to improve the emulation accuracy, the parameter space for training should be chosen larger than  for emulation.

By employing the GP emulation of $\delta U$, the full time cost of the GP-EC emulator for this case is on the order of milliseconds.  In comparison,  the original EC emulator takes $10^3$ seconds for each evaluation. 
The memory costs are  larger---from 10s up to 100 MB---than the GP-EC costs in Sec.~\ref{subsec:emulator_2_dim}, caused by the larger $N_b$ and  $N_\mathrm{gp}$ here.

\subsection{Training costs} \label{subsec:trainingcost}

The computing costs for training the emulators are another important factor to be considered. Since they are highly dependent on the complexity of the three-body problem, we concentrate on their scaling  with the training parameters and potential parallelizations of the training computations.  

The cost budget---at a single scattering energy---for the ``nonlinear-2'' emulator has three main parts: the first for directly solving the scattering problem at $N_b$ EC training points, the second for computing $\delta U$ at $N_{gp}$ GP training points, and the last for training the GPs using the GPy package. The first costs are proportional to $N_b$---on the order of 10 for the emulation in the three-dimensional space! These $N_b$ calculations are mutually independent and can  be readily parallelized. 

The second costs scale as $N_b^2\,N_{gp}$. They could also be significant but can be easily parallelized as well into $N_b^2\,N_{gp}$ independent calculations. 
The procedure of correcting the training wave functions' asymptotic behavior does not increase the second costs in any significant way. The two-body bound state problem can be either fully solved quickly---as implemented in this work---or by using a bound-state EC emulator~\cite{Konig:2019adq,Ekstrom:2019lss}. 
The GP training costs scale as $N_b^2\,N_{gp}^3$, however the calculations can be parallelized into $N_b^2$ independent ones.  

The ``nonlinear-1'' emulator has the same training budget structure, but the training calculations are simpler. The ``linear'' emulator costs  the least without using GP.

\section{Summary and outlook} \label{sec:summary}

\begin{figure}[tbh]
    \centering
    \includegraphics[width=\columnwidth ]{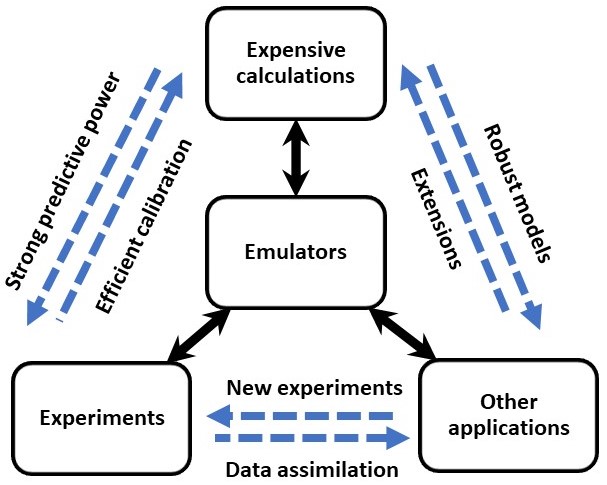}
    \caption{The emulator's role in the workflow for nuclear physics and other research areas. The dashed arrows indicate desirable connections enabled by the emulators.}
    \label{fig:emuator_in_workflow}
\end{figure}

In this work, we have developed fast emulators for quantum three-body scattering by combining the variational method and the EC concept. Different variants of the emulators are studied in anticipation of different application scenarios. Their accuracy and computing costs are summarized in Table~\ref{tab:results_summary}, and discussed in detail in Sec.~\ref{sec:3body_emulators}: the emulator errors in Figs.~\ref{fig:EmuErr_3body_B_2_LAM2_200_LAM4_300_gau}, 
\ref{fig:EmulatorInEmulator_RelError_B_2_LAM2_200_Nb_6_9_scalarGP_Gau}, \ref{fig:EmulatorInEmulator_RelError_B_2_Nb_14_scalarGP_Gau}, and~\ref{fig:EmulatorInEmulator_RelError_B_2_Nb_17_scalarGP_Gau}  (and  Figs.~\ref{fig:EmuErr_3body_B_10_LAM2_70_LAM4_80_gau}, \ref{fig:EmulatorInEmulator_RelError_B_10_LAM2_75_Nb_6_scalarGP_Gau}, \ref{fig:EmulatorInEmulator_RelError_B_10_LAM2_75_Nb_9_scalarGP_Gau}, \ref{fig:EmulatorInEmulator_RelError_B_10_Nb_14_scalarGP_Gau}, and~\ref{fig:EmulatorInEmulator_RelError_B_10_Nb_17_scalarGP_Gau} for the nuclear case), the emulator computing costs in the discussions of these figures, and their training costs in Sec.~\ref{subsec:trainingcost}.  (The codes to generate the results presented in this paper will be made public~\cite{BUQEYEgithub}.)

In Sec.~\ref{subsec:emulator_2_dim}, two different solutions are explored to expedite the EC emulators that get slowed down due to the nonfactorizable parametric dependence of the varying interactions within the Hamiltonian. One uses Lagrange functions to linearly decompose the interaction operators (see Eq.~\eqref{eq:g4decomp_Lag}), while the other---called a GP-EC emulator---employs GP emulation of the $\delta U$ matrix. The GP-EC emulators are again applied in Sec.~\ref{subsec:emulator_3_dim} to achieve fast emulation. The section also demonstrates a working procedure to correct the asymptotic behavior of the trial wave functions when building the  emulators.

Now, we comment on a few important and interesting generalizations and  their potential impact in nuclear physics and  beyond. 

To emulate $N$-$d$ and $d$-nucleus scattering, as well as  any processes that can be modeled as three-cluster systems, the current emulators need to be extended to include spin and isospin degrees of freedom, fermion statistics, and  higher partial waves (or even full scattering amplitude without partial wave decomposition). 
It will also be  useful to generalize the emulators for the energy region above the dimer break-up threshold  and to include scattering, nuclear excitation, and dimer breakup channels. Constructing  EC emulators for four- and higher-body scatterings is another interesting topic.

Our current estimates of the emulation accuracy and costs, as outlined in Table~\ref{tab:results_summary}, should be good guides for the future emulations of the \emph{realistic} calculations mentioned above.   
The essential question is how the numbers of trainings---determining both accuracy and costs---extrapolate from the present work to realistic calculations. In the EC emulation of the many-body nuclear structure calculation in Ref.~\cite{Ekstrom:2019lss},  $N_b \sim 60$ trainings were employed in a 16-dimensional parameter space of the chiral nucleon interaction theory to achieve percent-level emulation accuracy. 
For such $N_b$, which are only a few times of the largest $N_b = 17$ used here (in the ``nonlinear-2'' case), we expect the estimates in Table~\ref{tab:results_summary} to  hold within an order of magnitude.

Such emulators have the potential to create a new  workflow in nuclear physics research and in other fields, as implied by the diagram in Fig.~\ref{fig:emuator_in_workflow}.%
\footnote{We acknowledge discussions about potential broader impact of emulators at the INT Program 21-1b held online between April 19 and May 7, 2021~\cite{INT-Program-21-1b}.}
Although our focus is the three-body quantum scattering problem, the workflow diagram and its discussion could apply in other situations, such as higher-body scattering and many-body bound-state studies.

Calculations starting from more microscopic  physics
generally promise physics modeling with great predictive power, less model dependence, and more complete uncertainty quantification. 
These advantages are  desirable, but their expensive computational costs have frequently kept their interactions with other sectors (the other two blocks in Fig.~\ref{fig:emuator_in_workflow}) from growing stronger. This means a loss of novel research opportunities.  

 Developing efficient and accurate  emulators  will fundamentally change this situation. As the proxy of the expensive calculations, they can be coupled to data analysis and other applications, thanks to their low computational costs. Novel studies could emerge with the fostering of new interactions between different research areas.  For example, three-body emulators will enable direct applications of Faddeev scattering equations in experimental data analysis, experimental design, and the efficient calibration of theory.

The ``other applications'' in Fig.~\ref{fig:emuator_in_workflow} include collaborations with other theoretical approaches, 
either more or less microscopic  than the ``expensive calculations''.
For the three-nucleon emulators, they could be  Lattice-QCD calculations or a two-body study that treats the deuteron as a single particle degrees of freedom. 

In collaborations, ``expensive calculations'' provide  robust models for other approaches to compare with, and in return extend the range of their own applications.  An important part of working with  different approaches is in fact data analysis, i.e., constraining the macroscopic theories and models with the information from more microscopic ones. The three-nucleon emulators can be applied to extract three-nucleon interactions from Lattice-QCD calculations, an approach%
\footnote{This  approach is similar to that in Ref.~\cite{Eliyahu:2019nkz}. 
However instead of an emulator, the so-called Stochastic Variational Method was used there to solve the few-body problem.} that is different from current generalizations of the L\"{u}scher method~\cite{Jackura:2019bmu}. 
Similarly, matching three-cluster theories and many-body structure calculations can create a new method of {\it ab initio} nuclear scattering calculation, a desirable generalization of the recent two-cluster study~\cite{Zhang:2019odg}. 

Through the emulators of the ``expensive calculations'', the ``experiments'' and ``other applications'' are efficiently coupled as well. This allows the assimilation of experimental data into theoretical studies, which in return could create new experimental proposals. The three-nucleon emulators could provide a robust connection between  $N$-$d$ scattering measurements and the three-nucleon energy levels computed in Lattice-QCD. 

By merging variational methods with the EC principle, a new way to construct emulators is opened up. 
This strategy is fundamentally different from those based on GPs or neural networks, as used in machine learning. The EC emulators take into account the physics involved through variational methods, while the others learn the physics based on statistical methods. The new strategy may not have as broad applications, but whenever applicable,
the EC emulators could 
have much better accuracy for both interpolations and extrapolations. 

\emph{Note:} recently another type of three-nucleon scattering emulator was proposed in Ref.~\cite{Witala:2021xqm}. The emulator is based on a lowest-order perturbative treatment of the varying potential operators, in particular the contact three-nucleon interactions. The application scenario is the same as the ``linear'' case in Table~\ref{tab:results_summary}. The emulation accuracy is on the percent level and the time costs are seconds on a personal computer.  

\acknowledgments{
%We thank ?? for useful discussions.
X.Z and R.J.F were supported in part by the National Science Foundation
under Grant No.~PHY--1913069 and the CSSI program under award number OAC-2004601 (BAND Collaboration), and by the NUCLEI SciDAC Collaboration under Department of Energy MSU subcontract RC107839-OSU. X.Z. was   supported in part by the U.S. Department of Energy, Office of Science, Office of Nuclear Physics, under the FRIB Theory Alliance award DE-SC0013617.}

%\clearpage

\begin{appendix}

\section{Conventions for states}\label{app:conventions}
Single-particles states in coordinate and momentum space satisfy
\begin{align}
     \langle \myvec{p}'  | \myvec{p} \rangle &= \delta (\myvec{p} - \myvec{p}'), \\
     \langle \myvec{r}'  | \myvec{r} \rangle &= \delta (\myvec{r} - \myvec{r}'),  \\
      \langle \myvec{r} | \myvec{p} \rangle &= \frac{e^{i\myvec{p}\myvec{r} } }{\sqrt{\left(2\pi\right)^3}} . 
\end{align}
The partial wave bases are defined through   
\begin{align}
    \langle \myvec{p}' | p, \ell, m\rangle & = \frac{\delta(p'-p)}{p^2} Y_{\ell m }(\hat{p}')   , \\ 
    \langle \myvec{r}' | r, \ell, m\rangle & = \frac{\delta(r'-r)}{r^2} Y_{\ell m }(\hat{r}') . 
\end{align}
It can be  verified that 
\begin{align}
    \langle p', \ell', m' | p, \ell, m\rangle & = \frac{\delta(p'-p)}{p^2} \delta_{\ell \ell'} \delta_{m m'} , \\ 
    \langle \myvec{r} | p, \ell, m\rangle & =  \sqrt{\frac{2}{\pi}} i^\ell Y_{\ell m }(\hat{r}) j_\ell(p r) .
\end{align}

Let us turn to the states in the three-body system. The CM degree of freedom, which is always factorized from the full state, is kept implicit in the states. The states are now labeled by two kinematic variables---either coordinate or momentum---for the relative motions, as shown in Fig.~\ref{fig:3-body_layout} and discussed in the beginning of Sec.~\ref{subsec:FaddeevFormalism}. 
There are three equivalent sets of kinematic variables describing the same system. The variables are linearly related. For the case with the same mass for all particles, they are~\cite{Glockle:1983}:
\begin{align}
    \myvec{R}_2 & = -\frac{1}{2}\myvec{R}_1 +  \frac{3}{4} \myvec{r}_1 , \quad 
    \myvec{r}_2  = - \myvec{R}_1  -  \frac{1}{2} \myvec{r}_1 , \notag \\ 
    \myvec{R}_3 & = -\frac{1}{2}\myvec{R}_1 -  \frac{3}{4} \myvec{r}_1  ,   \quad 
    \myvec{r}_3  =  \myvec{R}_1  -  \frac{1}{2} \myvec{r}_1 . \label{eq:kinRrrels1}
\end{align}
Importantly, for $R_1$ (or $r_1$) approaching $+\infty$, the other variables approach $+\infty$ as well. 

The index outside of a state, such as $i$ in $| {} \rangle_i$, identifies the particular choice of coordinate system. 
Thus, $|\myvec{R}_1, \myvec{r}_1  \rangle_1$ means using the coordinate system with particle $1$ as spectator, with $\myvec{r}_1$ and $\myvec{R}_1$ for the relative coordinates between particle 2 and 3 and between 1 and the CM of 2 and 3, respectively. In contrast, $|\myvec{R}_2, \myvec{r}_2  \rangle_1$ has $\myvec{r}_2$ and $\myvec{R}_2$ for the relative coordinates between particle 2 and 3 and between 1 and the CM of 2 and 3 (\emph{not between particle 1 and 3 and between 2 and the CM of 1 and 3}). Of course,  $|\myvec{R}_\mu, \myvec{r}_\mu \rangle_\mu$ with $\mu = 1, 2, 3$ are the same state. The momentum space states are defined in the same way. In order to simplify the notation, the $|\myvec{R}_1, \myvec{r}_1 \rangle_1$ basis state will have index $1$ omitted, and the other states, such as $  | \myvec{P}_\gamma, g_\gamma \rangle $, rely on the index $\gamma$ to label the coordinate system, whenever doing so doesn't cause confusion.

\section{Three-body scattering wave functions} \label{app:Faddeev}

Detailed discussions of the Faddeev equations can be found in  Ref.~\cite{Glockle:1983}. 
Here we focus on computing scattering wave functions based on half-off-shell transition amplitudes, $X$, and the asymptotic behavior of the wave functions in coordinate space. 

We start with the  $V_4 =0$ case.
The Faddeev components are defined as 
\begin{align}
    | \psifd_{\alpha, \mu } \rangle & \equiv G_0 V_\mu | \psifull_\alpha^{(+)} \rangle  , \\ 
    \sum_\mu | \psifd_{\alpha, \mu } \rangle & =  G_0 \left(\sum_\mu V_\mu\right) | \psifull_\alpha^{(+)} \rangle = |\psifull_\alpha^{(+)} \rangle ,
\end{align}
with $\mu = 1, 2, 3$. To get the expressions for $\psifd_{\alpha,\mu}$, we can now multiply Eq.~\eqref{eq:UcoupledW3body3}  by $G_0$ from the left. For $\alpha =1$, we have
\begin{align}
    |\psifd_{1,1}\rangle & = \frac{1}{2} G_0 \left(U_{21}+ U_{31}- U_{11} \right) |\psifree_1\rangle  \   \notag \\ 
    & = G_0 |\myvec{P}_1, \hat{g}_1 \rangle + G_0 t_1 G_0 U_{11} G_0 |\myvec{P}_1, \hat{g}_1 \rangle , \label{eq:FDcompWo3BVdef1}\\
     |\psifd_{1,2}\rangle & = \frac{1}{2} G_0 \left(U_{11}+ U_{31}- U_{21} \right) |\psifree_1\rangle \   \notag \\ 
     & = G_0 t_2 G_0 U_{21} G_0 |\myvec{P}_1,\hat{g}_1 \rangle  ,  \label{eq:FDcompWo3BVdef2} \\ 
     |\psifd_{1,3}\rangle & = \frac{1}{2} G_0 \left(U_{11}+ U_{21}- U_{31} \right) |\psifree_1\rangle \   \notag \\ 
     & = G_0 t_3 G_0 U_{31} G_0 |\myvec{P}_1,\hat{g}_1 \rangle   . \label{eq:FDcompWo3BVdef3}  
\end{align}
These expressions can be generalized to~\cite{Watsonbook1967} 
\begin{align}
    |\psifd_{\alpha,\mu} \rangle & = \delta_{\alpha\mu}G_0 |\myvec{P}_\alpha, \hat{g}_\alpha \rangle + G_0 t_\mu G_0 U_{\mu\alpha} G_0 |\myvec{P}_\alpha, \hat{g}_\alpha \rangle . 
\end{align}

When $V_4 \neq 0$, the aforementioned FCs cannot be summed to get the full wave function. A fourth FC $| \psifd_{1,4} \rangle \equiv G_0 V_4 |  \psifull_1^{(+)} \rangle $ can be temporarily introduced. By repeating the derivations that leads to Eq.~\eqref{eq:FDcompWo3BVdef1}, \eqref{eq:FDcompWo3BVdef2} and~\eqref{eq:FDcompWo3BVdef3}, 
we get the same expressions for $\psifd_{1,\mu}$ for $\mu = 1,2,3$ and in addition, 
\begin{align}
     |\psifd_{1,4}\rangle & = \frac{1}{3} G_0 \left(U_{11}+ U_{21} + U_{31} - 2 U_{41} \right) |\psifree_1\rangle \notag \\ 
     & = G_0 t_4 G_0 U_{41} G_0 |\myvec{P}_1,\hat{g}_1 \rangle \  \notag \\
     &  = \bigg( G_0 t_4 + G_0 t_4  \sum_\beta G_0 t_\beta G_0 U_{\beta 1} \bigg)  G_0 |\myvec{P}_1,\hat{g}_1 \rangle . 
\end{align}
In the last step, the following identity is invoked~\cite{Glockle:1983}:
\begin{align}
    U_{4\alpha} & = G_0^{-1}  +  \sum_{\gamma} t_\gamma G_0 U_{\gamma \alpha}  .  \label{eq:UcoupledW3body2}
\end{align}
This suggests a new way to represent the full wave function, by adding all the pieces of $|\psifd_{1,4} \rangle $ to the other three FCs. We finally have Eq.~\eqref{eq:FC_distinguishable} for general channel $\alpha$. 

For identical bosons, the symmetrized FC is (see Sec.~\ref{subsec:id_bosons})
\begin{widetext}

\begin{align}
    |\psifd_\mu \rangle = \sum_{\alpha} |\psifd_{\alpha,\mu} \rangle= (1+G_0 t_4)\, G_0 |\myvec{P}_\mu, \hat{g}_\mu \rangle + \sum_\alpha (1+G_0 t_4)\, G_0 t_\mu G_0 U_{\mu\alpha} G_0 |\myvec{P}_\alpha, \hat{g}_\alpha \rangle
\end{align}
In momentum space with e.g., $\mu =1$, 

\begin{align}
    \langle \myvec{P}_1, \myvec{q}_1  |\psifd_{\vecPini,1} \rangle   = \, &  \tilde{Z}_d(\myvec{P}_1, \myvec{q}_1)\delta(\myvec{P}_1 -\vecPini) + \tilde{Z}_{14}(\myvec{P}_1,\myvec{q}_1)\,\tau_4(E)\, Z_{41}(\vecPini) + \tilde{Z}_d(\myvec{P}_1, \myvec{q}_1) \, \hat{\tau}(E-\frac{P_1^2}{2\mu^1})\, X (\myvec{P}_1, \vecPini) \notag \\ 
    & +  \tilde{Z}_{14}(\myvec{P}_1,\myvec{q}_1)\,\tau_4(E)\, \int d \myvec{P}\, Z_{41}(\myvec{P}) \, \hat{\tau}(E-\frac{P^2}{2\mu^1})\, X (\myvec{P}, \vecPini) ,  \label{eq:FDinpspaceW3BV}
\end{align}
\end{widetext}
with 
\begin{align}
    \tilde{Z}_d (\myvec{P}_1,\myvec{q}_1) 
     \equiv\, &  \frac{\hat{g}(\myvec{q}_1)}{E+i\epsilon-\frac{P_1^2}{2\mu^1}-\frac{q_1^2}{2\mu_1} }  , \\
    \tilde{Z}_{14} (\myvec{P}_1,\myvec{q}_1)  &\equiv  \langle \myvec{P}_1, \myvec{q}_1  | G_0(E)|g_4 \rangle  . 
\end{align}    
Other quantities, including $Z_{41}$ and $\tau_4$, are defined in Sec.~\ref{subsec:FaddeevFormalism}.

The asymptotic behavior of the FCs at large particle-dimer separation, as shown in Eq.~\eqref{eq:WFbelowbreakupAsymWo3BV}, is determined by their singularities in momentum space in Eq.~\eqref{eq:FDinpspaceW3BV}, including the $\delta(\myvec{P}_1 - \vecPini)$ term and the pole of  $\hat{\tau}(E-P_1^2/2\mu^1)$ in the third term when $P_1  = \Pini$. The other terms proportional to $\tau_4$ decrease exponentially with $R_1$ and thus do not contribute to the asymptotic behavior, below the break-up threshold.     

We can isolate these singularities by subtracting the FC with the asymptotic piece defined in Eq.~\eqref{eq:FD_asymp_reg_pspace}. Within $\psifdasym$, the large $P_1$ dependence of the $P_1 = \Pini$ pole term is further regularized using the $\tilde{\Lambda}$ parameter. In coordinate space, the subtraction projected to s-waves is
\begin{align}
\langle {R}_1, {r}_1|\psifdasym_{\vecPini, 1}\rangle 
   & =  \frac{\mathcal{N}}{\sqrt{v} } \frac{u_B({r}_1) }{ r_1 R_1}  \bigl(- e^{-i \Pini R_1} + e^{2i\delta_0} e^{i\Pini R_1} 
   \notag \\
   & \quad\null - 2 i \sin\delta_0 e^{i\delta_0} e^{-\tilde{\Lambda}R_1 } \bigr)     . 
   \label{eq:FD_asymp_reg_rspace} 
\end{align}
As can be seen, $\tildepsifdasym$ approaches to the FC's asympototic behavior at large $R_1$, but meanwhile becomes finite when $R_1 \to 0$, a boundary condition for a physical solution, thanks to the $\tilde{\Lambda}$-dependent regularization. 
%\end{widetext}

\section{Emulators} \label{app:emulators}
The following formula is used to compute $\delta U$ in constructing the EC emulators with varying  $V_{1,2,3}$ and $V_4$: 
\begin{align}
   \langle \tilde{\psifull}^{(i)} | E &- H |  \tilde{\psifull}^{(j)}\rangle 
   =  \langle \psifullsub_{(i)} | V^{(j)} - V |  \psifullsub_{(j)}\rangle \notag \\ 
   & \null + 3\, \langle \psifullsub_{(i)} | V^{(j)}_2 + V^{(j)}_3 + V^{(j)}_4 |  \psifdasym_{(j), 1}\rangle \notag \\ 
   & \null - 3\, \langle \psifullsub_{(i)} | V_2 + V_3 + V_4 |  \tildepsifdasym_{(j), 1}\rangle \notag \\
   & \null - 3\, \frac{\Pini^2 + \tilde{\Lambda}^2}{2\mu^1} \langle \psifullsub_{(i)} | \Delta\psifdasym_{(j), 1}  -  \Delta\tildepsifdasym_{(j), 1}\rangle \notag \\ 
   & \null + \langle \tildepsifullasym_{(i)} | V^{(j)} - V |  \psifullsub_{(j)}\rangle \notag \\ 
   & \null + 3\,\langle \tildepsifullasym_{(i)} | V^{(j)}_2 + V^{(j)}_3 + V^{(j)}_4 |  \psifdasym_{(j), 1}\rangle \notag \\ 
   & \null - 3\, \langle \tildepsifullasym_{(i)} | V_2 + V_3 + V_4 |  \tildepsifdasym_{(j), 1}\rangle \notag \\ 
   & \null - 3\, \frac{\Pini^2 + \tilde{\Lambda}^2}{2\mu^1} \langle \tildepsifullasym_{(i)} | \Delta\psifdasym_{(j), 1}  -  \Delta\tildepsifdasym_{(j), 1}\rangle . \label{eq:deltaU_varying_V2body}
\end{align}
$\tilde{\psifull^{(i)}}$ is defined in Eq.~\eqref{eq:modified_basis_def}. $V$ and $V^{(j)}$ are the full potential including both two and three-body interactions for the emulation point and the $j$th training point (labelled by $(j)$ in sub or superscripts) respectively. And $ \Delta\psifdasym_{(j), 1} $ is the difference between the regularized asymptotic FC in Eq.~\eqref{eq:FD_asymp_reg_pspace}  and the non-regularized one (i.e., the regularized one but with $\tilde{\Lambda} \to +\infty$):
\begin{align}
     \langle P_1, q_1 |\Delta\psifdasym_{(j), 1}\rangle  & \equiv  \psidimer_{B,(j)}(q_1) \frac{8\pi\mu^1}{P_1^2 + \tilde{\Lambda}^2}  X_{(j)}(\Pini, \Pini) , \\
      \langle P_1, q_1 | \Delta\tildepsifdasym_{(j), 1}\rangle  & \equiv  \psidimer_{B}(q_1) \frac{8\pi\mu^1}{P_1^2 + \tilde{\Lambda}^2}  X_{(j)}(\Pini, \Pini)  ,     \\
    \text{with} \  \psidimer_{B,(j)}(q_1) & \equiv \frac{\hat{g}_{(j)}({q})}{-B -\frac{q^2}{2\mu_1}}  , \ \psidimer_{B}(q_1)  \equiv \frac{\hat{g}({q})}{-B -\frac{q^2}{2\mu_1}} .
\end{align}
In coordinate space, they become 
\begin{align}
     \langle R_1, r_1| \Delta\psifdasym_{(j), 1} \rangle  & = -\psidimer_{B,(j)}(r_1) e^{i\delta_{(i)}} \sin{\delta_{(i)}}  \sqrt{\frac{2}{\pi}}  \frac{e^{-\tilde{\Lambda}R_1 }}{\Pini R_1}, \\ 
     \langle R_1, r_1| \Delta\tildepsifdasym_{(j), 1}\rangle  & =  -\psidimer_{B}(r_1) e^{i\delta_{(i)}} \sin{\delta_{(i)}}\sqrt{\frac{2}{\pi}} \frac{e^{-\tilde{\Lambda}R_1 }}{\Pini R_1}.
\end{align}
Note $\psidimer_{B}(q_1)$ and $\psidimer_{B}(r_1)$, for example, are the dimer bound-state wave function in momentum and coordinate space.

To derive Eq.~\eqref{eq:deltaU_varying_V2body},  we make use of Eq.~\eqref{eq:modified_basis_def} and  
\begin{align}
    (E- H) | \tilde{\psifull}^{(j)}  \rangle & = (E- H) | \psifull^{(j)} \rangle - (E- H) | \psifullasym_{(j)} \rangle
    \notag \\ 
    & \quad\null + (E- H) | \tildepsifullasym_{(j)} \rangle \notag \\ 
    & = (V^{(j)} - V ) \left( | \psifullsub_{(j)} \rangle + | \psifullasym_{(j)} \rangle \right) 
    \notag \\
    & \quad\null - (E - H ) \left( | \psifullasym_{(j)} \rangle - | \tildepsifullasym_{(j)} \rangle \right) . \label{eq:deltaU_varying_V2body_step1}
\end{align}
Because 
\begin{align}
    (E- H_0 - V_1^{(j)}) | \psifdasym_{(j), 1}\rangle & = \frac{\Pini^2 + \tilde{\Lambda}^2}{2\mu^1} | \Delta\psifdasym_{(j),1}\rangle ,  \\
     (E- H_0 - V_1) | \tildepsifdasym_{(j), 1}\rangle & = \frac{\Pini^2 + \tilde{\Lambda}^2}{2\mu^1} | \Delta\tildepsifdasym_{(j),1}\rangle  , 
\end{align}
we have 
\begin{align}
    (E - H_0 ) | \psifullasym_{(j)} \rangle & = (E - H_0 ) \sum_\mu | \psifdasym_{(j)}\rangle \notag\\
    &= \sum_\mu V_\mu^{(j)} |\psifdasym_{(j), \mu} \rangle  \notag \\
    & \quad \null + \frac{\Pini^2 + \tilde{\Lambda}^2}{2\mu^1} \sum_\mu \Delta\psifdasym_{(j),\mu} , 
\end{align}
\begin{align}
    (E - H_0 ) | \tildepsifullasym_{(j)} \rangle & = (E - H_0 ) \sum_\mu | \tildepsifdasym_{(j)}\rangle \notag\\
    &= \sum_\mu V_\mu |\tildepsifdasym_{(j), \mu} \rangle \notag \\
    & \quad\null + \frac{\Pini^2 + \tilde{\Lambda}^2}{2\mu^1} \sum_\mu \Delta\tildepsifdasym_{(j),\mu} .  
\end{align}
Plugging the above formula into Eq.~\eqref{eq:deltaU_varying_V2body_step1} gives Eq.~\eqref{eq:deltaU_varying_V2body}.

\end{appendix}

%\bibliography{bayesian_refs, Other_refs}
%merlin.mbs apsrev4-1.bst 2010-07-25 4.21a (PWD, AO, DPC) hacked
%Control: key (0)
%Control: author (8) initials jnrlst
%Control: editor formatted (1) identically to author
%Control: production of article title (-1) disabled
%Control: page (0) single
%Control: year (1) truncated
%Control: production of eprint (0) enabled
%

\clearpage

%%%%%%%%%% Merge with supplemental materials %%%%%%%%%%
\setcounter{equation}{0}
\setcounter{figure}{0}
\setcounter{table}{0}
\setcounter{section}{0}
\setcounter{page}{1}
\makeatletter
\renewcommand{\theequation}{S\arabic{equation}}
\renewcommand{\thefigure}{S\arabic{figure}}
\makeatother

\onecolumngrid
\begin{center}
  \textbf{\large Supplementary Material for Fast emulation of quantum three-body scattering}\\[.4cm]
  Xilin Zhang,$^{1,2}$ and  R.~J.~Furnstahl $^2$ \\[.1cm]
  $^1${\itshape Facility for Rare Isotope Beams, Michigan State University, MI 48824, USA} \\ 
  $^2${\itshape Department of Physics, The Ohio State University, Columbus, Ohio 43210, USA} 
\end{center}

\twocolumngrid

\section{Additional details for the exact solutions of the three-body scattering problem}

In the two body sector, the interaction form factor in coordinate space is 
\begin{align}
    g(\myvec{r})  = g({r}) =  C\,\LAMtwo^2  \exp\left[-\frac{\LAMtwo^2 r^2}{2 }\right]  . \label{eq:gauFF2def_r_space}
\end{align}  

The $C$  in Eq.~\eqref{eq:gauFF2def_r_space} and~\eqref{eq:gauFF2def} defining the Gaussian form factor is chosen such that at low-energy, the scattering length and effective range terms in the low energy effective range expansion (i.e., $p \cot\delta = -1/a_0 +1/2\, r_0\, p^2 + ...$) become  
\begin{align}
    -\frac{1}{a_0} & = - \frac{\LAMtwo}{\sqrt{\pi}} \left(1 + \lambda^{-1} \right)   , \label{eq:a0_for_Gau} \\ 
    r_0 &  = - \frac{2}{\LAMtwo} \left(\frac{1}{a_0 \LAMtwo} - \frac{2}{\sqrt{\pi}} \right) . \label{eq:r0_for_Gau} 
\end{align}

The two-body scattering $T$-matrix at general energy $e_2$, 
%
%\begin{widetext}
\begin{align}
    \tau^{-1}(e_2) & =\lambda^{-1} - \langle g | G_0^{(2b)}(e_2) | g \rangle \\  
    \langle g | G_0^{(2b)}(e_2) | g \rangle &  = (4\pi)\left(\frac{C}{\LAMtwo}\right)^2 (-\LAMtwo \mu_1) \notag \\
    & \times \left.\bigl[\sqrt{\pi} - \pi \sqrt{D}\, e^{D}\, \erfc\bigl(\sqrt{D}\bigr)\bigr] \right\vert_{D = - \frac{2\mu_1 e_2^+}{\LAMtwo^2}} 
\end{align}
%\end{widetext}
%
Note $G_0^{(2b)}$ is the two-body free Green's function here. 
\begin{align}
    e_2^+ & \equiv e_2 + i 0^+  , \\ 
    \erfc\left(\sqrt{D}\right) & = 1 - \erf\left(- i \frac{k}{\LAMtwo}\right) = 1 + i \erfi\left(\frac{k}{\LAMtwo}\right) . 
\end{align}  
The definitions of the error functions can be found in Ref.~\cite{NIST:DLMF}.
Based on the expressions for the $T$-matrix, we can recover the relationships between effective range parameters and $\LAMtwo, \lambda$ as just discussed above.  

To get the dimer bound state properly normalized, the $\bsren$ parameter is determined through ($B$ as the dimer's binding energy)
\begin{align}
  \frac{1}{\bsren^2} & = \langle g | \left[G_0^{(2b)}( - B)\right]^2 | g \rangle \notag \\
  &  = (4\pi)\left(\frac{C}{\LAMtwo}\right)^2 \left(\frac{\mu_1}{\LAMtwo}\right) \left(2\mu_1\right) \Bigl[-\sqrt{\pi} 
   \notag \\
 & \null + \frac{\pi}{2 \sqrt{D_B}} (1+2 D_B) \, e^{D_B}\, \erfc(\sqrt{D_B})\Bigr] \Bigr\vert_{D_B = \frac{2\mu_1 B}{\LAMtwo^2}} 
\end{align}

For the three-body potential, its form factor in coordinate space is  
\begin{widetext}
\begin{align}
    \langle \myvec{R}_1, \myvec{r}_1 | g_4 \rangle  & = \int d\myvec{P}_1 d\myvec{q}_1 \frac{e^{i\left(\myvec{P}_1\myvec{R}_1 + \myvec{q}_1\myvec{r}_1\right)}}{(2\pi)^3} \frac{1}{\sqrt{M \Lambda_4^4}} \exp\left[-\frac{M E_4}{2 \Lambda_4^2}\right]   =  \frac{\Lambda_4^4}{\sqrt{M }} \frac{2 \sin(2\phi)}{3 R_1 r_1} L^2 \exp\left(-\frac{\Lambda_4^2 L^2}{2}\right) \  \text{with} \notag  \\ 
  K^2 & = ME_4 = q_1^2 + \frac{3}{4}P_1^2  ,  \quad  L^2 = \frac{4}{3} R_1^2 + r_1^2 \quad \text{and} \quad \sin\phi \equiv \frac{r_1}{L}  
\end{align}

When solving Eq.~\eqref{eq:UcoupledSepW3bodyID}, we use the following analytical results. 
\begin{align}
    Z_{21}(\myvec{P}_2, \myvec{P}_1) & = \frac{M\bsren^2}{M(E+i\epsilon)-  P_1^2 - P_2^2  -  {P}_1 {P}_2 x } \left(\frac{C}{\LAMtwo}\right)^2\exp\left[-\frac{1}{\LAMtwo^2}\left(\frac{5}{8}P_1^2 +\frac{5}{8}P_2^2 + P_1 P_2 x  \right)\right]  \label{eq:Z21gau} \\ 
    Z_{21}( {P}_2,  {P}_1) & = \frac{1}{2} \int_{-1}^{1} d x Z_{21}(\myvec{P}_2, \myvec{P}_1) \notag \\ 
    & = (-)\frac{\bsren^2}{2}\frac{M}{P_1 P_2}\left(\frac{C}{\LAMtwo}\right)^2\exp\left[-\frac{5}{8\LAMtwo^2}\left(P_1^2 + P_2^2\right) \right] \left\{e^{\tilde{D}} \left[\Gamma\left(0, \tilde{D}-\tilde{d}\right)- \Gamma\left(0, \tilde{D}+\tilde{d}\right) \right]\right\}  \notag \\ 
\text{with}\quad    &  \tilde{D} = \frac{P_1^2 + P_2^2 - ME^+}{\LAMtwo^2}, \quad \tilde{d}=\frac{P_1 P_2}{\LAMtwo^2}, \quad \text{and} \quad x \equiv \frac{\myvec{P}_1\cdot \myvec{P}_2}{P_1 P_2}  
\end{align}    
The definition of the incomplete Gamma function, $\Gamma(0, x)$ can be found in \cite[Eq.~8.2.2]{NIST:DLMF}. 

In addition, 
\begin{align}
    Z_{14}(\myvec{P})  & =   {}_1\langle \hat{g}, \myvec{P} | G_0(E) | g_4 \rangle 
     = - 4\pi  M \bsren\, \frac{C}{\LAMtwo} \frac{1}{\sqrt{M}\Lambda_4^2} \exp\left(-\frac{3 P^2}{8\Lambda_4^2}\right)\, \frac{1}{2\sqrt{t}}\left[\sqrt{\pi} - 
    \pi e^{D\,t} \sqrt{D\,t}\erfc\left(\sqrt{D\,t}\right) \right] \\ 
   \text{with}\ t & \equiv \frac{1}{2}\left(\frac{1}{\LAMtwo^2} +\frac{1}{\Lambda_4^2} \right) \quad \text{and} \quad  D\equiv \frac{3}{4} P^2 - M E^+ 
\end{align}
and 
\begin{align}
    \tau_4^{-1} 
    = & \lambda_4^{-1} + \frac{4\pi^3}{3\sqrt{3}} \frac{M^2}{\Lambda_4^4} \left[\frac{1}{t^2} + \frac{E^+}{t} + e^{- t E } E^2 \Gamma\left(0, - t E^+\right) \right], \text{with} \ t \equiv \frac{M}{\Lambda_4^2} . 
\end{align}
Note $\Gamma\left(0, - t E^+\right) = E_1(-t E^+)$ \cite[Eq.~8.19.1]{NIST:DLMF} 
%is \texttt{$ss.exp1(-t E^+)$} in SciPython,
which cares about the $i 0^+$ in $E^+ \equiv E + i 0^+$. 

\end{widetext}

\section{Computing $\delta U$ for emulators in three-dimensional parameter space} 

\setcounter{equation}{14}

In order to compute the overlaps integrals in Eq.~\eqref{eq:deltaU_varying_V2body} that involve the two-body potentials, we need to apply the symmetrizer operator on a particular FC, e.g., $\psifd_1$, to get the full wave function $\psifull$.  

The symmetrizer is defined as $1 + \perm  =  1 +  P_{12}P_{23} + P_{13} P_{23}$, with $P_{ij}$ exchanging particle $i$ and $j$. See Eq.~(3.344) and its discussions in Ref.~\cite{Glockle:1983}, but note the notation differences, $P \leftrightarrow q_{\text{Ref.\,\cite{Glockle:1983}}}$, $q \leftrightarrow p_{\text{Ref.\,\cite{Glockle:1983}}}$.    

In momentum space, it can be shown for the s-waves, 
\begin{widetext}
\begin{align}
    & {}_1\langle P, q, (\ell_p, \ell_q) L | \perm | P', q', (\ell_p', \ell_q') L' \rangle_1 
    = \int_{-1}^1 d x \frac{\delta(q - \Pi)}{q^2 } \frac{\delta(q' - \Pi')}{q'^2 } \quad \text{when}\ \ell_p = \ell_q = \ell_p' = \ell_q' = L = L' = 0 \  \text{with} \notag \\ 
   &  \Pi   \equiv \sqrt{\frac{P^2}{4} +P'^2 + P P' x } ,\quad \text{and}  \quad    \Pi' \equiv \sqrt{\frac{P'^2}{4} +P^2 + P P' x }  .  \label{eq:Phat-in-p}  
\end{align}
$\ell_p$, $\ell_q$ and $L$ are the angular momenta between particle and dimer, between the particles within the dimer, and the total. Here we restrict all of them to s waves.  

Similarly  in coordinate space, 
\begin{align}
     & {}_1\langle R, r, (\ell_R, \ell_r) L | \perm | R', r', (\ell_R', \ell_r') L' \rangle_1   = \left(\frac{4}{3}\right)^3 \int_{-1}^1 d x \frac{\delta(r - \Pi)}{r^2 } \frac{\delta(r' - \Pi')}{r'^2 } \quad \text{when}\ \ell_R = \ell_r = \ell_R' = \ell_r' = L = L' = 0 \ \text{with} \notag \\ 
    & \Pi   \equiv \left(\frac{4}{3}\right) \sqrt{\frac{R^2}{4} +R'^2 + R R' x } , \quad \text{and} \quad  
    \Pi'   \equiv \left(\frac{4}{3}\right) \sqrt{\frac{R'^2}{4} +R^2 + R R' x } . \label{eq:Phat-in-r}   
\end{align}

Based on Eq.~\eqref{eq:Phat-in-p} and~\eqref{eq:Phat-in-r}, we can get the following formula which can allow us to compute the other two FCs based on say $\psifd_1$. (In the following as in the main texts,  $ | P, q \rangle_1$ is the state  projected to the total $L = 0$ and $\ell_p = \ell_q = 0$ partial wave, since we only need these states in this work.) I.e.,
\begin{align}
    |\psifull \rangle & = \left(1 + \perm\right)  | \psifd_1 \rangle  , \\ 
    _1\langle P, q|\psifull \rangle  & = \psifd_1(P,q)   + \int d P' d q' P'^2 q'^2  {}_1\langle P, q| \perm | P', q' \rangle_1 \psifd_1(P',q') \notag  \\ 
    & = \psifd_1(P,q)  + \frac{2}{P q} \int_{|\frac{P}{2} -q |}^{\frac{P}{2} + q } d P' P' \psifd_1(P', \Pi')  \quad
  \text{with}\  \Pi'   \equiv \sqrt{q^2 + \frac{3}{4} (P^2 - P'^2) } . \label{eq:fullWF_from_FD1_in_p}
\end{align}
Similarly, in coordinate  space, 
\begin{align}
    _1\langle R, r|\psifull \rangle  & = \psifd_1(R,r)  + \int d R' d r' R'^2 r'^2  {}_1\langle R, r| \perm | R', r' \rangle_1 \psifd_1(R',r')  \notag  \\ 
    & = \psifd_1(R,r)    + \frac{4}{3} \frac{2}{R r} \int_{|\frac{R}{2} -\frac{3}{4} r |}^{\frac{R}{2} + \frac{3}{4} r } d R' R' \psifd_1(R', \Pi')   \quad 
\text{with} \    \Pi'  \equiv \sqrt{r^2 + \frac{4}{3} (R^2 - R'^2) } \label{eq:fullWF_from_FD1_in_r} 
\end{align}
\end{widetext}

It is worth emphasizing that Eq.~\eqref{eq:fullWF_from_FD1_in_p} and~\eqref{eq:fullWF_from_FD1_in_r} apply for all the wave functions satisfying the symmetry properties of three identical bosons. I.e.,  $\psifull$ and $\psifd$ in these equations can be the asymptotic ones, the subtracted ones, and the full $\tilde{\psifull}$ and its FCs. 

In this work, Eq.~\eqref{eq:fullWF_from_FD1_in_p} is used to compute the full subtracted wave function, $\psifullsub$ from $\psifdsub_1$; the corresponding wave function in coordinate space is then computed via fast Fourier transformation. Meanwhile, Eq.~\eqref{eq:fullWF_from_FD1_in_r} is employed to calculate $\psifullasym$ in coordinate space, by using the asymptotic FC $\psifdasym$ which is known analytically. These wave functions can be plugged into Eq.~\eqref{eq:deltaU_varying_V2body} to evaluate the two-body potential related pieces in $\delta U$. 

Meanwhile, the three-body force's contributions in the $\delta U$ calculations are much simpler, since the form factor in $V_4$ is invariant with permuting the particle: $ \langle \psifull' | g_4 \rangle \langle g_4 |\psifull  \rangle = \langle \psifd_1 ' | (1+\perm)|  g_4 \rangle \langle g_4| (1+\perm) | \psifd_1 \rangle = 9 \langle \psifd_1 ' | g_4 \rangle \langle g_4| \psifd_1 \rangle$. The overlap integrals can be carried out either in the momentum or in  coordinate space.

\section{Results for the nuclear case}

\begin{figure*}
    \centering
    \includegraphics{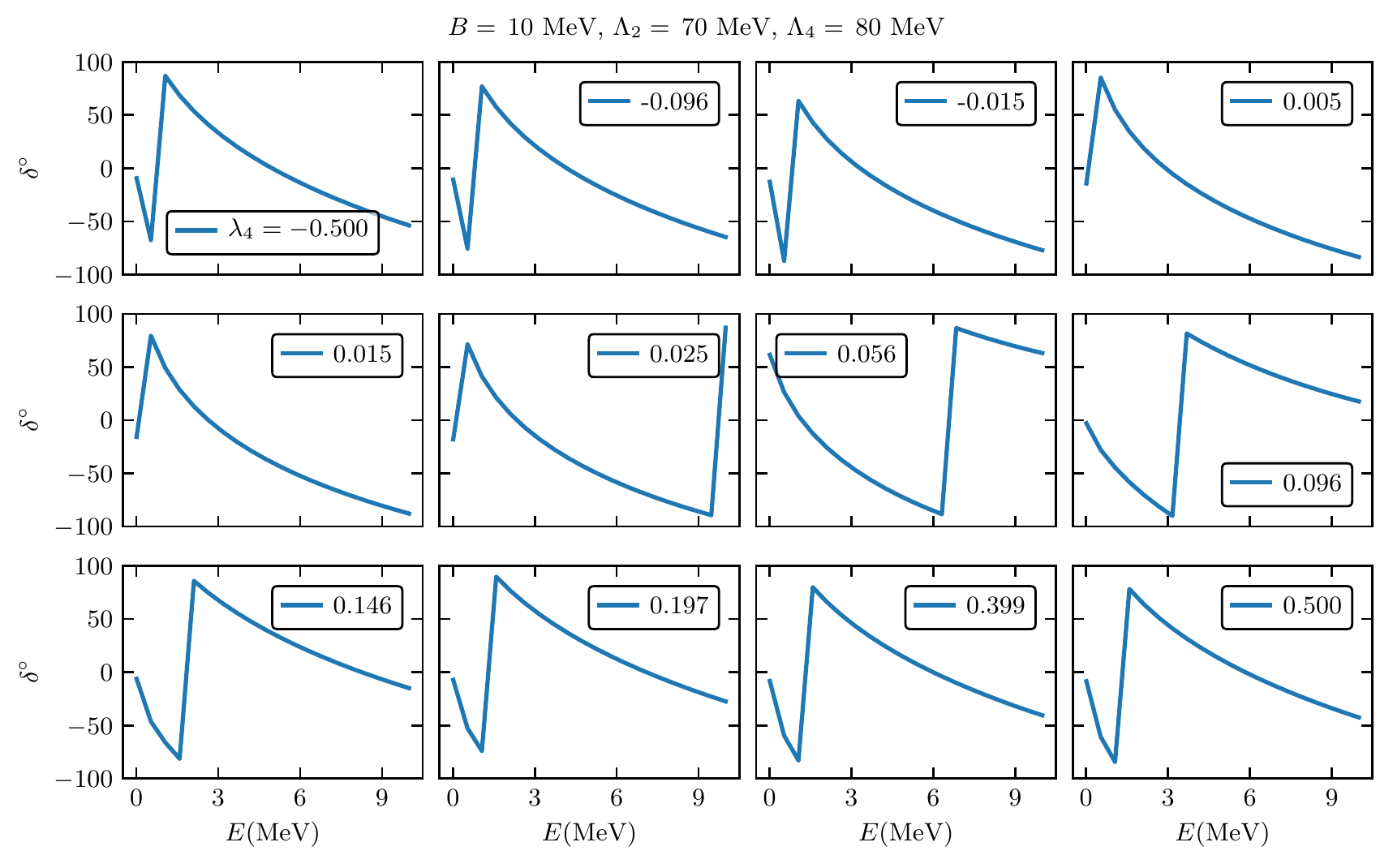}
    \caption{Particle-dimer s-wave scattering phase shift vs. scattering energy $E$ for the nuclear case with $B=10$ MeV, $\LAMtwo=70$ MeV and $\Lambda_4 = 80$ MeV. }
    \label{fig:3body_B_10_LAM2_70_LAM4_80_gau}
\end{figure*}

\begin{figure*}
    \centering
    \includegraphics{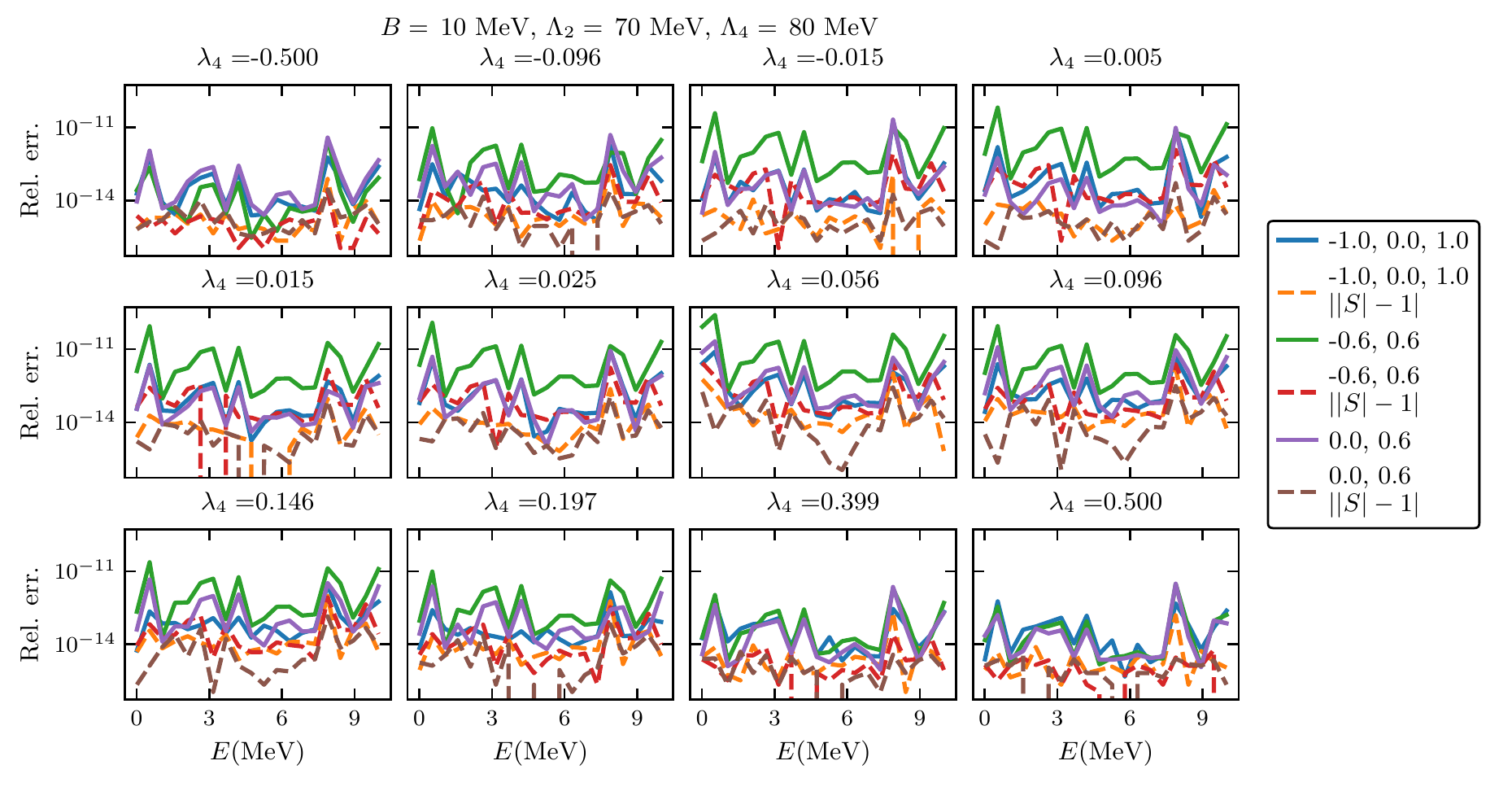}
    \caption{
    The EC emulator errors for various $\lambda_4$ values (shown at the top of each panel), for which the exact phase shifts have been plotted in Fig.~\ref{fig:3body_B_10_LAM2_70_LAM4_80_gau}. Three different training sets have been used for constructing the EC emulators. Their $\lambda_4$ values are shown in the legends. The solid curves are the relative errors of the $S$ matrix, while the dashed curves plot the unitarity violation.}
    \label{fig:EmuErr_3body_B_10_LAM2_70_LAM4_80_gau}
\end{figure*}

\begin{figure*}
    \centering
    \includegraphics{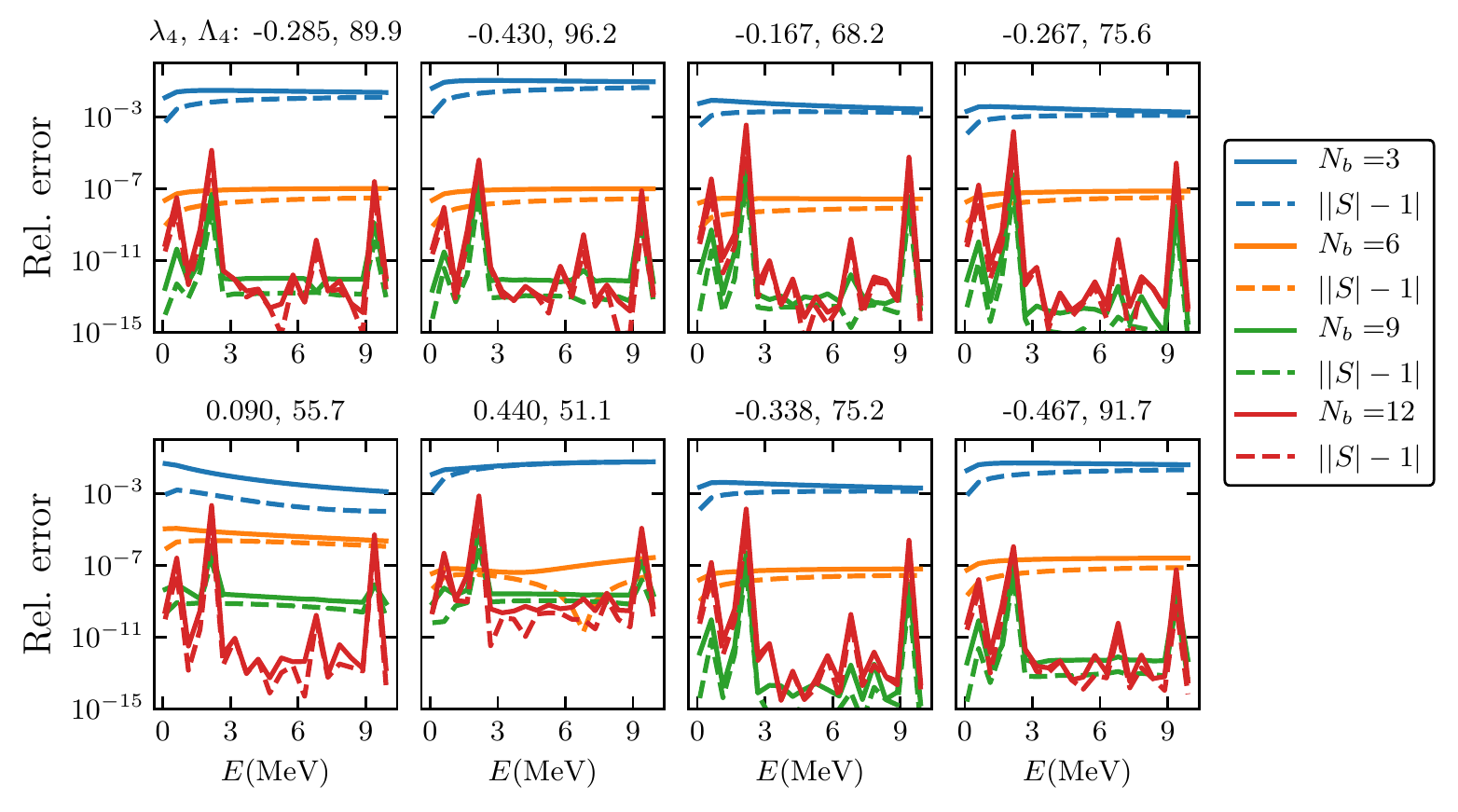}
    \caption{
    The EC emulator errors for a few test points in the two-dimensional parameter space for the nuclear case with $\LAMtwo = 75$ MeV.  The solid and dashed curves in the same color are the $S$ matrix's relative errors and the unitarity violation errors respectively with a fixed $N_b$ (see the legend).      }
    \label{fig:Emulator_RelError_Expamples_B_10_LAM2_75_Gau}
\end{figure*}

\begin{figure*}
    \centering
    \includegraphics{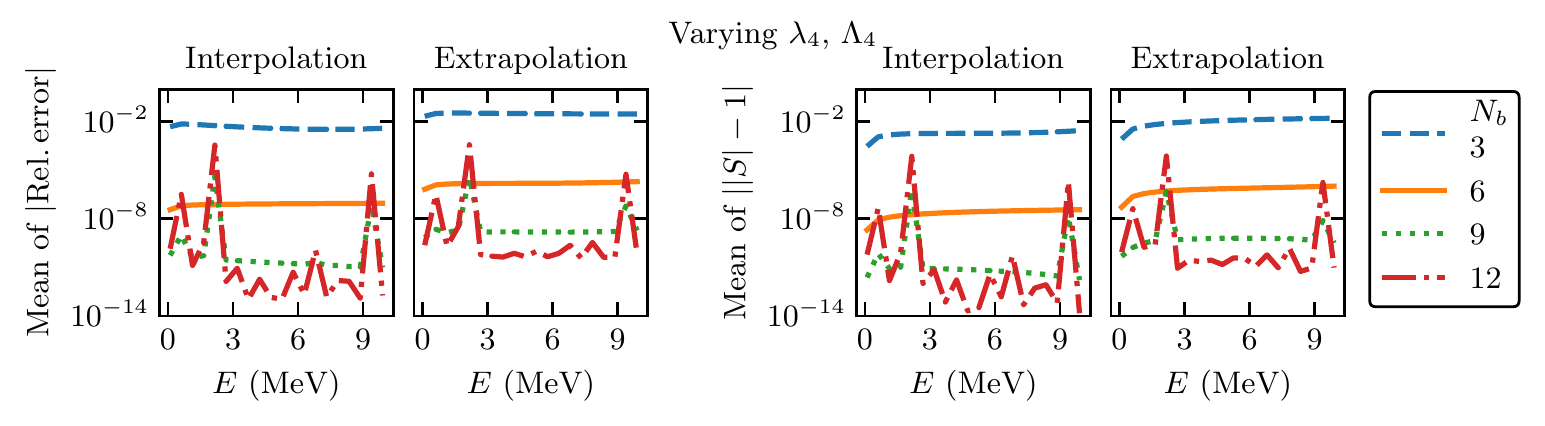}
    \caption{The mean values of the EC emulator errors in the two-dimensional parameter space for the nuclear case with the same training sets as in Fig.~\ref{fig:Emulator_RelError_Expamples_B_10_LAM2_75_Gau} (see the legend). $\LAMtwo$ is fixed to 75 MeV.  The left two panels are the $S$'s relative errors. The right two are the unitarity violation errors. In total, 200 points have been uniformly sampled  and tested.  The number of interpolation points grows with $N_b$. For $N_b = 12$, the interpolation and extrapolations have similar number of points. }   \label{fig:Emulator_RelError_B_10_LAM2_75_Gau}
\end{figure*}

\begin{figure*}
    \centering
    \includegraphics[width=\textwidth]{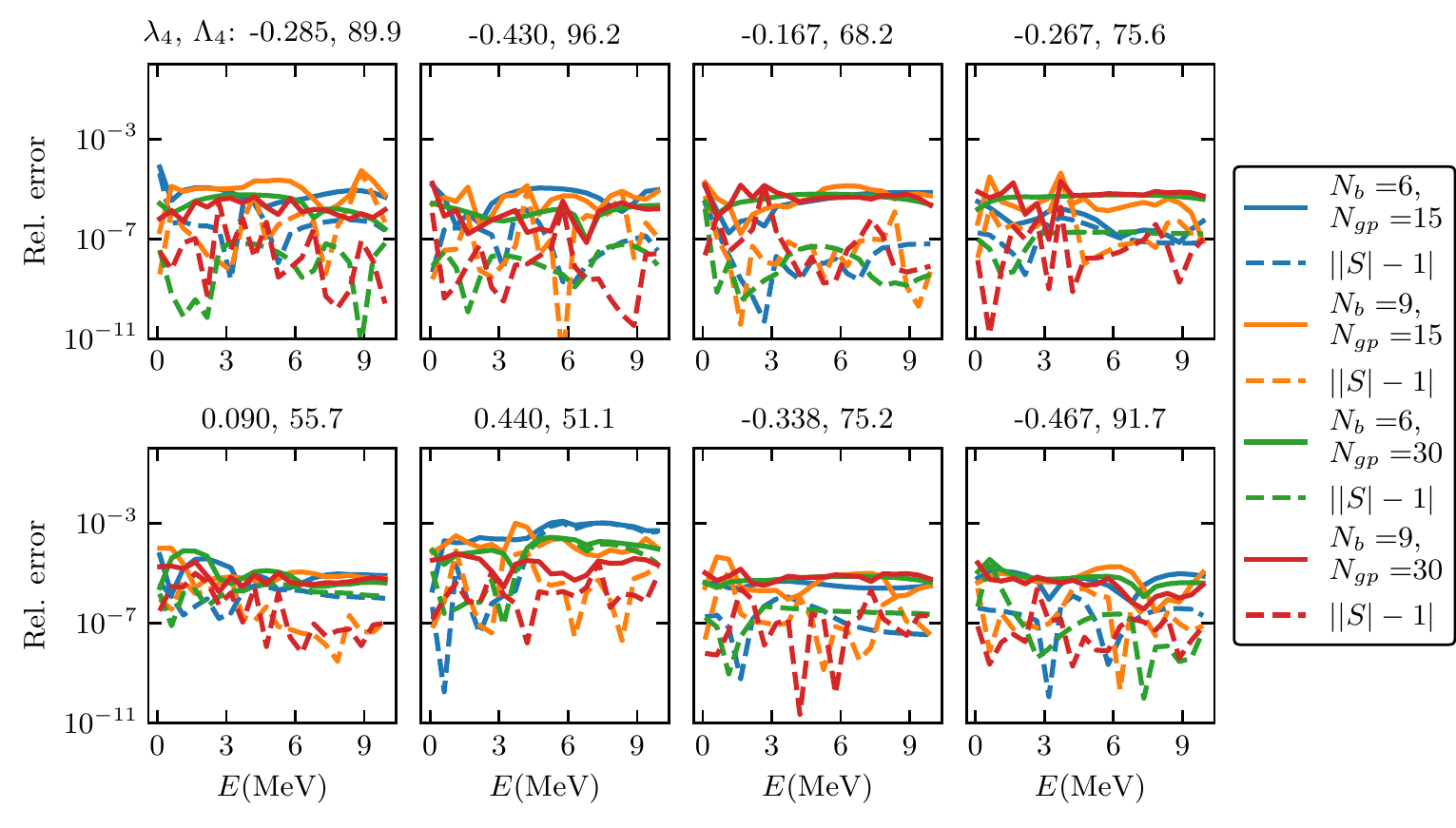}
    \caption{
    The GP-EC emulator errors for a few test points (the same as those in Fig.~\ref{fig:Emulator_RelError_Expamples_B_10_LAM2_75_Gau}) in the two-dimensional parameter space  for the nuclear case. $\LAMtwo = 75$ MeV. The emulators implement GP interpolation within the EC emulators.  The solid and dashed curves in the same color are the $S$  errors and the unitarity violation errors respectively with a pair of fixed $N_b$ and $N_{gp}$ (see the legend).
    }
    \label{fig:EmulatorInEmulator_RelError_Expamples_B_10_LAM2_75_Gau}
\end{figure*}

\begin{figure*}
    \centering
    \includegraphics[width=\textwidth]{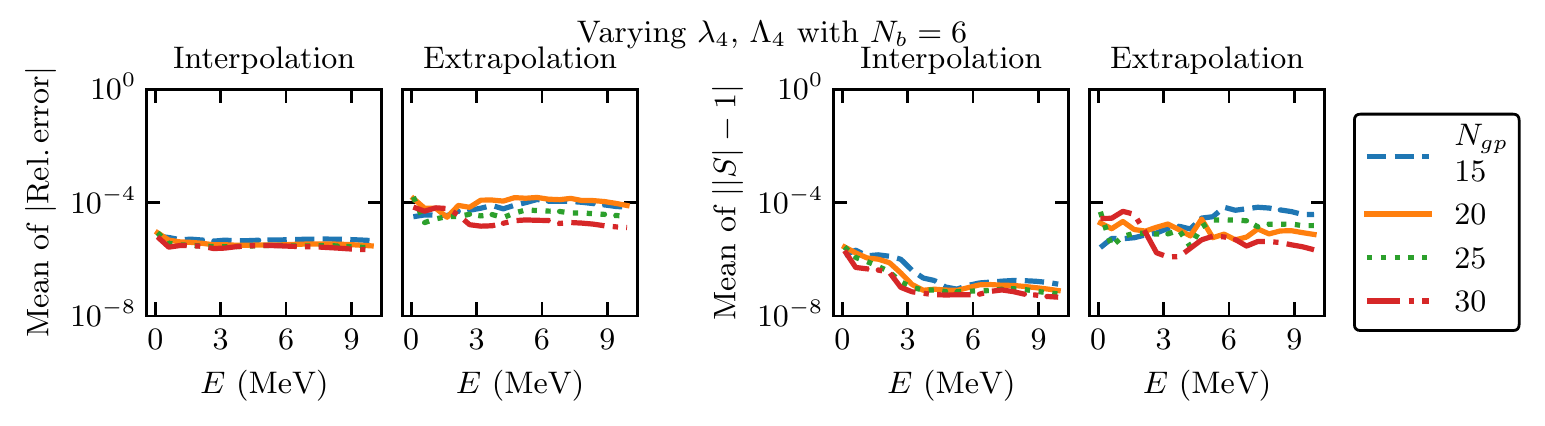}
    \caption{The mean of the GP-EC emulator  errors in the two-dimensional parameter space for the nuclear case based on the same 200 points tested in Fig.~\ref{fig:Emulator_RelError_B_10_LAM2_75_Gau}, with $N_b = 6$ and four different $N_{gp}$. Interpolation vs extrapolation are defined in the same way as in Figs.~\ref{fig:Emulator_RelError_B_10_LAM2_75_Gau} and~\ref{fig:Emulator_RelError_B_2_LAM2_200_Gau}.}
    \label{fig:EmulatorInEmulator_RelError_B_10_LAM2_75_Nb_6_scalarGP_Gau}
\end{figure*}

\begin{figure*}
    \centering
    \includegraphics[width=\textwidth]{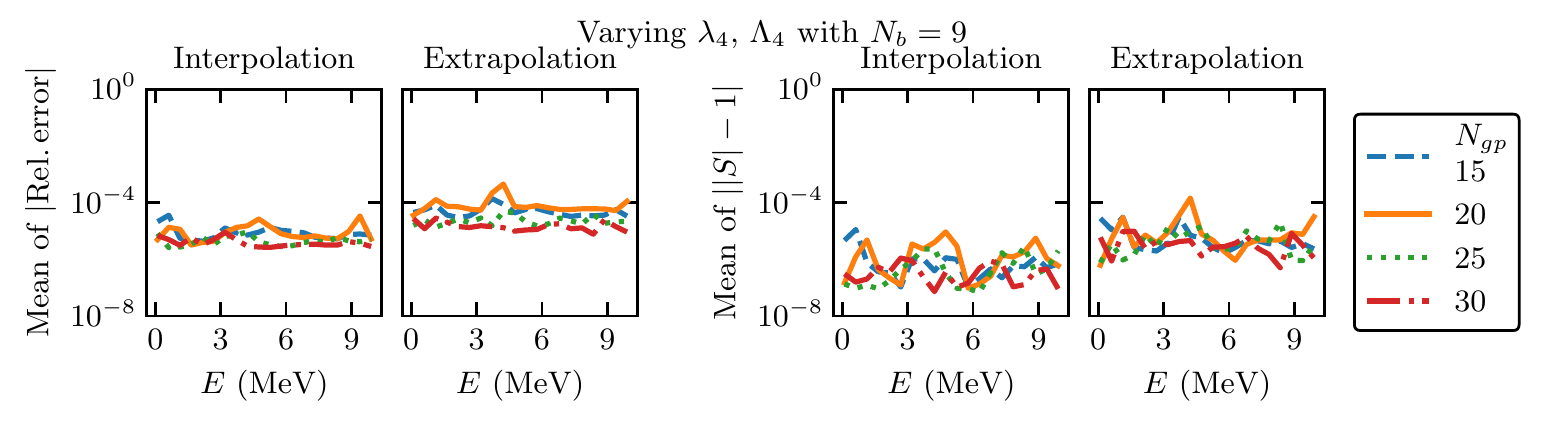}
    \caption{The same as Fig.~\ref{fig:EmulatorInEmulator_RelError_B_10_LAM2_75_Nb_6_scalarGP_Gau} but with a larger $N_b = 9$. }
    \label{fig:EmulatorInEmulator_RelError_B_10_LAM2_75_Nb_9_scalarGP_Gau}
\end{figure*}

\begin{figure*}
    \centering
    \includegraphics[width=\textwidth]{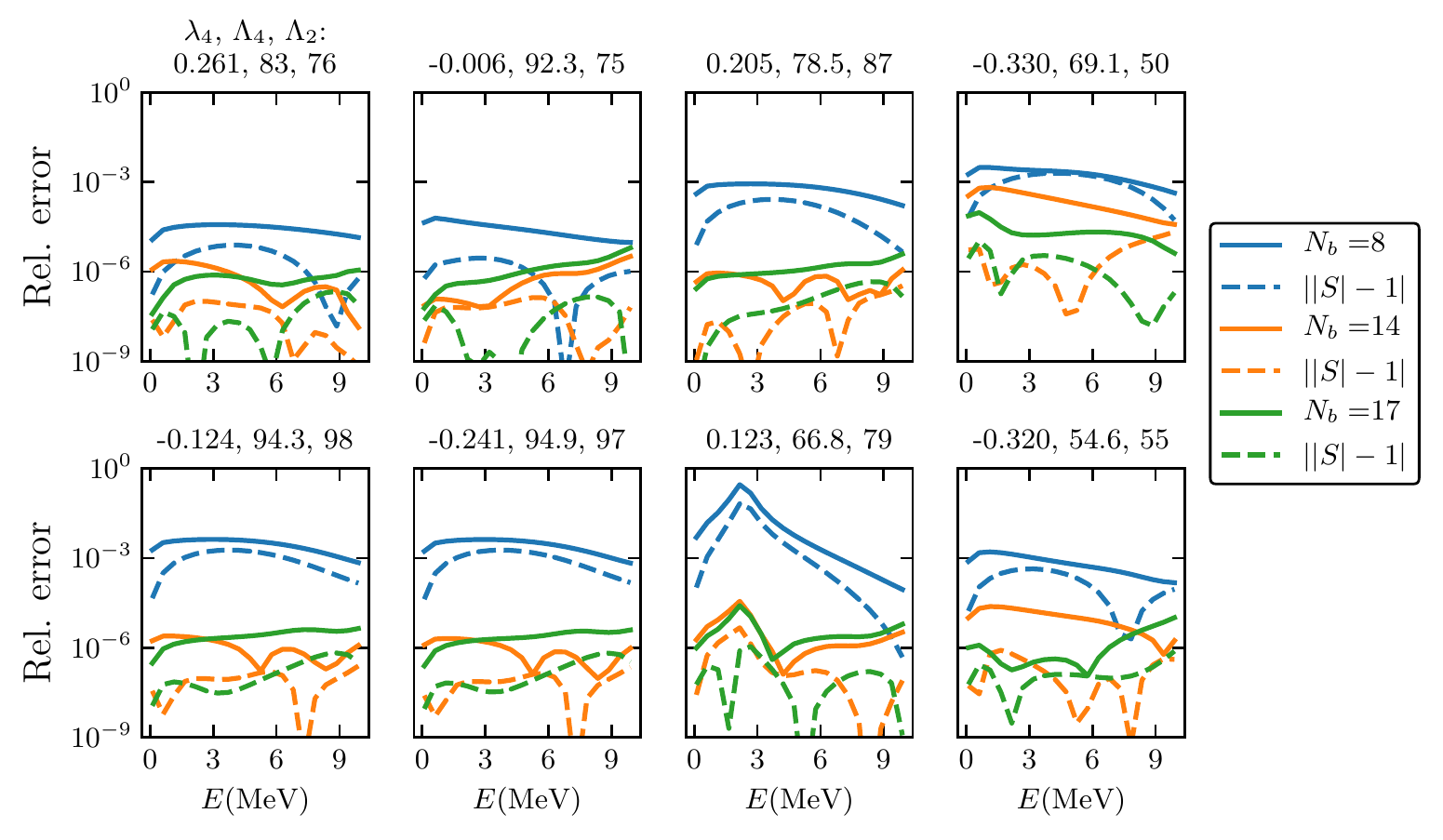}
    \caption{
    The EC emulator errors for a few test points in the three-dimensional parameter space for the nuclear case with increasing number of $N_b$.
The solid and dashed curves in the same color are the $S$ matrix's relative errors and the unitarity violation errors respectively with a fixed $N_b$ (see the legend). }
    \label{fig:Emulator_RelError_Expamples_B_10_Gau}
\end{figure*}

\begin{figure*}
    \centering
    \includegraphics[width=\textwidth]{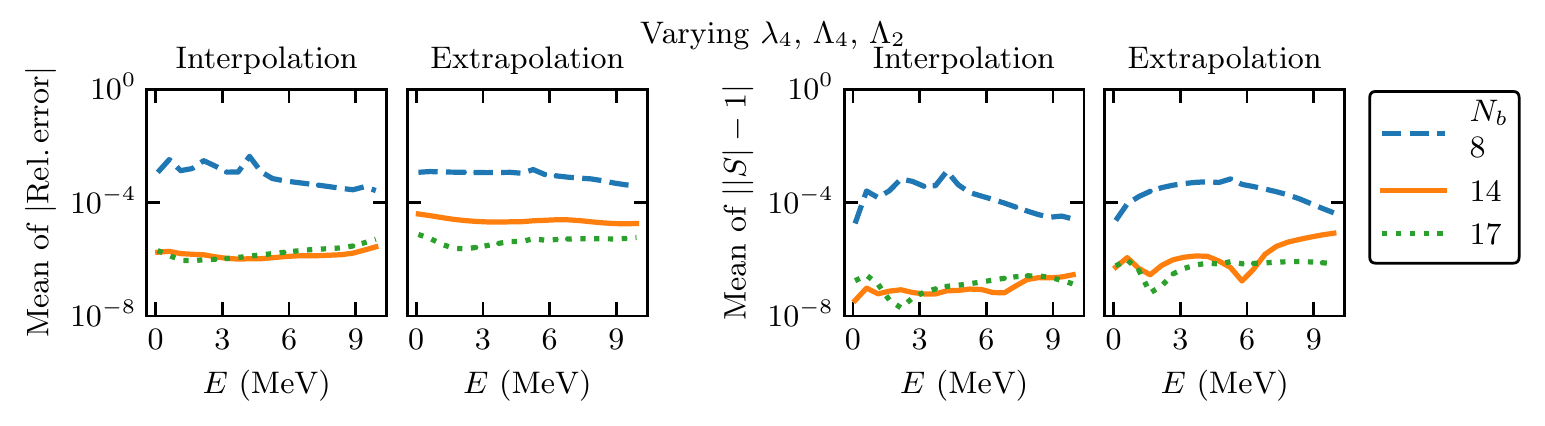}
    \caption{The mean of the EC emulator errors  in the three-dimensional parameter space for the nuclear case based on a sample of 1000 test points uniformly sampled in the parameter space. The same three different training sets as in Fig.~\ref{fig:Emulator_RelError_Expamples_B_10_Gau} are used.} 
    \label{fig:Emulator_RelError_B_10_Gau}
\end{figure*}

\begin{figure*}
    \centering
    \includegraphics[width=\textwidth]{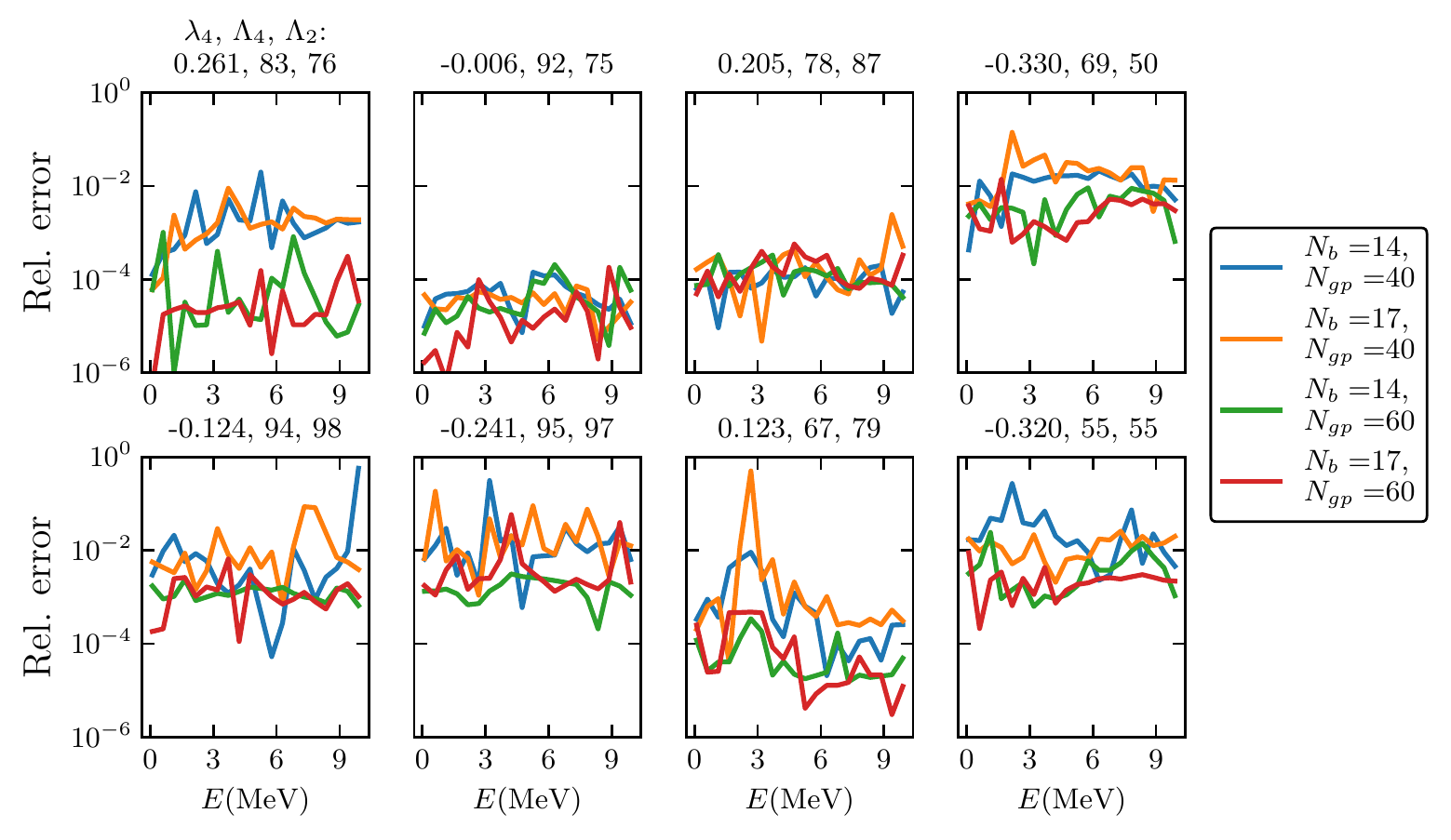}
    \caption{The $S$'s relative errors of the GP-EC emulator for a few test points in the three-dimensional parameter space---the same as those in Fig.~\ref{fig:Emulator_RelError_Expamples_B_10_Gau}---for the nuclear case. Four  training sets with different combinations of $N_b$ and $N_{gp}$ have been shown.}
    \label{fig:EmulatorInEmulator_RelError_Expamples_B_10_Gau}
\end{figure*}

\begin{figure*}
    \centering
    \includegraphics[width=\textwidth]{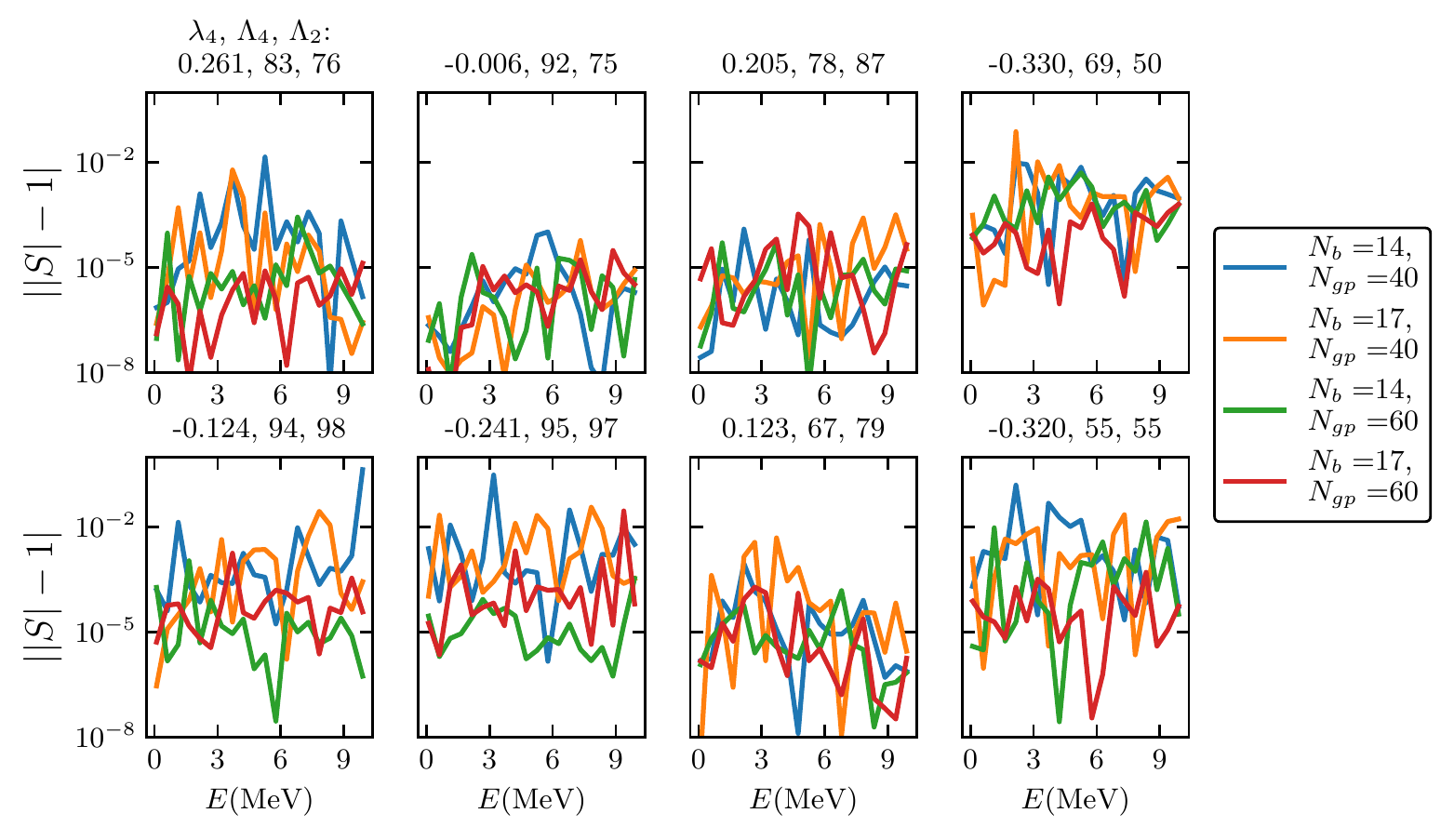}
    \caption{The same as Fig.~\ref{fig:EmulatorInEmulator_RelError_Expamples_B_10_Gau} but for the unitarity violation errors of the GP-EC emulators.}
    \label{fig:EmulatorInEmulator_UnitarityError_Expamples_B_10_Gau}
\end{figure*}

\begin{figure*}
    \centering
    \includegraphics[width=\textwidth]{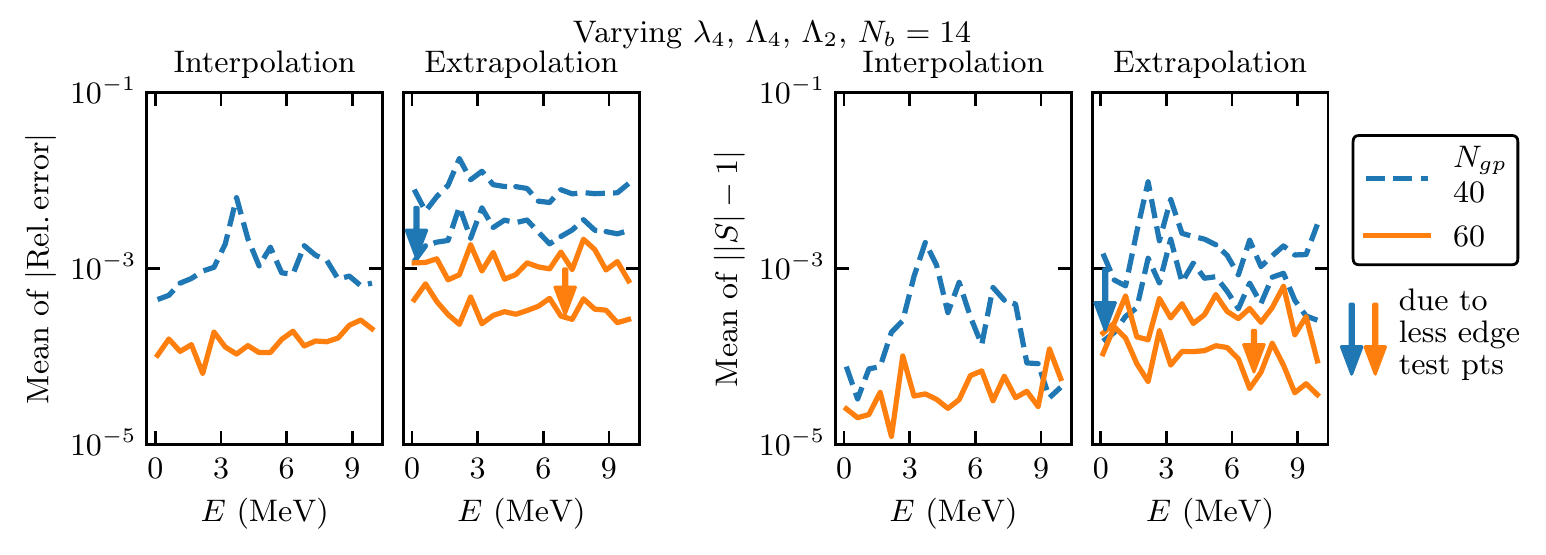}
    \caption{
    The mean of the GP-EC emulator errors vs $E$  (left two for the $S$ relative errors and the right two for the unitarity violation errors) in the three-dimensionalparameter space for the nuclear case, with $N_b = 14$ and different $N_{gp}$.  Two different calculations are done: the first based on the same 1000 test point sample as used in Fig.~\ref{fig:Emulator_RelError_B_10_Gau} with $\LAMtwo$ and $\Lambda_4$ in $[50, 100]$ MeV (266 interpolation and 734 extrapolation points) and the second with $\LAMtwo$ and $\Lambda_4$ in a smaller $[55, 95]$ MeV range (260 interpolation and 380 extrapolation points). As can be seen, reducing the number of the edge test points reduces the errors in the extrapolation sample to a similar level as those in the interpolation sample. }
    \label{fig:EmulatorInEmulator_RelError_B_10_Nb_14_scalarGP_Gau}
\end{figure*}

\begin{figure*}
    \centering
    \includegraphics[width=\textwidth]{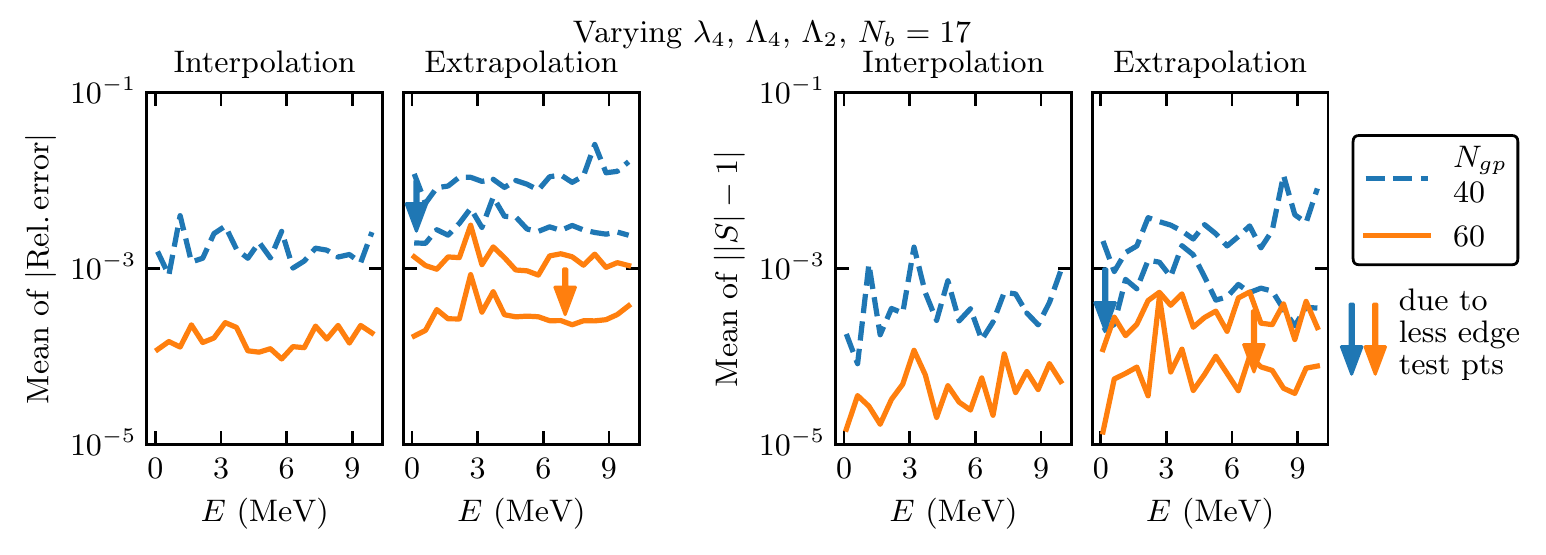}
    \caption{The same as in Fig.~\ref{fig:EmulatorInEmulator_RelError_B_10_Nb_14_scalarGP_Gau} but with $N_b = 17$.
   For the full sample calculation with $\LAMtwo$ and $\Lambda_4$ in $[50, 100]$ MeV, there are 340 interpolation and 660  extrapolation points, while for the smaller range calculations 328 interpolation and 312 extrapolation points. Similar to Fig.~\ref{fig:EmulatorInEmulator_RelError_B_10_Nb_14_scalarGP_Gau}, reducing the number of the edge test points reduces the errors in the extrapolation sample to the same level as those in the interpolation sample. }
    \label{fig:EmulatorInEmulator_RelError_B_10_Nb_17_scalarGP_Gau}
\end{figure*}

We also explore a second parameter set that mimics the nuclear case with binding energy 10 MeV, with $ |\lambda_4| \leq 0.5$ and $\LAMtwo$ and $\Lambda_4$ between $50$ and $100$ MeV. The figures in this section are in parallel to those for the nucleon case in the main text. The results in both cases are qualitatively similar. 

\end{document}